\pdfoutput=1
\documentclass[12pt]{scrartcl}
\usepackage{hyphenat}
\usepackage[T1]{fontenc}         
\usepackage[utf8]{inputenc}
\usepackage[british]{babel}
\areaset{158mm}{238mm}
\usepackage{setspace}
\usepackage[all]{xypic}
\usepackage{scrlayer-scrpage}
\usepackage{tikz}
\usetikzlibrary{cd}

\usepackage{extarrows}
\usetikzlibrary{arrows}

\usepackage{amsmath}
\usepackage{mathtools}
\usepackage{mathrsfs}
\usepackage{physics}

\usepackage{amssymb}
\usepackage[mathscr]{eucal}
\usepackage{atbegshi}
\usepackage[unicode=true,linktoc=all]{hyperref}
\usepackage{cite}
\usepackage{bm}
\usepackage{amsthm}
\usepackage{slashed}
\usepackage{tikz,pgf}
\usetikzlibrary{shapes}
\usetikzlibrary{calc}
\usetikzlibrary{decorations.pathmorphing}
\usetikzlibrary{decorations.pathreplacing,shapes.misc}
\usetikzlibrary{positioning}
\usetikzlibrary{decorations.markings}
\tikzset{->-/.style={decoration={
  markings,
  mark=at position .5 with {\arrow{>}}},postaction={decorate}}}
\tikzset{-<-/.style={decoration={
  markings,
  mark=at position .5 with {\arrow{<}}},postaction={decorate}}}

\usepackage{xcolor}
  \definecolor{rblue}{RGB}{81, 49, 193}
  \definecolor{rorange}{RGB}{255, 147, 40}
  \definecolor{rgreen}{RGB}{176, 233, 0}

\renewcommand{\tilde}{\widetilde}

\usetikzlibrary{decorations.pathmorphing}

\usepackage{lscape}

\newcommand{\bea}{\begin{equation}\begin{aligned}}
\newcommand{\eea}{\end{aligned}\end{equation}}
\newcommand{\beas}{\begin{equation*}\begin{aligned}}
\newcommand{\eeas}{\end{aligned}\end{equation*}}
\newcommand{\be}{\begin{equation}}
\newcommand{\ee}{\end{equation}}

\newcommand{\mathbbm}[1]{\mathds{#1}}

\def\p{\pi}

\def\ZZ{\mathcal{Z}_{\Sigma_{g},\mathcal{L}}(X)}

\DeclareMathOperator{\sign}{sign}

\tikzset{pics/.cd,
handle/.style={code={
\draw[fill=gray!20]  (-2,0) coordinate (-left) 
to [out=260, in=60] (-3,-2) 
to [out=240, in=110] (-3,-4) 
to [out=290,in=180] (0,-6) 
to [out=0,in=250] (3,-4) 
to [out=70,in=300] (3,-2) 
to [out=120,in=280] (2,0)  coordinate (-right);
\pgfgettransformentries{\tmpa}{\tmpb}{\tmp}{\tmp}{\tmp}{\tmp}
\pgfmathsetmacro{\myrot}{-atan2(\tmpb,\tmpa)}
\draw[rotate around={\myrot:(0,-2.5)}] (-1.2,-2.4) to[bend right]  (1.2,-2.4);
\draw[fill=white,rotate around={\myrot:(0,-2.5)}] (-1,-2.5) to[bend right] (1,-2.5) 
to[bend right] (-1,-2.5);
}}}

\numberwithin{equation}{section}

\vspace{0cm}

\title{\vspace{-1cm} Non-invertible Symmetries of Class $\cS$ Theories}

\author{Vladimir Bashmakov,$^1$ Michele Del Zotto,$^{1,2}$ \\ Azeem Hasan,$^{2}$ and Justin Kaidi$^{3,4}$
\\[1cm]
	\small\slshape$1$ Department of Physics and Astronomy, Uppsala University,  \\[-0.2cm] 
	\small\slshape Box 516, SE-75120 Uppsala, Sweden\\
	\small\slshape$2$ Mathematics Institute, Uppsala University,  \\[-0.2cm] 
	\small\slshape Box 480, SE-75106 Uppsala, Sweden\\
	\small\slshape$3$ Department of Physics,\\[-0.2cm]
    \small\slshape University of Washington, Seattle, WA, 98195, USA\\
	\small\slshape$4$ Kavli Institute for the Physics and Mathematics of the Universe,\\[-0.2cm]
	\small\slshape University of Tokyo, Kashiwa, Chiba 277-8583, Japan
	}
\date{}

\def\sfA{{\mathsf A}}
\def\sfB{{\mathsf B}}

\def\sfS{{\mathsf S}}
\def\sfT{{\mathsf T}}
\def\mS{{\mathsf S}}
\def\mT{{\mathsf T}}
\def\ZZ{{\mathbb Z}}
\def\RR{{\mathbb R}}
\def\NN{{\mathbb N}}
\def\CC{{\mathbb C}}

\def\cL{{\cal L}}

\def\cP{{\cal P}}

\def\cB{{\cal B}}
\def\cT{{\cal T}}

\def\cC{{\cal C}}
\def\cS{{\cal S}}

\def\cN{{\cal N}}

\def\half{ {1\over 2}}
\def\p{\partial}
\def\tauYM{\tau_{\mathrm{YM}}}
\def\no{\nonumber}

\newcommand{\beaa}{\begin{eqnarray}}
\newcommand{\eeaa}{\end{eqnarray}}
\usepackage{dsfont}

\definecolor{dgreen}{rgb}{0, 0.55, 0}
\usepackage{pifont}
\usepackage{amssymb}
\newcommand{\cmark}{\ding{51}}%
\newcommand{\xmark}{\ding{55}}%
\definecolor{GreenYellow}{cmyk}{0.6,0,1.,0}%%%PANTONE GREEN
\definecolor{Red}{cmyk}{0,1.,1.,0}%%%PANTONE RED
\newcommand{\red}[1]{\textcolor{Red}{#1}}
\newcommand{\green}[1]{\textcolor{GreenYellow}{#1}}

\newtheorem{theorem}{Theorem}

\newtheorem{proposition}{Proposition}
\usetikzlibrary{shapes.multipart}
\usetikzlibrary{decorations.pathreplacing,calligraphy}
\usetikzlibrary{decorations.pathmorphing}
\tikzset{snake it/.style={decorate, decoration=snake}}
\usepackage{ulem}
\definecolor{darkred}{rgb}{0.8,0.1,0.1}
\hypersetup{colorlinks=true, linkcolor=darkred, citecolor=dgreen, linktoc=page}

\begin{document}

\maketitle

\paragraph{\hspace{7cm}\large{Abstract}}
\vspace{-1cm}
\begin{abstract}

\noindent We study the non-invertible symmetries of class $\cS$ theories obtained by compactifying the type $\mathfrak{a}_{p-1}$ 6d (2,0) theory on a genus $g$ Riemann surface with no punctures. After setting up the general framework, we describe how such symmetries can be classified up to genus 5. Of central interest to us is the question of whether a non-invertible symmetry is \textit{intrinsic}, i.e. whether it can be related to an invertible symmetry by discrete gauging. We then describe the higher-dimensional origin of our results, and explain how the Anomaly and Symmetry TFTs, as well as $N$-ality defects, of class $\cS$ theories can be obtained from compactification of a 7d Chern-Simons theory. Interestingly, we find that the Symmetry TFT for theories with intrinsically non-invertible symmetries can only be obtained by coupling the 7d Chern-Simons theory to topological gravity.

\end{abstract}

\vfill{}
--------------------------

November 2022

PREPRINT UUITP 50/22
\thispagestyle{empty}

\newpage
\setcounter{tocdepth}{2}
\tableofcontents

\section{Introduction}
The study of generalized global symmetries  \cite{Gaiotto:2010be,Kapustin:2013qsa,Kapustin:2013uxa,Aharony:2013hda,Gaiotto:2014kfa,Sharpe:2015mja} is undergoing a surprising evolution. 
Recent progress has taken us beyond the paradigm of group-like symmetry to ``categorical'' symmetry, the structure of which is encoded in a fusion (higher-)category. A particularly interesting class of such symmetries are those which are ``non-invertible.'' Non-invertible symmetries have long been known in $(1+1)$-dimensions \cite{verlinde1988fusion,Petkova:2000ip,Fuchs:2002cm,Bhardwaj:2017xup,Chang:2018iay,Lin:2022dhv,Komargodski:2020mxz,Tachikawa:2017gyf, Frohlich:2004ef, Frohlich:2006ch, Frohlich:2009gb, Carqueville:2012dk, Brunner:2013xna, Huang:2021zvu, Thorngren:2019iar, Thorngren:2021yso, Lootens:2021tet, Huang:2021nvb, Inamura:2022lun}, but it was not until fairly recently that they were realized in higher-dimensional theories, with several different constructions now known \cite{Kaidi:2021xfk,Choi:2021kmx, Koide:2021zxj,Choi:2022zal,Apruzzi:2021nmk,Arias-Tamargo:2022nlf,Hayashi:2022fkw,Roumpedakis:2022aik,Bhardwaj:2022yxj,Kaidi:2022uux,Choi:2022jqy,Cordova:2022ieu,Antinucci:2022eat,Bashmakov:2022jtl,Damia:2022rxw,Damia:2022bcd,Choi:2022rfe,Lu:2022ver,Bhardwaj:2022lsg,Bartsch:2022mpm,Lin:2022xod,Apruzzi:2022rei,GarciaEtxebarria:2022vzq, Benini:2022hzx, Wang:2021vki, Chen:2021xuc, DelZotto:2022ras,Bhardwaj:2022dyt,Brennan:2022tyl,Delmastro:2022pfo, Heckman:2022muc,Freed:2022qnc,Niro:2022ctq,Kaidi:2022cpf,Mekareeya:2022spm,vanBeest:2022fss,Antinucci:2022vyk,Chen:2022cyw}.\footnote{\ For further constructions of generalised symmetries in higher-dimensional theories see e.g. \cite{DelZotto:2015isa,Cordova:2018cvg,Benini:2018reh,Eckhard:2019jgg,Bergman:2020ifi,Morrison:2020ool,DelZotto:2020sop,Albertini:2020mdx,Bah:2020uev,DelZotto:2020esg,Bhardwaj:2020phs,Cordova:2020tij,Apruzzi:2020zot,BenettiGenolini:2020doj,DeWolfe:2020uzb,Gukov:2020btk,Iqbal:2020lrt,Hidaka:2020izy,Brennan:2020ehu,Closset:2020scj,Closset:2020afy,Apruzzi:2021phx,Apruzzi:2021vcu,Hosseini:2021ged,Cvetic:2021sxm,Buican:2021xhs,Iqbal:2021rkn,Braun:2021sex,Cvetic:2021maf,Closset:2021lhd,Bhardwaj:2021wif,Hidaka:2021mml,Lee:2021obi,Lee:2021crt,Hidaka:2021kkf,Apruzzi:2021mlh,Bah:2021brs,Closset:2021lwy,DelZotto:2022fnw,Cvetic:2022uuu,Beratto:2021xmn,DelZotto:2022joo}.} This paper continues the study of non-invertible symmetries in a particular class of supersymmetric $(3+1)$d theories, namely those of class $\cS$ \cite{Gaiotto:2009we,Gaiotto:2009hg}.

A particular way of realising non-invertible symmetries, which we will heavily use in this paper, is by half-space gauging \cite{Choi:2021kmx}. Let us illustrate the technique in the specific case of $SU(2)$ $\cN=4$ super-Yang Mills (SYM) theory. We begin by defining an operation $\sigma$ which gauges the $\ZZ_2^{(1)}$ one-form symmetry of the $SU(2)$ theory in half of space-time, with Dirichlet boundary conditions at the interface. As discussed in \cite{Kaidi:2021xfk,Choi:2021kmx}, this gives rise to an interface between the $SU(2)$ and $SO(3)_+$ theories, both at the same value of the Yang-Mills coupling $\tauYM$. Because the theories on the left and the right of the interface are in general distinct, $\sigma$ itself does not correspond to a defect in a single theory.  However, $\cN=4$ SYM is known to enjoy an $SL(2,\ZZ)$ duality, which in particular contains an $S$-duality operation denoted by $\sfS$. This operation acts on the global form $SU(2)$ in the same way as $\sigma$, and also acts on the coupling. Composing the two operations gives the configuration in Figure \ref{fig:Ndef}. From this we see that the combined transformation $\cN_\sfS:= \mS \sigma$ gives rise to a symmetry in the $SU(2)$ theory, as long as we choose the couplings on both sides to match, i.e.
\bea
\tauYM = - 1/ \tauYM \hspace{0.5 in} \Rightarrow \hspace{0.5 in} \tauYM = i~. 
\eea
Since it involves a half-space gauging, this symmetry is non-invertible by construction.

For this particular non-invertible symmetry, there is known to be an alternative construction: namely, we can gauge an invertible symmetry with mixed anomaly in the $SO(3)_-$ theory at $\tauYM= i$ \cite{Kaidi:2021xfk}. In general, if one begins with a theory with  invertible, anomalous symmetries and then does appropriate topological manipulations (in practice, discrete gaugings) to obtain a non-invertible symmetry in a different global variant of the theory \cite{Kaidi:2021xfk}, the resulting non-invertible symmetry is referred to as \textit{non-intrinsic} \cite{Kaidi:2022uux}: one can always find a different global variant of the theory in which the consequences of the non-invertible symmetry can be reinterpreted in terms of those of an invertible one. In the context of class $\cS$, a particular family of such non-intrinsic non-invertible symmetries was identified in \cite{Bashmakov:2022jtl}.

In contrast, non-invertible symmetries which cannot be related to anomalous invertible symmetries via topological manipulations are referred to as \textit{intrinsically} non-invertible symmetries. One of the main goals of this work is to classify non-invertible symmetries of general theories of class $\cS$, and to establish criteria for when they are intrinsic. Perhaps unsurprisingly, we will find that intrinsic non-invertibility is the generic situation.  

Since all of the relevant tools for the study of general class $\cS$ can be developed already at the level of 4d $\cN=4$ theories, we will now describe this case in full detail. The reader who has studied this introduction should have no trouble following the more technical discussions of the main text. 

\begin{figure}[!tbp]
\begin{center}
\[\hspace*{-0.8 cm}\begin{tikzpicture}[baseline=19,scale=0.8]
\draw[  thick] (0,-0.2)--(0,3.2);
\draw[  thick] (3,-0.2)--(3,3.2);
  
          \shade[line width=2pt, top color=red,opacity=0.4] 
    (0,0) to [out=90, in=-90]  (0,3)
    to [out=0,in=180] (3,3)
    to [out = -90, in =90] (3,0)
    to [out=190, in =0]  (0,0);
    
\node[left] at (-0.7,1.5) {$SU(2)[\tauYM]$};
    \node at (1.5,1.5) {$SO(3)_+[\tauYM]$};
    \node[right] at (3.7,1.5) {$SU(2)[-{1\over \tauYM}]$};
      \node[left] at (0,-0) {$\sigma$};
          \node[right] at (3,-0) {$\sfS$};  
\end{tikzpicture}
\hspace{0.35 in}\Rightarrow\hspace{0.35 in}
\begin{tikzpicture}[baseline=19,scale=0.8]
\draw[red, thick] (0,-0.2)--(0,3.2);
  \node[left] at (-0.7,1.5) {$SU(2)[\tauYM]$};
    \node[right] at (+0.7,1.5) {$SU(2)[-{1\over \tauYM}]$};
      \node[left] at (0,0) {$\cN_\sfS$};
\end{tikzpicture}
\]
\caption{At $\tauYM=i$, the $SU(2)$ theory has a non-invertible defect $\cN_{\mathsf{S}}$, which can be understood as the composition of a defect $\sigma$ implementing gauging of the $\ZZ_2^{(1)}$ one-form symmetry, together with an invertible $\sfS$ defect. }
\label{fig:Ndef}
\end{center}
\end{figure}

\subsection{Global variants of $\mathfrak{su}(p)$ SYM}

In a general 4d $\cN=4$ theory, one can try to identify non-invertible symmetries descending from the $\mS$ and $\mS\mT$ transformations at $\tau = i$ and $e^{2\pi i \over 3}$. This exercise was carried out in \cite{Kaidi:2022uux} for all gauge algebras, including the exceptional and non-simply-laced cases, with one family of exceptions: the theories $\mathfrak{su}(N)$ for $N>4$. The reason that this set of theories is the most difficult is that in this case the one-form symmetry, and likewise the number of global variants, grows with $N$. Thus the straightforward technique of enumerating global variants and identifying combinations of $\{\mS, \mT, \sigma, \tau\}$ (with $\sigma$ introduced previously and $\tau$ corresponding to stacking with appropriate SPT phase) leaving a particular variant unchanged is no longer tractable. However, if we restrict to $N=p$ prime, then it turns out that significant progress can be made. This was hinted at in \cite{Kaidi:2022uux}, and will be discussed from a completely different perspective now.

From now on we restrict to $N=p$ prime. It will prove useful to first understand how the number of global variants grows with $p$. To specify a global variant we must specify the charge lattice, as well as the invertible phases with which we stack. Let us begin with the former. In order to specify the charge lattice, we begin by choosing a single non-trivial point $(e,m) \in \ZZ_p \times \ZZ_p$ (all physically realized charge lattices must contain the trivial point $(0,0)$), and then allow all other points $(e',m')$ which satisfy the mutual locality condition \cite{Aharony:2013hda}
\bea
e m' - m e' = 0 \,\,\,\mathrm{mod}\,\,p~. 
\eea
Note that it suffices to specify only a \textit{single} point. One might have naively expected that, after specifying the initial point $(e,m)$, there could be multiple choices $(e'_1,m'_1)$ and $(e_2',m_2')$ for the second point, which would give different charge lattices. But for $p$ prime this is not the case: if $(e'_1,m'_1)$ and $(e_2',m_2')$ satisfy the mutual locality condition with $(e,m)$, then they are also mutually local to one another, and hence the charge lattice is independent of the choice of the second point.\footnote{Concretely, we begin by assuming that $e \neq 0$ mod $p$ and consider the mutual locality conditions
\bea
e m'_1 - m e'_1 = 0 \,\,\,\mathrm{mod}\,\,p~, \hspace{0.6 in} e m'_2- m e'_2 = 0\,\,\,\mathrm{mod}\,\,p~.
\eea
Multiplying the first equation by $e'_2$ and the second equation by $e'_1$, and then subtracting the two gives
\bea
e(e'_2 m'_1 - e'_1 m'_2) = 0 \,\,\,\mathrm{mod}\,\,p~. 
\eea
Thus $(e'_1,m'_1)$ and $(e_2',m_2')$ are mutually local as well. If on the other hand $e = 0\,\,\,\mathrm{mod}\,\,p$, then $m \neq 0 \,\,\,\mathrm{mod}\,\,p$ (since we must specify a non-trivial point) and we may repeat the argument by first multiplying by $m'_2$ and  $m'_1$ and then subtracting.
}

Furthermore, to specify the lattice we may restrict to pairs $(e,m)$ such that $e$ and $m$ are coprime, i.e. $\mathrm{gcd}(e,m)=1$. Indeed, say that we instead try to specify the lattice by a point $(\tilde e, \tilde m)$ such that $\mathrm{gcd}(\tilde e,\tilde m)=d$. Then there exist coprime integers $( e, m)$ such that $(\tilde e, \tilde m) = d ( e,  m)$, and we see that $(\tilde e, \tilde m)$ and $( e,  m)$ are mutually local, 
\bea
\tilde e  m - m \tilde e = d ( e  m -  m  e ) = 0~. 
\eea
Thus the lattice specified by $(\tilde e, \tilde m)$ contains $( e,  m)$, and is in fact identical to the one specified by $( e,  m)$. 

Of course, since we are working modulo $p$ we must be careful about what we mean by coprime. For example, though the pairs $(1,2)$ and $(2,1)$ both involve coprime integers, as elements of $\ZZ_3$ the two can be related by $(1,2)= 2 \cdot (2,1)$. To avoid overcounting, we must therefore mod out by the multiplicative group  $\ZZ_p^\times$, which is of order $p-1$ (it does not contain the zero element). 

We may now count the number of allowed charge lattices, which is the number of coprime pairs $(e,m)$ modulo $\ZZ_p^\times$. In general, the number of $k$-plets $(n_1,\dots, n_k)$ such that $n_i$ are all mutually coprime and coprime to $N$ is counted by the Jordan totient function 
\bea
J_k(N) = N^k \prod_{p|N} \left(1- {1 \over p^k} \right)~.
\eea
In the current case we have $k=2$ and $N=p$ prime, so the number of coprime pairs is simply $J_2(p) = p^2-1$. Dividing by $\ZZ_p^\times$ then gives the number of allowed charge lattices, namely ${J_2(p)\over p-1} = p+1$. 

For each charge lattice, we may further stack with $k$ copies of the invertible phase\footnote{For $p=2$ this should be replaced by ${\pi k\over 2} \int \cP(B)$ with $\cP(B)$ the Pontryagin square of $B$.} 
\bea
\label{eq:g1SPT}
{2 \pi k \over p} \int {p+1\over 2} B \cup B~, \hspace{0.5 in} k = 0,\dots, p-1
\eea
where $B$ is the background gauge field for the $\ZZ_p^{(1)}$ one-form symmetry. The theories with different $k$ count as different global variants. We thus arrive at the final count for the number of global variants $d(p)$ for $\mathfrak{su}(p)$ SYM, namely 
\bea
d(p) = p \times {J_2(p)\over p-1} = p(p+1)~. 
\eea

\subsection{Modular orbits of global variants}

\begin{figure}[!tbp]
\begin{center}
\begin{tikzpicture}[baseline=0,scale = 0.7, baseline=-10]
 \node[below] (1) at (0,0) {$SU(2)_0$};
 
     \node[below] (1) at (0,-1) {$\left(\begin{matrix} 0 & 1 \\ 1 & 0 \end{matrix}\right)$};

 \draw [blue,{Latex[length=2.5mm]}-{Latex[length=2.5mm]}] (0.1,0) to (0.1,2);
  \node[right]  at (0.1,1) {$\color{blue}{\tau}$};
  \draw [dgreen,{Latex[length=2.5mm]}-{Latex[length=2.5mm]}] (-0.1,0) to (-0.1,2);
  \node[left]  at (-0.1,1) {$\color{dgreen}{\mT}$};
  \node[above] (4) at (0,2) {$SU(2)_1$};
  
    \node[above] (4) at (0,3) {$\left(\begin{matrix} 0 & 1 \\ 1 & 1 \end{matrix}\right)$};
  
   \node[below] (2) at (8,0) {$SO(3)_{+,0}$};
   
     \node[below] (2) at (8,-1) {$\left(\begin{matrix} 1 & 0 \\ 0 & 1 \end{matrix}\right)$};

 \draw [blue,{Latex[length=2.5mm]}-{Latex[length=2.5mm]}] (8,0) to (8,2);
 \node[left] at (8,0.5) {$\color{blue}{\tau}$};
  \node[above] (5)  at (8,2) {$SO(3)_{+,1}$};
  
  \node[above] (5) at (8,3) {$\left(\begin{matrix} 1 & 1 \\ 0 & 1 \end{matrix}\right)$};

 \node[below] (3) at (16,0) {$SO(3)_{-,0}$};
 
   \node[below] (3) at (16,-1) {$\left(\begin{matrix} 1 & 0 \\ 1 & 1 \end{matrix}\right)$};
   
 \draw [blue,{Latex[length=2.5mm]}-{Latex[length=2.5mm]}] (15.9,0) to (15.9,2);
  \draw [dgreen,{Latex[length=2.5mm]}-{Latex[length=2.5mm]}] (16.1,0) to (16.1,2);
 \node[left]  at (15.9,1) {$\color{blue}{\tau}$};
  \node[right]  at (16.1,1) {$\color{dgreen}{{\mS}}$};
  \node[above] (6) at (16,2) {$SO(3)_{-,1}$};
    \node[above] (6) at (16,3) {$\left(\begin{matrix} 1 & 1 \\ 1 & 0 \end{matrix}\right)$};

 \draw[dgreen,{Latex[length=2.5mm]}-{Latex[length=2.5mm]}] (1.2,-0.5) to (6.5,-0.5);
   \node[] at (3.85,0) {$\color{dgreen}\mS$};
    \draw[dgreen,{Latex[length=2.5mm]}-{Latex[length=2.5mm]}] (9.3,-0.5) to (14.7,-0.5);
   \node[] at (11.85,0) {$\color{dgreen}\mT$};
   
     \draw[dgreen,{Latex[length=2.5mm]}-{Latex[length=2.5mm]}] (1.2,2.5) to (6.5,2.5);
      \node[above] at (3.9,2.5) {$\color{dgreen}\mS$};
       \draw[dgreen,{Latex[length=2.5mm]}-{Latex[length=2.5mm]}] (9.3,2.5) to (14.7,2.5);
   \node[above] at (12,2.5) {$\color{dgreen}\mT$};

 \draw[blue,{Latex[length=2.5mm]}-{Latex[length=2.5mm]}] (0,-1) to[out=-20,in=200] (8,-1);
\node[] at (4,-1.5) {$\color{blue} \sigma$};
              
   \draw[blue,{Latex[length=2.5mm]}-{Latex[length=2.5mm]}] (2,2.2) to (14.8,-0.3);
\node[above] at (9.4,0.8) {$\color{blue}{\sigma}$};
                
 \draw[blue,{Latex[length=2.5mm]}-{Latex[length=2.5mm]}] (8,3) to[out=20,in=160] (16,3);
 \node[above] at (12,3.8) {$\color{blue}{\sigma}$};   
\end{tikzpicture} 
\caption{Labelling of global variants of $\mathfrak{su}(2)$ SYM by matrices in $SL(2, \ZZ_2) / \ZZ_2^\times = SL(2, \ZZ_2)$. They are defined such that modular transformations $\mathsf{S}$ and $\mathsf{T}$ act from the left. The subscript $0,1$ denotes the number of copies of the invertible phase ${\pi \over 2}\int \cP(B)$ that we stack with. }
\label{fig:su2wmatrices}
\end{center}
\end{figure}
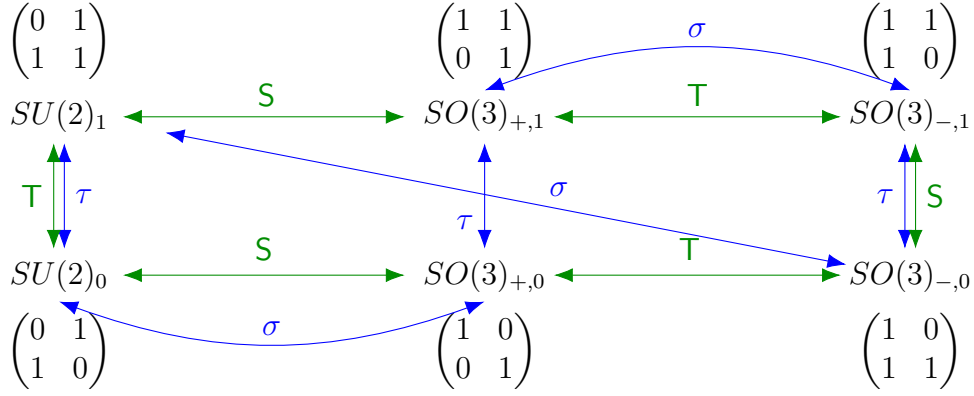

We now notice a small ``coincidence,'' namely that\footnote{This formula can also be found in \cite{Gaiotto:2014kfa}. We will give a geometric interpretation of it later on. }
\bea
\label{eq:numglobalvarsg1}
d(p) = p(p+1) = {|SL(2, \ZZ_p)| \over |\ZZ_p^\times|} ~. 
\eea
What this means is that each global variant of $\mathfrak{su}(p)$ SYM can be labelled by a ray-matrix $M \in SL(2, \ZZ_p)$. Given one such label, we may act on it with a modular transformation $\mathsf{S}, \mathsf{T} \in SL(2, \ZZ)$ to obtain a unique new ray-matrix $M'$. In fact, transitivity of the group $SL(2, \ZZ_p)$ assures us that \textit{every} global variant can be obtained in this way---in other words, there are no disconnected modular orbits for $\mathfrak{su}(p)$ SYM.\footnote{See \cite{bergman2022holography} for similar, more refined statements.} With this in mind, we choose our labels as follows. We first choose one global variant to assign the identity matrix; for concreteness we take this to be $PSU(p)_{0,0}$ (which for $p=2$ is $SO(3)_{+,0}$).\footnote{The first subscript denotes the discrete theta angle, while the second subscript denotes the number of copies of the invertible phase (\ref{eq:g1SPT}) that are stacked with the theory.} We then label each other global variant by the particular element of the modular group that is needed to map from $PSU(p)_{0,0}$ to that global variant,
\bea
F(\mS, \mT): \hspace{0.2 in} M \mapsto F(\mS, \mT)^T M ~, \hspace{0.6 in} F(\mS, \mT) \in SL(2, \ZZ)~.
\eea
 in the conventions 
\bea
\mathsf{S} = \left(\begin{matrix}0 & -1 \\ 1 & 0 \end{matrix} \right)~, \hspace{0.5 in} \mathsf{T} = \left(\begin{matrix}1 & 1 \\ 0 & 1 \end{matrix} \right)~
\eea
For example, for $p=2$ we obtain the labelling shown in Figure \ref{fig:su2wmatrices}.

The topological operations $\sigma$ and $\tau$ generate $SL(2,\ZZ_p)$ and should likewise be realizable as a matrix action on the global variants, but they should not act on the left, lest they fail to commute with the modular transformations. Instead, the topological operations act on the global variants from the \textit{right}, 
\bea
G(\sigma, \tau): \hspace{0.2 in} M \mapsto  M G(\sigma, \tau) ~, \hspace{0.6 in} G(\sigma, \tau)  \in SL(2, \ZZ_p)~,
\eea
with $\sigma$ and $\tau$ given by the same matrices as $\sfS$ and $\sfT$ respectively, as may be checked in the explicit example of $\mathfrak{su}(2)$ in Figure \ref{fig:su2wmatrices}. 

There are two main takeaways from this discussion, both of which will generalize to other class $\cS$ theories:
\begin{enumerate}
\item Every global variant can be mapped to every other by the action of both $\mS, \mT \in SL(2, \ZZ)$ and $\sigma, \tau \in SL(2,\ZZ_p)$. 

\item Every action of the modular group $F(\mS, \mT)$ can be undone by an appropriate topological manipulation $G(\sigma, \tau)$. This is true independent of the global variant $M$ under consideration. 
\end{enumerate}

As we have mentioned before, the first of these implies that there are no disconnected components in the orbit of $\mS, \mT \in SL(2, \ZZ)$, nor in the orbit of $\sigma, \tau \in SL(2,\ZZ_p)$. On the other hand, the second of these means that, for any $F(\mS, \mT)$ leaving a certain value of the coupling fixed (i.e. $F(\mS, \mT) = \mathsf{S}$ or $\mS\mT$ in the current case), we will \textit{always} get a symmetry by dressing with the appropriate topological manipulation $G(\sigma, \tau)$. This statement holds regardless of the global variant under consideration.  The topological manipulation will generically contain factors of $\sigma$ which make the symmetry non-invertible, and so we see that non-invertible symmetries are, in some sense, the norm. 

From this point of view, instead of asking ``For what global forms $M$ do we have a non-invertible symmetry?'' we should really ask ``For what global forms $M$ do we \textit{not} have a non-invertible symmetry?'' In other words, we ask if/when the non-invertible symmetry becomes invertible. By point 1 above, we see that if any single global variant has an invertible symmetry, then the potential non-invertible symmetry in all other global variants is non-intrinsic.

\subsection{Invariant global forms}
We now focus on the identification of global forms with invertible symmetries. A symmetry $F(\mS, \mT)$ is invertible in the global form $M$ if it acts as\footnote{We could also allow for operations such as rescaling of the background fields, but since we will set the background fields to zero in a moment anyways we do not elaborate on these here.}
\bea
F(\mS, \mT)^T M = \lambda M \tau^n~, \hspace{0.5 in}\lambda \in \ZZ_p^\times~, \,\,\,\,n \in \ZZ_p~. 
\eea
The factor of $\lambda$ is allowed since $M$ is in the first place defined only modulo $\ZZ_p^\times$. The factor of $\tau^n$ captures potential anomalies of the invertible symmetry. If we are only concerned with whether or not a symmetry is invertible, and not whether it is anomalous, then it suffices to set all background gauge fields to zero, in which case the factor of $\tau^n$ drops out, and we may redefine the label $M$ to be simply the first column of the matrix $M$ we were using before (e.g. $SU(2) \leftrightarrow \binom{0}{1}, SO(3)_+ \leftrightarrow \binom{1}{0}, SO(3)_- \leftrightarrow \binom{1}{1}$). We are then left with a simple eigenvalue equation 
\bea
\label{eq:genus1eigenvalue}
F(\mS, \mT)^T M = \lambda M~, \hspace{0.5 in} \lambda \in \ZZ_p^\times~. 
\eea
The global form $M$ has an invertible symmetry if and only if this equation is satisfied. 

Let us begin by fixing to $F(\mS, \mT) = \mS$. In this case the possible eigenvalues $\lambda$ are given by the characteristic polynomial
\bea
\mathrm{det} (\mS - \lambda \mathds{1}) = \lambda^2 + 1 = 0\,\,\,\mathrm{mod}\,\,p~. 
\eea
We must now ask if this equation admits a solution in $\ZZ_p^\times$. The answer to this question is a standard result in the theory of quadratic residues---the number of solutions is given by 
\bea
\label{eq:sqrtminus1}
\#\,\,\mathrm{solutions} = \left\{\begin{matrix} 1 & & p=2 \\ 1+(-1)^{p-1 \over 2} && \mathrm{otherwise} \end{matrix}  \right. 
\eea
Whenever a $\lambda$ in $\ZZ_p^\times$ exists, there automatically exists a vector with coefficients in $\ZZ_p$ satisfying (\ref{eq:genus1eigenvalue}), and hence an invariant global form. 
For $p=2$ we see that there is one solution, and hence one global form which is invariant under $\mS$. This is none other than the $SO(3)_-$ theory, which is indeed invariant under $\mS$ \cite{Aharony:2013hda}.\footnote{Recall that we are turning off the background fields for this analysis. In the presence of background fields $SO(3)_-$ transforms under $\sfS$ by an invertible phase, indicating a mixed `t Hooft anomaly between the invertible $\sfS$ symmetry and the $\ZZ_2^{(1)}$ one-form symmetry.} For $p \in 4\NN+1$, we see that there are two global forms left invariant, and otherwise there are no global forms left invariant. Thus only for $p>2$ and $p \not\in 4 \NN+1$ is the $\mS$ symmetry (dressed with the appropriate topological manipulation $G(\sigma, \tau)$) intrinsically non-invertible.\footnote{A similar statement can be found in Appendix C of \cite{Choi:2022zal}.}

We may carry out a similar analysis for $F(\mS, \mT) = \mS\mT$. In this case the possible eigenvalues $\lambda$ are given by the characteristic polynomial
\bea
\mathrm{det} (\mS\mT - \lambda \mathds{1}) = \lambda^2 - \lambda + 1 = 0\,\,\,\mathrm{mod}\,\,p~. 
\eea
We again ask for which $p$ this admits a solution in $\ZZ_p^\times$. The answer is given by 
\bea
\#\,\,\mathrm{solutions} = \left\{\begin{matrix} 0 & & p=2 \\1  & & p=3 \\  1+(-3|p) && \mathrm{otherwise} \end{matrix}  \right.
\eea
where $(n|p)$ is the Legendre symbol. We conclude that there is no global form left invariant for $p=2$, one global form left invariant for $p=3$ (identified as the $PSU(3)_1$ theory in \cite{Kaidi:2022uux}), two global forms invariant for $p\in 3\NN+1$, and none otherwise. Thus only for $p>3$ and  $p\not\in 3\NN+1$ is the $\mS\mT$ symmetry (dressed with the appropriate topological manipulation $G(\sigma, \tau)$) intrinsically non-invertible.

We have now understood the full spectrum of invertible modular symmetries in the global variants of $\cN=4$ $\mathfrak{su}(p)$ SYM for any prime $p$. This in particular tells us whether non-invertible symmetries descending from $\mS$ and $\mS\mT$ can be intrinsic.  We summarize the results for the first few primes in Table \ref{tab:N4intrinsic}. 

\begin{table}[!tp]
\begin{center}
\begin{tabular}{c|cccccccccc}
$p$ & 2 & 3 & 5 & 7 & 11 & 13 & 17 & 19 & 23& 29
\\\hline
$\mS$ intrinsic? & \red{\xmark} & \green{\cmark} & \red{\xmark} & \green{\cmark} & \green{\cmark} & \red{\xmark} &  \red{\xmark}& \green{\cmark}&\green{\cmark}& \red{\xmark}
\\
$\mS\mT$ intrinsic?  & \green{\cmark} & \red{\xmark} & \green{\cmark} & \red{\xmark} & \green{\cmark} & \red{\xmark} & \green{\cmark}& \red{\xmark}& \green{\cmark}&\green{\cmark}
\end{tabular}
\caption{For any $p$, there are global variants such that $\mS$ and $\mS\mT$ give rise to non-invertible symmetries (upon appropriate dressing with topological manipulations $G(\sigma, \tau)$). We may then ask if these non-invertible symmetries are intrinsic or not. This may be answered by asking if there exist global forms left invariant by $\mS$ or $\mS\mT$. The results are shown for the first few primes. }
\label{tab:N4intrinsic}
\end{center}
\end{table}%

\subsection{Higher-dimensional point of view} 

The 4d $\cN=4$ theories that we have been discussing thus far are the simplest examples of theories of class $\cS$. In particular, they can be obtained by compactifying the type $\mathfrak{a}_{p-1}$ 6d (2,0) theory on a torus. An important point, to be elaborated on further in the main text, is that the 6d (2,0) theory is not actually a well-defined theory, but rather a ``relative theory,'' i.e. a theory living on the boundary of a non-trivial TFT in 7d \cite{Witten:2009at,Freed:2012bs,Monnier:2017klz}. 

It may come as a surprise that the 4d $\cN=4$ theories, which are well-defined, can be obtained via torus compactification of the 6d (2,0) theory, which is relative. In particular, when the coupled 6d-7d system is compactified on a torus, one obtains a coupled 4d-5d system, and one might worry that this indicates that the 4d theory is relative as well. This however is not the case. To see this, note that the theory in the 7d bulk $W_7$ is given schematically by 
\bea 
S_{7d} = {p \over 4 \pi} \int_{W_7} c \wedge d c ~, 
\eea
which upon compactification on $T^2$ with $H_1(T^2,\ZZ_p)$ generated by the usual $\sfA, \sfB$ cycles gives 
\bea
\label{eq:BFtheoryg1}
S_{5d} = {p \over 2 \pi} \int_{W_5} b \wedge d \widehat b~ 
\eea
for $b := \oint_\sfA c$ and $\widehat b := \oint_\sfB c$.  In Section \ref{sec:topops}, this BF theory will be seen to be closely related to---though not in general identical to---the Symmetry TFT (SymTFT) of the 4d $\cN=4$ theory. 

Crucially, unlike the 7d CS theory, the 5d BF theory admits topological boundary conditions. It is this fact which allows the 4d $\cN=4$ theory to be well-defined---indeed, one may place the 5d BF theory on an interval with a topological boundary condition on one end and the dimensionally-reduced 6d (2,0) theory on the other end, and then shrink the slab to obtain a well-defined 4d $\cN=4$ theory. In this picture, the choice of global structure of the 4d $\cN=4$ theory (e.g. $SU(2)$ versus $SO(3)_+$ versus $SO(3)_-$) is captured by the choice of topological boundary.

 The way that these topological boundary conditions are realized in 7d is by taking the 7d manifold to be of the form $W_7 = V_3 \times X_4$, where $\partial V_3 = T^2$ and the product is not necessarily trivial. A particularly simple class of three-manifolds $V_3$ with torus boundary are solid tori, i.e. genus-1 handlebodies. As will be discussed in Section \ref{sec:topops}, the number of distinct genus-1 handlebodies (at the level of homology over $\ZZ_p$) is given by $p+1$, matching precisely with the number of distinct charge lattices of the 4d $\cN=4$ theory. We may further consider inserting charge $q \in \ZZ_p$ Wilson surfaces along the non-contractible cycle of the solid torus, which capture the background gauge fields for the one-form symmetry as well as the possible stackings with SPT phases. We thus obtain a picture in which the choice of handlebody and longitudinal Wilson surfaces determines completely the 5d topological boundary conditions, and hence the global variant of the 4d $\cN=4$ theory. 
 
 In this higher-dimensional language, the question of intrinsic versus non-intrinsic non-invertible symmetry becomes a question of whether the isometries of $T^2$ extend to isometries of the solid torus $V_3$. When the isometries do extend, then all non-invertible symmetries are non-intrinsic, and the BF theory given in (\ref{eq:BFtheoryg1}) is the full SymTFT for the boundary theory (at least, the full SymTFT capturing the one-form symmetry). On the other hand, when the isometries do not extend, the non-invertible symmetries are intrinsic, and the SymTFT is \textit{not}  (\ref{eq:BFtheoryg1}), but rather a gauging of this theory by an outer autormorphism symmetry, as discussed in \cite{Kaidi:2022cpf}. Since this automorphism symmetry shuffles the fields $b$ and $\widehat b$, the 7-dimensional realization of it should shuffle the $\sfA$- and $\sfB$-cycles of the torus. In other words, this additional discrete gauging in 5d corresponds to a \textit{sum over geometries} in 7d. Obtaining the full SymTFT for theories with intrinsically non-invertible symmetries thus involves a (topological) quantum gravity computation in higher dimensions.
 
 Let us close by saying that the theory obtained by gauging the outer automorphism symmetry of the BF theory is not in general a gauge theory, even for a higher-group symmetry \cite{Johnson-Freyd:2021tbq}. In general, it is a difficult question when exactly the result {is} a gauge theory, or more precisely a Dijkgraaf-Witten (DW) theory. By recalling that a symmetry is non-intrinsically non-invertible if and only if the SymTFT is a DW theory \cite{Kaidi:2022cpf}, our previous results on intrinsic non-invertiblity (summarized for example in Table \ref{tab:N4intrinsic}) immediately allow us to answer this question: 
 
  \begin{theorem}
  \label{thm:1}
 Gauging the $\ZZ_2^{\mathrm{EM}}$ electro-magnetic duality symmetry acting as $F=\sfS$ in $(4+1)$d $\ZZ_p^{(2)}$ gauge theory gives a spin Dijkgraaf-Witten theory if and only if $p \in 4 \NN+1$.
 \end{theorem}
 
 \noindent
 and likewise 
 
   \begin{theorem}
   \label{thm:2}
 Gauging the $\ZZ_3$ traility symmetry acting as $F = \mathsf{ST}$ in $(4+1)$d $\ZZ_p^{(2)}$ gauge theory gives a spin Dijkgraaf-Witten theory if and only if $p=3$ or $p \in 3 \NN+1$.
 \end{theorem}
 
 \noindent 
 The reader who is mainly interested in this higher-dimensional perspective may skip to Section \ref{sec:topops}, which is largely self-contained.

\subsection{Goals and organization} 

The main goal of this paper is to extend the analysis above to a more general family of class $\cS$ theories.\footnote{Previous works studying global variants and invertible symmetries of class $\cS$ theories include \cite{Tachikawa:2013hya,Bhardwaj:2021pfz,Bhardwaj:2021ojs,Bhardwaj:2021wif,Bhardwaj:2021mzl}.}  As will be reviewed in Section \ref{sec:newclassS}, class $\cS$ theories are obtained by reducing the 6d (2,0) theory of type $\mathfrak{g} \in \{\mathfrak{a}, \mathfrak{d},\mathfrak{e}\}$ on a Riemann surface $\Sigma_{g,n}$ of genus $g$ with $n$ punctures. In this work we will mainly focus on the subset of theories obtained by taking the 6d (2,0) theory to be of type $\mathfrak{a}_{p-1}$ with $p$ prime, and with no punctures $n=0$. We will denote the resulting theories by $\cT^{p,g,0}_L[\Omega]$, where $\Omega$ is the period matrix for the Riemann surface and $L$ is a partial label of the global variant, to be discussed more below. In the case of $g=1$, the theories $\cT^{p,1,0}_L[\tauYM]$ are precisely the $\cN=4$ $\mathfrak{su}(p)$ theories discussed above, with coupling $\tauYM$ and global variant labelled by $L$. 

The first step in identifying (non-)invertible symmetries will be to identify the points $\Omega$ such that the Riemann surface has an enhanced symmetry. This is analogous to identifying the points $\tauYM = i$ or $e^{2 \pi i \over 3}$ at genus 1. Once we have obtained these points, together with the generators $F \in Sp(2g, \ZZ)$ of the enhanced symmetries, the second step is to identify a topological manipulation $G$ which can undo the action of this symmetry on the global form (but not on the period matrix).

 At genus 1, we saw that for \textit{any} $F$ and global variant $M$, there always existed a topological manipulation $G$ which could undo the action of $F$ on $M$, thereby turning $F$ into a (potentially non-invertible) symmetry of the global variant $M$. In that sense, the second step just mentioned was actually unnecessary. 
At higher genus, we will see that effectively the same is true. There are several steps to proving this: 
\begin{enumerate}
\item Show that the full set of topological manipulations $G$ form $Sp(2g, \ZZ_p)$,
\item Show that the global variants of genus $g$ class $\cS$ are labelled by elements in $Sp(2g, \ZZ_p)$ (modulo some appropriate factor),
\item Show that the global variants are acted on from the left by $F \in Sp(2g, \ZZ)$, and from the right by $G \in Sp(2g, \ZZ_p)$.
\end{enumerate}
These statements will be the subject of Section \ref{sec:newclassS}. 

Having done this, we may then simply focus on the first task, namely identification of the points of enhanced symmetry on a genus $g$ Riemann surface. As one might imagine, this exercise has already been carried out in the math literature up to genus 5 \cite{gottschling1961fixpunkte,gottschling1961fixpunktuntergruppen,gottschling1967uniformisierbarkeit,kuribayashi1986automorphism,kuribayashi1990automorphism1,kuribayashi1990automorphism2}, and we will borrow results from there. The math results also include the characteristic polynomials of the generators of the enhanced symmetries, which allow us to evaluate the number of global forms in a way similar to that described above. 

In Section  \ref{sec:genus2} we carry out the analysis at genus 2 in significant detail, in order to illustrate the many new features that arise at $g>1$. In Section \ref{sec:non-invert-symm} we do a similar, somewhat abridged, analysis for genus $g=3,4,$ and $5$. Finally in Section \ref{sec:topops}, we give a higher-dimensional perspective on all of the above results, including a geometric origin of the topological boundary conditions and condensation defects of the 5d SymTFT. 

For the reader's convenience, we also include four appendices. In Appendix \ref{app:matrices} we give explicit matrix representations for the generators of all enhanced symmetries for genus 2 Riemann surfaces. In Appendix \ref{app:arithmetic} we review various arithmetic facts that are necessary for determining when a global form is invariant under an enhanced symmetry. In Appendix \ref{app:handlebodies}, we review some basic facts about handlebodies necessary for understanding the passage from the 6d-7d system to the 4d-5d system. Finally in Appendix \ref{app:altCF} we give an alternative form for the condensation defects $\cC_{F}(X_4)$, first introduced in Section \ref{sec:condensationdefects}, implementing automorphism symmetries of the 5d TFT.

\section{Non-invertible symmetries for class $\cS$ theories}
\label{sec:newclassS}

The main goal of this work is to identify the spectrum of invertible versus non-invertible symmetries in certain theories of class $\cS$. To introduce our notations we begin with a brief review of the class $\cS$ construction.

\subsection{A review of class $\cS$}
\label{sec:classSintro}

\begin{figure}[t]
    \centering
    {\begin{tikzpicture}[baseline=35]
 \shade[top color=red!40, bottom color=red!10,rotate=90]  (0,-0.7) -- (2,-0.7) -- (2.6,-0.2) -- (0.6,-0.2)-- (0,-0.7);
 \draw[thick,rotate=90] (0,-0.7) -- (2,-0.7);
\draw[thick,rotate=90] (0,-0.7) -- (0.6,-0.2);
\draw[thick,rotate=90]  (0.6,-0.2)--(2.6,-0.2);
\draw[thick,rotate=90]  (2.6,-0.2)-- (2,-0.7);
\draw[thick,rotate=90]  (2.6,-0.2)-- (2.6,3);
\draw[thick,rotate=90,dashed]  (0.6,-0.2)-- (0.6,3);
\draw[thick,rotate=90]  (0,-0.7)-- (0,2.5);
\draw[thick,rotate=90,]  (2,-0.7)-- (2,2.5);
\node[below] at (1, -0) {$|\mathfrak{a}_{N-1}\rangle$};
\node[below] at (-1.5,1.5) {7d TQFT};
\end{tikzpicture}}
    \caption{The 6d (2,0) theory of type $\mathfrak{a}_{N-1}$ should be thought of as a state $|\mathfrak{a}_{N-1}\rangle$ in the Hilbert space of a non-trivial 7d TQFT. This TQFT does {not} admit topological boundary conditions.}
    \label{fig:6drel}
\end{figure}
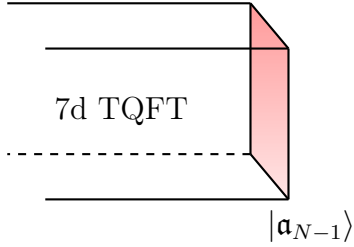

The class $\mathcal{S}$ construction begins with a 6d (2,0) theory of type $\mathfrak{g} \in \{\mathfrak{a}, \mathfrak{d}, \mathfrak{e}\}$ on a closed, compact six-manifold $X_6$, which is reduced on a genus $g$ Riemann surface with $n$ punctures $\Sigma_{g,n}$ to obtain a four-dimensional theory on $X_4 \subset X_6$. Despite naive expectations, the four-dimensional theories obtained in this way depend on more data than just the Riemann surface $\Sigma_{g,n}$ and the algebra $\mathfrak{g} $. The origin of this additional data lies in the fact that the 6d (2,0) theory is not itself a well-defined theory (at least not for the cases of $\mathfrak{g}=\mathfrak{a}_{p-1}$, which will be the focus of this work), but rather a \textit{relative theory} \cite{Witten:2009at,Freed:2012bs}. This means that the 6d (2,0) theory should be thought of as living on the boundary of a non-trivial bulk TQFT on a seven-manifold $W_7$ with $\p W_7 = X_6$. This  seven-dimensional TQFT was identified as a certain Wu-Chern-Simons theory in \cite{Monnier:2017klz}. The basic setup is illustrated in Figure \ref{fig:6drel}.

In this picture, the 6d (2,0) theory is identified with a state $| \mathfrak{a}_{N-1}\rangle$ in the Hilbert space of the bulk TQFT. In order to specify this state, one must first fix a basis for the Hilbert space, which is specified by a maximal isotropic sublattice $\cL \in H_3(X_6, \ZZ_N)$ \cite{Tachikawa:2013hya}; details will be given in Section \ref{sec:6d20review}. For the present purposes it suffices to say that a maximal isotropic sublattice is a maximal set of $M_3 \in H_3(X_6, \ZZ_N)$ such that
\bea
\langle M_3, M_3'\rangle = 0 ~, \hspace{0.5 in}\,\,M_3,M_3'\in \cL~,
\eea
where $\langle \cdot, \cdot \rangle$ is the standard intersection pairing of 3-cycles in 6d.

In the context of class $\cS$, we are interested in six-manifolds of the form $X_6 = \Sigma_{g,n} \times X_4$. In this case the group $H_3(X_6, \ZZ_N)$ splits via the K{\"u}nneth formula, 
\bea
H_3(X_6, \ZZ_N) \cong H_1(\Sigma_{g,n}, \ZZ_N) \otimes H_2(X_4, \ZZ_N)~, 
\eea
subject to the additional assumption (which we will make throughout) that $H_{1,3}(X_4, \ZZ)$ are trivial. There is an analogous splitting of the maximal isotropic sublattice $\cL$ as 
\bea
\label{eq:latticesplits}
\cL &=& L \otimes H_2(X_4, \ZZ_N) ~, \hspace{0.5 in} L \subset H_1(\Sigma_{g,n}, \ZZ_N)~. 
\eea
We will likewise split $M_3 \in H_3(X_6, \ZZ_N) $ into elements $\gamma \in H_1(\Sigma_{g,n}, \ZZ_N)$ and $M_2 \in H_2(X_4, \ZZ_N)$ via $M_3= \gamma \otimes M_2$, on which the 6d intersection pairing decomposes as a product 
\bea
\langle M_3, M_3' \rangle = \langle \gamma, \gamma' \rangle \times (M_2,M_2')~. 
\eea
Here the pairing $ \langle \gamma, \gamma' \rangle$ is the usual (antisymmetric) intersection pairing between 1-cocycles on $\Sigma_{g,n}$, whereas $(M_2,M_2')$ is a symmetric pairing between 2-cocycles in four dimensions.

As first discussed in \cite{Tachikawa:2013hya}, in order to specify the four-dimensional theory we must specify not only the Riemann surface $\Sigma_{g,n}$ on which we compactify, but also a lattice $L \in H_1(\Sigma_{g,n}, \ZZ_N) $. 
This corresponds to the specification of the particular {charge lattice} of the four-dimensional theory. Without specifying $L$, the four-dimensional theory remains relative. 

Furthermore, to fully specify the 4d theory we should also specify a particular representative of the non-trivial classes of $L^\perp \otimes H_2(X_4, \ZZ_N)$, with $L^\perp:= H_1(\Sigma_{g,n}, \ZZ_N)/L$. The choice of representatives in  $L^\perp$ determines the SPT phases with which the 4d theory is stacked, while the choice of elements in $H_2(X_4, \ZZ_N)$ determines the background fields for the corresponding one-form symmetries. In total, we will denote the representative by $\cB$. The geometric origin of this extra data will be explained in Section \ref{sec:topops}. 

 The well-defined four-dimensional theory obtained by specifying the above additional data will be denoted by $\cT^{N,g,n}_L[\Omega, \cB]$ where $\Omega$ is the period matrix of $\Sigma_{g,n}$, capturing the coupling constants of the theory. The argument $\cB$ collectively represents the data about background fields for the one-form symmetries of the theory, and will be suppressed unless important to the discussion. When we want to discuss the class $\cS$ theory without specifying the global variant (the analog of $\cN=4$ $\mathfrak{su}(N)$ SYM) then we will drop the $L$ subscript and write $\cT^{N,g,n}[\Omega]$.

Let us briefly review the example of the $\mathfrak{a}_1$ 6d (2,0) theory on $\Sigma_{2,0} = T^2$. It is well-known that this gives 4d $\cN=4$ $\mathfrak{a}_1$ super-Yang Mills. Taking the homology of the torus to be spanned by the usual $\sfA$ and $\sfB$ cycles,
\bea
\label{eq:torusH1}
H_1(T^2, \ZZ_2) = \left\{1,\, \sfA,\, \sfB,\, \sfA + \sfB \right\}~, \hspace{0.5 in} \langle \sfA, \sfB\rangle = 1
\eea
we identify three maximal isotropic sublattices, 
\bea
\label{eq:su2maxlat}
L_{\sfA} = \{1,\, \sfA\}~, \hspace{0.5 in}L_{\sfB} = \{1,\, \sfB\}~, \hspace{0.5 in}L_{\sfA+\sfB} = \{1,\, \sfA+\sfB\}~. 
\eea
From this we expect three distinct charge lattices for the 4d theory, which we denoted in the introduction by
\bea
L_{\sfA}  \leftrightarrow SU(2)~, \hspace{0.5in}L_{\sfB}  \leftrightarrow SO(3)_+~, \hspace{0.5in}L_{\sfA+\sfB}  \leftrightarrow SO(3)_-~. 
\eea
The data about invertible phases is then given by specifying a representative for the non-trivial class of $L^\perp$. For example, we have $L^\perp_\sfA = \{ [1], [\sfB] \}$, where the class $[1]:= \{1,\sfA\}$ and $[\sfB] := \{\sfB, \sfB+\sfA\}$, and the SPT phase is specified by choosing the particular representative $\sfB$ (giving $SU(2)_0$) or $\sfB+\sfA$ (giving $SU(2)_1$).

One of the main utilities of the 6d point of view is that it is now easy to see how the global variants map into one another under Montonen-Olive duality transformations of the 4d $\cN=4$ theory. Indeed, the Montonen-Olive duality group is realized in the class $\cS$ construction as the modular group $SL(2, \ZZ)$ of the Riemann surface $T^2$, which can be generated by operations $\sfT$ and $\sfS$ acting on the elements of $H_1(T^2, \ZZ_p)$ via,
\beaa
\label{eq:TSdef1}
&\sfT:& \mathsf{A} \rightarrow \mathsf{A} ~, \hspace{0.8 in} \sfS:\,\,\, \mathsf{A} \rightarrow \mathsf{B}~, 
\no\\
&\vphantom{.}& \mathsf{B} \rightarrow \mathsf{A}+ \mathsf{B} \hspace{0.88 in} \,\,\, \mathsf{B} \rightarrow -\mathsf{A}~.
\eeaa
Alternatively, when realized on the vector $(\sfB, \sfA)^T$, these take the form of matrices
\bea
\label{eq:TSdef2}
\sfT = \left( \begin{matrix} 1 & 1 \\ 0 & 1 \end{matrix} \right)~,  \hspace{0.5 in} \sfS = \left(\begin{matrix} 0 & -1 \\ 1 & 0 \end{matrix} \right)~. 
\eea
We thus see that the $\sfS$ operation exchanges $L_{\sfA}$ and $L_{\sfB}$, and hence the $SU(2)$ and $SO(3)_+$ charge lattices, whereas it leaves $L_{\sfA+\sfB}$, and hence the $SO(3)_-$ charge lattice, unchanged. On the other hand, $\sfT$ leaves the $SU(2)$ charge lattice unchanged, whereas it exchanges the $SO(3)_+$ and $SO(3)_-$ charge lattices. Tracking the action of $\sfS$ and $\sfT$ on the representatives of $L^\perp$ allows us to reproduce the results in  Figure \ref{fig:su2wmatrices}.

\subsection{Non-invertible symmetries and the modular group}

We may now generalize the construction of non-invertible symmetries in 4d $\cN=4$ theories, reviewed in the introduction and Figure \ref{fig:Ndef}, to arbitrary class $\cS$ theories $\cT^{p,g,n}_L[\Omega]$. Our starting point will be the application of an element $F$ of the relevant modular group $\mathrm{Mod}(\Sigma_{g,n},p)$. For the case of no punctures, i.e. $n=0$, this is $\mathrm{Mod}(\Sigma_{g,0},p)=Sp(2g, \ZZ_p)$. An element of the modular group will in general act on $L$ and $\Omega$, but will never act on $p,g,$ or $n$. Having done a modular transformation, the next step is to do an appropriate topological operation $G:=G(\sigma, \tau)$, to be explained below, which acts on $L$ but not on $\Omega$. The situation is then as in Figure \ref{fig:moregeneralNdef}.

In order to have a non-invertible symmetry, we require the existence of a solution to the following equations
\bea
\label{eq:maineqs}
GF(L)=L~, \hspace{0.8 in}F(\Omega)=\Omega~. 
\eea
The second equation is conceptually straightforward, and amounts to identifying the fixed points of the modular group. The former instructs us to look for a topological manipulation $G$ which can undo the action of $F$ on the global variant $L$. We distinguish between three conceptually distinct cases: 

\begin{itemize}
\item $F(L)= L$ and $G$ trivial:  In this case the symmetry is \textit{invertible} and \textit{non-anomalous}.\footnote{Here we are referring to mixed anomalies between $F$ and the one-form symmetry; there could still be self-anomalies of $F$, along the lines of \cite{Witten:1995gf,Hsieh:2019iba,Hsieh:2020jpj}.}
\item $F(L)= L$ and $G(L) = L$ with $G$ non-trivial: In this case $F$ again corresponds to an invertible symmetry, but the partition function transforms by a counterterm which is removed by $G$. In other words, $F$ is \textit{invertible} and \textit{anomalous}.
\item $F(L)\neq L$: In this case the symmetry is \textit{non-invertible}. There are two subcases: 
\begin{itemize}
\item When $F(L) \neq L$, but there exists at least one $L' \in H_1(\Sigma_{g,n}, \ZZ_N)$ such that $F(L') = L'$, then the symmetry is \textit{non-intrinsically} non-invertible.
\item When $F(L) \neq L$, and there is no $L' \in H_1(\Sigma_{g,n}, \ZZ_N)$  such that $F(L') = L'$, then the symmetry is \textit{intrinsically} non-invertible.
\end{itemize}

\end{itemize}

\begin{figure}[!tbp]
\begin{center}
\[\hspace*{-0.8 cm}\begin{tikzpicture}[baseline=19,scale=0.8]
\draw[  thick] (0,-0.2)--(0,3.2);
\draw[  thick] (3,-0.2)--(3,3.2);
  
          \shade[line width=2pt, top color=red,opacity=0.4] 
    (0,0) to [out=90, in=-90]  (0,3)
    to [out=0,in=180] (3,3)
    to [out = -90, in =90] (3,0)
    to [out=190, in =0]  (0,0);
    
\node[left] at (-0.7,1.5) {$\cT^{N,g,n}_L[\Omega]$};
    \node at (1.5,1.5) {$\cT^{N,g,n}_{F(L)}[F(\Omega)]$};
    \node[right] at (3.7,1.5) {$\cT^{N,g,n}_{GF(L)}[F(\Omega)]$};
      \node[left] at (0,-0) {$F$};
          \node[right] at (3,-0) {$G$};  
\end{tikzpicture}
\hspace{0.35 in}\Rightarrow\hspace{0.35 in}
\begin{tikzpicture}[baseline=19,scale=0.8]
\draw[red, thick] (0,-0.2)--(0,3.2);
  \node[left] at (-0.7,1.5) {$\cT^{N,g,n}_L[\Omega]$};
    \node[right] at (+0.7,1.5) {$\cT^{N,g,n}_{GF(L)}[F(\Omega)]$};
      \node[left] at (0,0) {$\cN_F$};
\end{tikzpicture}
\]
\caption{When $GF(L)=L$ and $F(\Omega)=\Omega$ admit solutions, then the composite $\cN_F:= F G$ gives a non-invertible symmetry of the  class $\cS$ theory. }
\label{fig:moregeneralNdef}
\end{center}
\end{figure}

The constraints in (\ref{eq:maineqs}) were already discussed in detail in the case of $(g,n) = (1,0)$ in the introduction. In that case, we observed that for $N=p$ prime, given an $F$ and $\Omega$ satisfying $F(\Omega) = \Omega$, there \textit{always} existed a $G$ which could undo the action of $F$ on any global variant $L$. Hence any such $F$ automatically gave a symmetry, regardless of the global form, with the (non-)invertibility stemming from whether $G$ contained discrete gaugings $\sigma$.
We will now see that this is again the case for class $\cS$ theories of type $\cT^{p,g,0}_L[\Omega]$. The first step in showing this is to understand the spectrum of global variants of theories of class $\cS$.

\subsection{Global variants of theories of class $\cS$}
\label{sec:globalvariantsofclassS}

We restrict to the case of $N=p$ prime and no punctures $n=0$, and denote the generators of $H_1(\Sigma_{g,0},\ZZ_p)$ as $\sfA_I$ and $\sfB_I$ for $I=1,\dots,g$, with intersection pairing 
\beaa
\label{eq:H1basis}
\langle \sfA_I , \sfA_J\rangle = \langle \sfB_I , \sfB_J\rangle = 0 ~, \hspace{0.6 in} \langle \sfA_I, \sfB_J \rangle = \delta_{IJ}~.
\eeaa
This intersection pairing is preserved by matrices $F$ acting as 
\bea
\label{eq:actionofphi}
\binom{\vec{\sfB}}{\vec{\sfA}} \rightarrow F \binom{\vec{\sfB}}{\vec{\sfA}}
\eea
as long as the matrices satisfy 
\bea
F^T  \mathfrak{I} F = \mathfrak{I}~, \hspace{0.5 in}\mathfrak{I} = \left(\begin{matrix} 0 & - \mathds{1}_{g\times g}   \\ \mathds{1}_{g\times g}  & 0\end{matrix}\right)~.
\eea
The matrices $F$ should have integer entries so that they map properly normalized cycles amongst themselves, and hence the $F$ are matrices in $Sp(2g, \ZZ)$.

It will be useful for us to assign to each element $\sfA_I$ and $\sfB_I$ a $2g$-vector, 
\bea
v_{\sfB_1} = \left(\begin{matrix} 1\\ 0_{2g-1} \end{matrix}  \right)~, \hspace{0.3 in} v_{\sfB_2} = \left(\begin{matrix} 0\\1 \\0_{2g-2} \end{matrix}  \right)~,\hspace{0.3 in}  \dots~, \hspace{0.3 in} v_{\sfB_g} = \left(\begin{matrix} 0_{g-1} \\ 1 \\ 0_{g} \end{matrix}  \right)~, 
\no\\
v_{\sfA_1} = \left(\begin{matrix} 0_g \\ 1 \\ 0_{g-1} \end{matrix}  \right)~, \hspace{0.3 in} v_{\sfA_2} = \left(\begin{matrix} 0_{g+1}  \\1\\ 0_{g-2} \end{matrix}  \right)~,\hspace{0.3 in}  \dots~, \hspace{0.3 in} v_{\sfA_g} = \left(\begin{matrix} 0_{2g-1} \\ 1 \end{matrix}  \right)~,
\eea
where $0_n$ represents an array of $n$ copies of $0$. Likewise $v_{m \sfA_I + n \sfB_J} = m\, v_{\sfA_I}+n\,v_{\sfB_J}$. In terms of these vectors the intersection pairing may then be represented as 
\bea
\label{eq:usefulintpairing}
\langle \sfA_I, \sfB_J \rangle = v_{\sfA_I }^T \,\mathfrak{I} \,v_{\sfB_J}~.
\eea

A maximal isotropic sublattice $L$ is spanned by precisely half of the above generators, i.e. $g$ distinct elements $\gamma_i$ for $i=1,\dots, g$. We may then unambiguously label such an $L$ by a $2g\times g$ matrix whose $g$ columns are the $2g$-vectors corresponding to each of the generators of the lattice. We denote this $2g \times g$ matrix by $K_L = (v_{\gamma_1}, \dots, v_{\gamma_g})$. Similarly, the background gauge-field is specified by a representative $\cB$ of $L^\perp \otimes H^2(X_4, \ZZ_p)$, which can be labelled by a $2g \times g$ matrix $K_\cB$, where now the coefficients of the vector take values in $H^2(X_4, \ZZ_p)$. Given a particular global variant, specified by $L$ and the representative of $L^\perp \otimes H^2(X_4, \ZZ_p)$, we may  assign to it a $2g \times 2g$ matrix, 
\bea
\label{eq:matrixdef}
M_{L,\cB}:=(K_L\,\, K_\cB)~. 
\eea

A useful fact is that the matrices obtained in this way are elements of $Sp(2g,\ZZ_p)$. 
Indeed, denoting the generators of $L^\perp$ by $\gamma_{i+g}$ for $i=1, \dots, g$, we have 
\bea
M_{L,\cB}^T \, \mathfrak{I}\, M_{L,\cB} = \binom{K_L^T}{K_\cB^T}\, \mathfrak{I}\, (K_L\,\, K_\cB) = \left(\begin{matrix}v_{\gamma_{1}}^T \mathfrak{I} v_{\gamma_{1}} & \dots & v_{\gamma_{1}}^T \mathfrak{I} v_{\gamma_{2g}} \\ & \ddots & \\ v_{\gamma_{2g}}^T \mathfrak{I} v_{\gamma_{1}} & \dots & v_{\gamma_{2g}}^T \mathfrak{I} v_{\gamma_{2g}} \end{matrix} \right) = \mathfrak{I}~,
\eea
where in the last step we have used the presentation of the intersection pairing in (\ref{eq:usefulintpairing}), together with isotropicity and maximality of $L$.

Conversely, every element of $Sp(2g,\ZZ_p)$ defines a global variant. However, not all matrices denote \textit{distinct} global variants. Given a lattice $L$ spanned by $g$ elements $\gamma_i$ for $i=1,\dots,g$, it is clear that linear transformations of these elements give rise to equivalent bases for $L$. Similar statements hold for $L^\perp$, which is spanned by $\gamma_{i+g}$ for $i=1,\dots,g$. In contrast, transformations which mix the generators of $L$ and $L^\perp$ change the global form, and do not give redundancies. The total set of redundancies is then given by block diagonal matrices $\left(\begin{smallmatrix} Q & 0 \\ 0 & Q' \end{smallmatrix} \right) $ acting on $(\gamma_1, \dots, \gamma_{2g})^T$. By linearity of the labels $v_\gamma$, these matrices act in the same way on $v_\gamma$, and hence we have the identification under right-multiplication
\bea
\label{footnote:QQp}
M_{L,\cB} \cong M_{L,\cB} \left(\begin{matrix} Q & 0 \\ 0 & Q' \end{matrix} \right)~.
\eea
Importantly, the matrices $Q$ and $Q'$ appearing here are actually not independent---instead, $Q$ and $Q'$ are constrained to satisfy $Q'Q^T = \mathds{1}_{g\times g}$ such that the resulting matrix is symplectic.
We conclude that each redundancy is given by a single element in $GL(g,\ZZ_p)$.

In summary, we expect that the total number of global variants of type $\mathfrak{a}_{p-1}$ class $\cS$ on $\Sigma_{g,0}$ is 
\bea
d(g,p) := {|Sp(2g,\ZZ_p)|\over |GL(g, \ZZ_p)|}~. 
\eea
In the particular case of $g=1$, this reduces to $d(g,1) = |SL(2,\ZZ_p)|/|\ZZ_p^\times| = p(p+1)$, reproducing the result in (\ref{eq:numglobalvarsg1}). More generally, we have
\bea
\label{eq:primedim}
d(g,p) = p^{g^2} \prod_{m=1}^g {p^{2m}-1 \over p^g - p^{m-1}}  = p^{\half g(g+1)} \prod_{m=1}^g (p^m + 1)~,
\eea
where the first equality follows from the known formulas for the orders of the finite groups, and the second equality follows from some elementary algebra. This formula admits a simple physical interpretation: the factor $\prod_{m=1}^g (p^m + 1)$ counts the number of maximal isotropic sublattices of $H_1(\Sigma_{g,0}, \ZZ_N)$, while the factor $p^{\half g(g+1)}$ counts the number of invertible phases that can be stacked with the four-dimensional theory.\footnote{The latter follows from noting that all SPT phases are of the type $\int B_i \cup B_j$, with $i,j=1,\dots, g$. There are ${\half g(g+1)}$ distinct combinations, and we can allow for between 0 and $p-1$ copies of each.  
 }

\subsubsection{An example} Let us give a concrete example before moving on. Consider the case of $(g,p) = (2,2)$. We may begin by considering the maximally isotropic lattice $L = \mathrm{span}\{\sfA_1, \sfA_2\}$. The corresponding $4 \times 2$ matrix $K_L$ is then
\bea
K_L = (v_{\sfA_1} \,\,v_{\sfA_2} ) = \left(\begin{matrix} 0 & 0 & 1 & 0 \\ 0 & 0& 0 & 1 \end{matrix} \right)^T~.
\eea
The orthogonal lattice $L^\perp = H_1(\Sigma_{2,0}, \ZZ_p)/L$ admits four explicit choices of basis vectors, $\{\sfB_1, \sfB_2\}$, $\{\sfB_1+\sfA_1, \sfB_2\}$, $\{\sfB_1, \sfB_2+\sfA_2\}$, and $\{\sfB_1+\sfA_1, \sfB_2+\sfA_2\}$, with respective matrices $K_\cB$ given by
\bea
K_\cB = \left(\begin{matrix} 1 & 0 & 0& 0 \\ 0 & 1 & 0 & 0\end{matrix} \right)^T~,\,\,\, \left(\begin{matrix} 1 & 0 & 1 & 0 \\ 0 & 1 & 0 & 0\end{matrix} \right)^T~,\,\,\,\left(\begin{matrix} 1 & 0 & 0 & 0 \\ 0 & 1 & 0 & 1\end{matrix} \right)^T~,\,\,\,\left(\begin{matrix} 1 & 0 & 1 & 0 \\ 0 & 1 & 0 & 1\end{matrix} \right)^T~. 
\eea
The four corresponding global variants are thus labelled by the following $4\times 4$ matrices, 
\bea
M_{L, \cB} =  \left(\begin{matrix} 0 & 0 & 1 & 0  \\ 0 & 0 & 0 & 1\\1 & 0 & 0 & 0  \\ 0 & 1 & 0 & 0 \end{matrix} \right)~,\,\,\, \left(\begin{matrix} 0 & 0 & 1 & 0 \\ 0 & 0 & 0 & 1 \\ 1 & 0 & 1 & 0 \\ 0 & 1 & 0 & 0 \end{matrix} \right)~,\,\,\,\left(\begin{matrix} 0 & 0 & 1 & 0 \\ 0 & 0 & 0 & 1 \\ 1 & 0 & 0 & 0 \\ 0 & 1 & 0 & 1\end{matrix} \right)~,\,\,\,\left(\begin{matrix} 0 & 0 & 1 & 0 \\ 0 & 0 & 0 & 1 \\ 1 & 0 & 1 & 0 \\ 0 & 1 &0 & 1\end{matrix} \right)~,
\eea
each of which is easily confirmed to be an element of $Sp(4, \ZZ_2)$. A similar exercise can be done for the remaining choices of $L$. 

Of course, we are always free to change our choice of bases such that e.g. $L = \mathrm{span}\{\sfA_1+\sfA_2, \sfA_2\}$ instead of $\mathrm{span}\{\sfA_1, \sfA_2\}$, and this should not change the global form. On the other hand, mixing between elements of $L$ and $L^\perp$ will change the global form. Thus the redundancy in the $4 \times 4$ matrices given above is by right-multiplication 
\bea
M_{L, \cB} \cong  M_{L, \cB} \left(\begin{matrix} Q & 0 \\ 0 & Q' \end{matrix} \right)
\eea
with $Q$ and $Q'$ both $2 \times 2$ matrices in $GL(2,\ZZ_2)$. Requiring that the matrix is always symplectic fixes $Q'$ in terms of $Q$, as discussed above.

We note in closing that in our notation, for all $(g,p)$ the global form with $L = \mathrm{span}\{\sfB_1, \dots, \sfB_g\}$ and $L^\perp = \mathrm{span}\{\sfA_1, \dots, \sfA_g\}$ is assigned the identity matrix in $Sp(2g, \ZZ_p)$. 

\subsection{Modular orbits of global variants}

In the previous subsection, we showed how to assign to each $\cT^{p,g,0}[\Omega]$ a distinct matrix in $Sp(2g,\ZZ_p)/GL(g, \ZZ_p)$. The action of the modular group $Sp(2g, \ZZ_p)$ on the global variants follows from this definition. Indeed, recall that any $F\in Sp(2g, \ZZ_p)$ acts as in (\ref{eq:actionofphi}), namely via left-multiplication
\bea
F:\hspace{0.5 in}\binom{\vec{\sfB}}{\vec{\sfA}} \mapsto F \binom{\vec{\sfB}}{\vec{\sfA}} ~.
\eea
On the other hand, the vectors $v_{\sfA_I}$, $v_{\sfB_I}$ transform like a basis of charge vectors for the one-cycles, that is\footnote{For example, at genus 1 we have $\sfT: \sfA \mapsto \sfA$ and $v_\sfA= \binom{0}{1}$. In order to obtain the appropriate $\sfT$ transform of $v_\sfA$, we must take $\sfT^T v_\sfA = \left(\begin{smallmatrix} 1 & 1 \\ 0 & 1 \end{smallmatrix}\right)^T \binom{0}{1} = v_\sfA $, as opposed to $\sfT v_\sfA = v_\sfA+v_\sfB$.}
\bea
F:\hspace{0.5 in}( v_{\vec{\sfB}}\,,~ v_{\vec{\sfA}}) \,\mapsto\, F^T (v_{\vec{\sfB}}\,,~v_{\vec{\sfA}}) ~.
\eea
 This means that by construction modular transformations act via left-multiplication by $F^T$ on the matrix $M_{L,\cB}$ labelling the global form. By transitivity of $Sp(2g,\ZZ_p)$, we see that every global variant must be connected to every other via modular transformations---i.e. there are no disconnected modular orbits. 
 
 On the other hand, the topological manipulations $G \in Sp(2g, \ZZ_p)$ commute with modular transformations, and act on the global forms from the right. This fact, together with the fact that the transformations $G$ span $Sp(2g, \ZZ_p)$, will be made more apparent in Section \ref{sec:topops}. Thus in general we have the transformations 
 \bea
 \label{eq:maintransformationlaw}
 M_{L, \cB} \rightarrow F^T M_{L, \cB} G~. 
 \eea
From this we may draw two important conclusions: 
\begin{enumerate}
\item Every global variant can be mapped to every other by the action of both the modular group and the topological manipulations. 

\item Every action $F$ of the modular group can be undone by an appropriate topological manipulation $G$. This is true independent of the global variant $M_{\cL, \cB}$ under consideration. 
\end{enumerate}
 The problem of identifying (non-)invertible symmetries is thus reduced to the identification of points of enhanced symmetry for $\Sigma_{g,0}$. 
 
\section{Non-invertible symmetries for genus-2 class $\cS$}
\label{sec:genus2}

We now begin our analysis of points of enhanced symmetry for the case of genus 2. For pedagogy we will work with explicit matrix representations of all group elements here. For higher genus somewhat more abstract techniques will be required. 

\subsection{Genus 2 and the modular group $Sp(4, \ZZ_p)$}
We work with the basis of $H_1(\Sigma_{g,0}, \ZZ_p)$ given above, namely $\{\sfA_I, \sfB_I\}$ for $I=1,\dots,g$ satisfying (\ref{eq:H1basis}). For general $g$ the first cohomology group of $\Sigma_{g,0}$ is given by $H^1(\Sigma_{g,0}, \RR) = \RR^{2g}$, and admits a splitting into Dolbeault cohomology groups $H^1(\Sigma_{g,0}, \RR) = H^{(1,0)}(\Sigma_{g,0}) \oplus H^{(0,1)}(\Sigma_{g,0})$. A canonical basis of $H^{(1,0)}(\Sigma_{g,0}, \ZZ)$ is given by  holomorphic (1,0)-forms $\omega_I$ for $I=1,\dots, g$, with periods normalized as 
\bea
\oint_{\sfA_I} \omega_J = \delta_{IJ}~.
\eea
In terms of these forms we may define the {period matrix} $\Omega$ of $\Sigma_{g,0}$, 
\bea
\oint_{\sfB_I} \omega_J = \Omega_{IJ}~.
\eea
The matrices $F \in Sp(2g, \ZZ)$ then act on $\Omega$ as 
\bea
F: \hspace{0.2 in} \Omega \mapsto (A\Omega + B)(C\Omega+D)^{-1}~, \hspace{0.5 in} F = \left( \begin{matrix} A & B \\ C & D\end{matrix}\right)~.
\eea

At genus 2, the period matrix is a $2 \times 2$ matrix 
\bea
\Omega = \left( \begin{matrix} \tau_1 & \tau_2 \\ \tau_2 & \tau_3  \end{matrix}\right)~,\hspace{0.5 in} \mathrm{det}(\mathrm{Im}(\tau_i)) >0 ~, \hspace{0.2 in}\mathrm{Tr}(\mathrm{Im}(\tau_i)) >0~.
\eea
Note that the limit of $\tau_2 \rightarrow 0$ is a degeneration limit in which the genus-2 Riemann surface splits into two genus-1 Riemann surfaces with complex structure moduli $\tau_1$ and $\tau_3$. Indeed, we have an embedding of the corresponding modular groups $SL(2, \ZZ) \times SL(2, \ZZ) \subset Sp(4, \ZZ)$. 

\begin{figure}[tbp]
\begin{center}

\begin{tikzpicture}[scale=0.3]

\shade[top color=gray!40, bottom color=gray!10] (0.03,-2) 
to [out=0, in=180] (-6.43,-2) 
to [out=270, in=90] (-6.43,2)
to [out=180, in=0] (0.03,2)
to [out=90, in=270] (0.03,-2) ;

\shade[top color=gray!40, bottom color=gray!10,rotate = 90]  (-2,0) coordinate (-left) 
to [out=260, in=60] (-3,-2) 
to [out=240, in=110] (-3,-4) 
to [out=290,in=180] (0,-6) 
to [out=0,in=250] (3,-4) 
to [out=70,in=300] (3,-2) 
to [out=120,in=280] (2,0)  coordinate (-right);

\draw[rotate = 90]  (-2,0) coordinate (-left) 
to [out=260, in=60] (-3,-2) 
to [out=240, in=110] (-3,-4) 
to [out=290,in=180] (0,-6) 
to [out=0,in=250] (3,-4) 
to [out=70,in=300] (3,-2) 
to [out=120,in=280] (2,0)  coordinate (-right);

\shade[top color=gray!40, bottom color=gray!10,rotate = 270,yshift=-2.5in]  (-2,0) coordinate (-left) 
to [out=260, in=60] (-3,-2) 
to [out=240, in=110] (-3,-4) 
to [out=290,in=180] (0,-6) 
to [out=0,in=250] (3,-4) 
to [out=70,in=300] (3,-2) 
to [out=120,in=280] (2,0)  coordinate (-right);

\draw[rotate = 270,yshift=-2.5in]  (-2,0) coordinate (-left) 
to [out=260, in=60] (-3,-2) 
to [out=240, in=110] (-3,-4) 
to [out=290,in=180] (0,-6) 
to [out=0,in=250] (3,-4) 
to [out=70,in=300] (3,-2) 
to [out=120,in=280] (2,0)  coordinate (-right);

\draw[] (0,-2)--(-6.4,-2);
\draw[] (0,2)--(-6.4,2);

\pgfgettransformentries{\tmpa}{\tmpb}{\tmp}{\tmp}{\tmp}{\tmp}
\pgfmathsetmacro{\myrot}{-atan2(\tmpb,\tmpa)}
\draw[rotate around={\myrot:(0,-2.5)},yshift=1in,xshift=1.2in] (-1.2,-2.4) to[bend right]  (1.2,-2.4);
\draw[fill=white,rotate around={\myrot:(0,-2.5)},yshift=1in,xshift=1.2in] (-1,-2.5) to[bend right] (1,-2.5) 
to[bend right] (-1,-2.5);

\draw[rotate around={\myrot:(0,-2.5)},yshift=1in,xshift=-3.7in] (-1.2,-2.4) to[bend right]  (1.2,-2.4);
\draw[fill=white,rotate around={\myrot:(0,-2.5)},yshift=1in,xshift=-3.7in] (-1,-2.5) to[bend right] (1,-2.5) 
to[bend right] (-1,-2.5);

\begin{scope}[yshift=-4in,xshift=-1in]
\draw[thick] (-3.5,0)--(3.5,0);
\filldraw[color=white] (0,0) circle (25pt);
\draw[thick] (0,0) circle (25pt);
\node at (0,0) {$p$};
\draw[thick] (6.4,0) circle (80pt);
\draw[thick] (-6.4,0) circle (80pt);

\filldraw[color=white]  (9.1,0) circle (25pt);
\draw[thick]  (9.1,0) circle (25pt);
\node at (9.1,0) {$p$};

\filldraw[color=white]  (-9.1,0) circle (25pt);
\draw[thick]  (-9.1,0) circle (25pt);

\node at (-9.1,0) {$p$};
\end{scope}

\end{tikzpicture}\hspace{1 in}
\begin{tikzpicture}[scale=0.3]
\shade[top color=gray!40, bottom color=gray!10,rotate = 90]  (-2,0) coordinate (-left) 
to [out=260, in=60] (-3,-2) 
to [out=240, in=110] (-3,-4) 
to [out=290,in=180] (0,-6) 
to [out=0,in=250] (3,-4) 
to [out=70,in=300] (3,-2) 
to [out=120,in=280] (2,0)  coordinate (-right);

\draw[rotate = 90]  (-2,0) coordinate (-left) 
to [out=260, in=60] (-3,-2) 
to [out=240, in=110] (-3,-4) 
to [out=290,in=180] (0,-6) 
to [out=0,in=250] (3,-4) 
to [out=70,in=300] (3,-2) 
to [out=120,in=280] (2,0)  coordinate (-right);

\shade[top color=gray!40, bottom color=gray!10,rotate = 270]  (-2,0) coordinate (-left) 
to [out=260, in=60] (-3,-2) 
to [out=240, in=110] (-3,-4) 
to [out=290,in=180] (0,-6) 
to [out=0,in=250] (3,-4) 
to [out=70,in=300] (3,-2) 
to [out=120,in=280] (2,0)  coordinate (-right);

\draw[rotate = 270]  (-2,0) coordinate (-left) 
to [out=260, in=60] (-3,-2) 
to [out=240, in=110] (-3,-4) 
to [out=290,in=180] (0,-6) 
to [out=0,in=250] (3,-4) 
to [out=70,in=300] (3,-2) 
to [out=120,in=280] (2,0)  coordinate (-right);

\pgfgettransformentries{\tmpa}{\tmpb}{\tmp}{\tmp}{\tmp}{\tmp}
\pgfmathsetmacro{\myrot}{-atan2(\tmpb,\tmpa)}
\draw[rotate around={\myrot:(0,-2.5)},yshift=0.1in,xshift=-0.5in,rotate=90,scale=1.5] (-1.2,-2.4) to[bend right]  (1.2,-2.4);
\draw[fill=white,rotate around={\myrot:(0,-2.5)},yshift=0.1in,xshift=-0.5in,rotate=90,scale=1.5] (-1,-2.5) to[bend right] (1,-2.5) 
to[bend right] (-1,-2.5);

\draw[rotate around={\myrot:(0,-2.5)},yshift=0.1in,xshift=0.5in,rotate=270,scale=1.5] (-1.2,-2.4) to[bend right]  (1.2,-2.4);
\draw[fill=white,rotate around={\myrot:(0,-2.5)},yshift=0.1in,xshift=0.5in,rotate=270,scale=1.5] (-1,-2.5) to[bend right] (1,-2.5) 
to[bend right] (-1,-2.5);

\begin{scope}[yshift=-4in,xshift=0in]
\draw[thick] (0,4)--(0,-4);
\filldraw[color=white] (0,0) circle (25pt);
\draw[thick] (0,0) circle (25pt);
\node at (0,0) {$p$};
\draw[thick] (0,0) circle (115pt);

\filldraw[color=white]  (4,0) circle (25pt);
\draw[thick]  (4,0) circle (25pt);
\node at (4,0) {$p$};

\filldraw[color=white]  (-4,0) circle (25pt);
\draw[thick]  (-4,0) circle (25pt);
\node at (-4,0) {$p$};
\end{scope}
\end{tikzpicture}

\caption{Two weakly-coupled limits for class $\cS$ on $\Sigma_{2,0}$, with the corresponding generalized quivers. The trivalent junctions represent $T_p$ theories, while the circle represent $\mathfrak{a}_{p-1}$ gauge nodes.}
\label{fig:deglimits}
\end{center}
\end{figure}
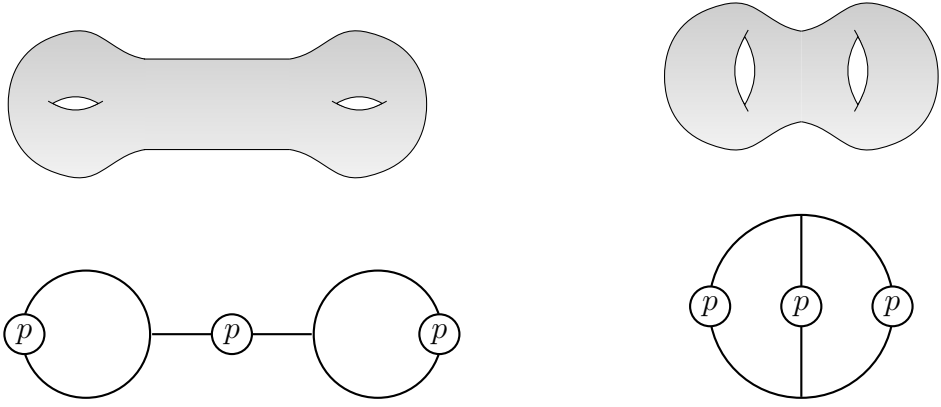

Each value of the period matrix $\Omega$ corresponds to a certain choice of couplings for the corresponding four-dimensional class $\cS$ theory, and in certain limits one may have a Lagrangian description of the theory as a (generalized) quiver theory, as shown in Figure \ref{fig:deglimits}. In the separating limit mentioned in the previous paragraph, in which e.g. the tube in the top left side of Figure \ref{fig:deglimits} is stretched infinitesimally thin and infinitely long, one retrieves two copies of the appropriate $\mathfrak{a}_{p-1}$ $\cN=4$ theory, which have the non-invertible symmetries studied in \cite{Kaidi:2022uux}. More generally, the points at which non-invertible symmetries arise will be strongly-coupled and non-Lagrangian.

For the purposes of this section it will be useful to have an explicit matrix representation of the modular group $Sp(4, \ZZ)$, which is given by
\bea
Sp(4, \ZZ) = \langle T_1,~ T_2,~ S_{12},~ W \rangle ~, 
\eea
in terms of the following four generators,
\bea
 T_1 = \left(\begin{matrix} 1 & 0 & 1 & 0 \\ 0 & 1 & 0 & 0 \\ 0 & 0 & 1 & 0 \\ 0 & 0 & 0 & 1 \end{matrix}\right),~ T_2 =\left(\begin{matrix} 1 & 0 & 0 & 0 \\ 0 & 1 & 0 & 1 \\ 0 & 0 & 1 & 0 \\ 0 & 0 & 0 & 1 \end{matrix}\right),~ S_{12} =  \left(\begin{matrix} 0 & 0 & 1 & 0 \\ 0 & 0 & 0 & 1 \\ -1 & 0 & 0 & 0 \\ 0 & -1 & 0 & 0 \end{matrix}\right),~ W=  \left(\begin{matrix} 1 & 0 & 0 & 1 \\ 0 & 1 & 1 & 0 \\ 0 & 0 & 1 & 0 \\ 0 & 0 & 0 & 1 \end{matrix}\right)\no~.
\eea
These generators act on the basis of one-cycles as,
\beaa
T_1 \left(\begin{matrix}\sfB_1 \\ \sfB_2\\ \sfA_1 \\ \sfA_2 \end{matrix} \right) = \left(\begin{matrix}\sfB_1 + \sfA_1 \\ \sfB_2\\ \sfA_1 \\ \sfA_2 \end{matrix} \right)~, \hspace{0.2 in} T_2 \left(\begin{matrix}\sfB_1 \\ \sfB_2\\ \sfA_1 \\ \sfA_2 \end{matrix} \right) &=& \left(\begin{matrix}\sfB_1  \\ \sfB_2+ \sfA_2\\ \sfA_1 \\ \sfA_2 \end{matrix} \right)~, \hspace{0.2 in} S_{12} \left(\begin{matrix}\sfB_1 \\ \sfB_2\\ \sfA_1 \\ \sfA_2 \end{matrix} \right)  = \left(\begin{matrix}\sfA_1 \\ \sfA_2\\ -\sfB_1 \\ -\sfB_2 \end{matrix} \right) ~,
\no\\ W \left(\begin{matrix}\sfB_1 \\ \sfB_2\\ \sfA_1 \\ \sfA_2 \end{matrix} \right)  &=& \left(\begin{matrix}\sfB_1 + \sfA_2\\ \sfB_2 + \sfA_1\\ \sfA_1 \\ \sfA_2 \end{matrix} \right)~.
\eeaa
We see that $T_1$ and $T_2$ act as separate $T$ transformations on the two sub-tori, whereas $S_{12}$ acts as a simultaneous $S$ transformation on both tori. The $W$ transformation is an intrinsically genus-2 transformation. 

\subsection{Fixed points of $Sp(4,\ZZ)$}

With an explicit basis of generators for $Sp(4, \ZZ)$, we may now search for elements admitting fixed points in their action on $\Omega$. As one might have anticipated, these results have already been obtained in the mathematics literature \cite{gottschling1961fixpunkte,gottschling1961fixpunktuntergruppen,gottschling1967uniformisierbarkeit}; here we simply quote them. There are two qualitatively distinct classes of fixed loci:

\begin{table}[tp]
\begin{center}
\begin{tabular}{c|ccc}
Order & Subgroup & Generators & $\Omega$
\\ \hline
10& $\ZZ_{10}$ &  $\phi$ &$ \left( \begin{smallmatrix} \varepsilon & \varepsilon + \varepsilon^{-2} \\ \varepsilon + \varepsilon^{-2}  &-\varepsilon^{-1} \end{smallmatrix} \right)$
\\
24 & $(\ZZ_2 \times \ZZ_6) \rtimes \ZZ_2$ & $M_1, M_2, M_3$ & ${i \over \sqrt{3}} \left( \begin{smallmatrix} 2 & 1 \\ 1 & 2 \end{smallmatrix} \right)$
\\
& $\ZZ_{12} \times \ZZ_2$ & $C, M_4$ & $ \left( \begin{smallmatrix} \rho & 0 \\ 0 & i \end{smallmatrix} \right)$
\\
32 & $(\ZZ_4 \times \ZZ_4)\rtimes \ZZ_2$ & $M_5, M_6, M_7$ & $ \left( \begin{smallmatrix} i & 0 \\ 0 & i \end{smallmatrix} \right)$
\\
48 & $GL(2,3)$ & $M_7, M_8$ & $ {1 \over 3}\left( \begin{smallmatrix} 1+2i \sqrt{2} & -1+i \sqrt{2}  \\ -1+i\sqrt{2} & 1+2i \sqrt{2} \end{smallmatrix} \right)$
\\
72 & $\ZZ_3 \times (\ZZ_6 \times \ZZ_2) \rtimes \ZZ_2$ &$ M_{7}, M_{9}, M_{10}$ & $\left( \begin{smallmatrix} \rho & 0 \\ 0 & \rho \end{smallmatrix} \right)$
\end{tabular}
\end{center}
\caption{Singular moduli of $Sp(4,\ZZ)$ occurring at isolated values of $\Omega$. Here we have defined $\rho := e^{2 \pi i \over 3}$ and $\varepsilon := e^{2 \pi i \over 5}$.}
\label{tab:g2singmodtab1}
\end{table}

 \paragraph{Isolated loci:} The first type of fixed loci are those of dimension zero, i.e. isolated fixed points. This is the only case which occurred at genus 1, where the fixed loci were the isolated points $\tauYM = i$ and $e^{2\pi i \over 3}$. The list of all such isolated fixed points, together with the stabilizing subgroup of $Sp(4,\ZZ)$, are collected in Table \ref{tab:g2singmodtab1}. An explicit matrix representation for the generators $\phi$, $C$, and $M_i$ for $i=1, \dots, 10$ is given in Appendix \ref{app:matrices}.\footnote{Note that the same results can be found in \cite{Nilles:2021glx}.} 
 
 Let us make some comments about these fixed points. First, the fixed points at $\Omega =  \left( \begin{smallmatrix} \rho & 0 \\ 0 & i \end{smallmatrix} \right), \left( \begin{smallmatrix} i & 0 \\ 0 & i \end{smallmatrix} \right)$, and $\left( \begin{smallmatrix} \rho & 0 \\ 0 & \rho \end{smallmatrix} \right)$ correspond to degeneration limits of the genus-2 Riemann surface. The (non-)invertible symmetries corresponding to these points are the same as those discussed in the introduction up to an additional symmetry corresponding to permutation of the two genus-1 Riemann surfaces. In particular, the question of intrinsic versus non-intrinsic invertibility as a function of the prime $p$ is the same as before. Thus we do not discuss these cases here. 
 
 The remaining three fixed points are intrinsically genus 2. The first fixed point is $\Omega =  \left( \begin{smallmatrix} \varepsilon & \varepsilon + \varepsilon^{-2} \\ \varepsilon + \varepsilon^{-2}  &-\varepsilon^{-1} \end{smallmatrix} \right)$ with $\varepsilon = e^{2 \pi i \over 5}$, which leads to an enhanced $\ZZ_{10}$ symmetry of the Riemann surface. This is the direct analog of the $\ZZ_6$ symmetry at $\tauYM = e^{2\pi i \over 3}$ at genus-1.\footnote{More generally, for every genus $g$ there is an isolated point at which there is an enhanced $\ZZ_{4g+2}$ symmetry.} Similarly at $\Omega = {i \over \sqrt{3}} \left( \begin{smallmatrix} 2 & 1 \\ 1 & 2 \end{smallmatrix} \right)$ we have an enhanced  $(\ZZ_2 \times \ZZ_6) \rtimes \ZZ_2$ symmetry, and at $\Omega = {1 \over 3}\left( \begin{smallmatrix} 1+2i \sqrt{2} & -1+i \sqrt{2}  \\ -1+i\sqrt{2} & 1+2i \sqrt{2} \end{smallmatrix} \right)$ an enhanced $GL(2, 3)$ symmetry. The latter is in fact the most symmetric (connected) genus-2 Riemann surface, known as the Bolza surface. 
 
 We should point out here that unlike for 4d $\cN=4$ SYM, the enhanced symmetries can now be non-Abelian. This can give rise to non-Abelian fusion rules involving non-invertible defects, as will be discussed further in Section \ref{sec:condensationdefects}. Note that while non-Abelian non-invertible fusion rules were identified previously for triality defects in \cite{Choi:2022zal}, the non-Abelian fusion rules here are of a slightly different nature: in particular, the fusion rules continue to be non-Abelian even modulo condensation defects. In the language of \cite{Bhardwaj:2022yxj}, the fusion rules are non-Abelian even at the level of \textit{local} fusion.  
 
 \paragraph{Extended loci:}  Unlike for genus 1, at higher genus there can also be entire loci of fixed points. At genus 2, these can be of complex dimension 1 or 2. We list the former in Table \ref{tab:g2singmodtab2}, and the latter in Table \ref{tab:g2singmodtab3}. Of these, the ones at $\Omega = \left( \begin{smallmatrix} i & 0 \\ 0 & \tau_3  \end{smallmatrix} \right),  \left( \begin{smallmatrix} \tau_1 & 0 \\ 0 & \tau_1 \end{smallmatrix} \right), \left( \begin{smallmatrix} \rho & 0 \\ 0 & \tau_3 \end{smallmatrix} \right),$ and $ \left( \begin{smallmatrix} \tau_1 & 0 \\ 0 & \tau_3  \end{smallmatrix} \right)$ correspond to degeneration limits, and will not be discussed further. Let us now comment on the remaining fixed loci. 
 
 First there are two loci of complex dimension one $\Omega = \left( \begin{smallmatrix} \tau_1 & \half \\ \half & \tau_1 \end{smallmatrix} \right)$ and $\left( \begin{smallmatrix} \tau_1 & \half \tau_1  \\ \half \tau_1 & \tau_1 \end{smallmatrix} \right)$ which have enhanced symmetries that are respectively $D_8$ and $D_{12}$. Second, there is a complex dimension two  locus $\Omega = \left( \begin{smallmatrix} \tau_1 & \tau_2 \\ \tau_2 & \tau_1 \end{smallmatrix} \right)$ with an enhanced $\ZZ_2 \times \ZZ_2$ symmetry. Note that the existence of such loci suggests the possibility of having continuous families of theories, parameterized by either 1 or 2 complex couplings, all of which have non-invertible symmetries. This is in contrast to the case of 4d $\cN=4$, for which the non-invertible symmetries only emerged at isolated points $\tauYM = i$ and $e^{2\pi i \over 3}$. 
 
\begin{table}[tp]
\begin{center}
\begin{tabular}{c|ccc}
Order & Subgroup & Generators & $\Omega$
\\ \hline
8& $\ZZ_2 \times \ZZ_4$ &  $C, N_1$ &$ \left( \begin{smallmatrix} i & 0 \\ 0 & \tau_3  \end{smallmatrix} \right)$
\\
 & $D_8$ & $M_7, N_2$ & $ \left( \begin{smallmatrix} \tau_1 & \half \\ \half & \tau_1 \end{smallmatrix} \right)$
\\
& $D_8$ & $M_7, N_3$ & $ \left( \begin{smallmatrix} \tau_1 & 0 \\ 0 & \tau_1 \end{smallmatrix} \right)$
\\
12 & $\ZZ_2 \times \ZZ_6$ & $C, N_4$ & $ \left( \begin{smallmatrix} \rho & 0 \\ 0 & \tau_3 \end{smallmatrix} \right)$
\\
 & $D_{12}$ & $N_5, N_6$ & $ \left( \begin{smallmatrix} \tau_1 & \half \tau_1  \\ \half \tau_1 & \tau_1 \end{smallmatrix} \right)$
\end{tabular}
\end{center}
\caption{Singular moduli of $Sp(4,\ZZ)$ occurring along complex one-dimensional loci.}
\label{tab:g2singmodtab2}
\end{table}%

\begin{table}[tp]
\begin{center}
\begin{tabular}{c|ccc}
Order & Subgroup & Generators & $\Omega$
\\ \hline
4& $\ZZ_2 \times \ZZ_2$ &  $C,P_1$ &$ \left( \begin{smallmatrix} \tau_1 & 0 \\ 0 & \tau_3  \end{smallmatrix} \right)$
\\
 & $\ZZ_2 \times \ZZ_2$ & $C,M_7$ & $ \left( \begin{smallmatrix} \tau_1 & \tau_2 \\ \tau_2 & \tau_1 \end{smallmatrix} \right)$
\end{tabular}
\end{center}
\caption{Singular moduli of $Sp(4,\ZZ)$ occurring along complex two-dimensional loci.}
\label{tab:g2singmodtab3}
\end{table}

\subsection{Invariant global forms}
\label{eq:genus2invglob}
By the results of Section \ref{sec:globalvariantsofclassS}, we know that each geometric symmetry $F$ gives rise to a symmetry of $\cT^{p,2,0}_L[\Omega]$, regardless of the choice of global variant $L$. 
Depending on $L$, the topological manipulation $G$ which is required to undo the action of $F$ on $L$ may involve a discrete gauging, and hence may give rise to a non-invertible symmetry. Indeed, this is the generic situation. What remains is only the question of whether the non-invertible symmetries are intrinsic or not. We now answer this question. We warn that the following analysis is somewhat technical; the reader only interested in the answer can refer to Tables \ref{table:genus2intrinsic} and \ref{tab:GL23intrinsic}. 

To answer the question of intrinsic non-invertibility, we search for global forms which are invariant under $F$. The existence of such global forms will depend delicately on the value of $p$. 
For an invariant global form $M_{\cL, \cB}$ to exist, we must have
\bea
\label{eq:maineqclassS}
F^T M_{L, \cB} = M_{L, \cB}\,P\,\chi(\tau)~,\hspace{0.5 in}  P = \left( \begin{matrix} Q & 0 \\ 0 & Q' \end{matrix} \right) ~,
\eea
where $Q$ and $Q'$ are elements of $GL(2, \ZZ_p)$ and the matrix $Q'$ is completely determined in terms of $Q$ by the equation $Q' Q^T = \mathds{1}_{g \times g}$, as explained around (\ref{footnote:QQp}). The reason for allowing the matrix $P$ on the right-hand side is that $M_{L, \cB}$ is, in the first place, defined only up to such transformations. The term $\chi(\tau)$ is a topological manipulation involving only invertible $\tau$ operations (and potentially automorphisms of the one-form symmetry), which captures `t Hooft anomalies of the invertible symmetry. As in the genus-1 analysis in the introduction, we will set the background fields to zero, which will allow us to drop the $\chi(\tau)$ operation and reduce $M_{L,\cB}$ to the $2g \times g$ matrix $K_L$. We thus aim to solve 
\bea
\label{eq:maineqclassS2}
F^T K_L = K_L Q~, \hspace{0.5 in} Q \in GL(g, \ZZ_p)~. 
\eea
The question is now whether any solution to this equation exists. For genus-1, the corresponding equation was a standard eigenvalue equation (\ref{eq:genus1eigenvalue}), and we could understand the existence of solutions by simply trying to solve the characteristic polynomial modulo $p$. The current case is somewhat more involved.

Given a matrix $Q$ over an algebraically closed field (such as $\CC$) we may always do a similarity transformation to put $Q$ in Jordan normal form, i.e. $Q = B^{-1} J B$. A matrix $J$ in Jordan normal form is a block diagonal matrix $\bigoplus_i J_i$ where each block $J_i$ has the form 
\bea
\label{eq:Jordannormaldef}
J_i = \left( \begin{matrix} \lambda_i & 1 & 0 & \dots & 0 \\ 0 & \lambda_i & 1 & \dots & 0 \\   0 & 0 & \lambda_i & \dots & 0 \\  &  & \vdots & & \\ 0 & 0 & \dots & 0 & \lambda_i  \end{matrix} \right) ~, 
\eea
i.e. it is an upper-triangular matrix with a single number $\lambda_i$ on the diagonal and ones on the superdiagonal. For the case of $Q$ being a $2\times 2$ matrix, there are two possible Jordan decompositions, 
\bea 
\label{eq:genus2Jposs}
J= \left(\begin{matrix} \lambda_1 & 0 \\ 0 & \lambda_2 \end{matrix} \right)~,\,\,\,\, \left(\begin{matrix} \lambda_1 & 1 \\ 0 & \lambda_1\end{matrix} \right)~. 
\eea

In the current case we are concerned with $Q$ defined over the finite field $\ZZ_p$, which in general is not algebraically closed. However, by lifting to the splitting field $\ZZ_p[\lambda_i]$, it is again always possible to put $Q$ in Jordan normal form (of course, if the $\lambda_i$ are all valued in $\ZZ_p$, then the splitting is possible in $\ZZ_p$ itself). We may thus write 
\bea Q = B^{-1} J B~, \hspace{0.5 in} Q \in GL(g, \ZZ_p)~,\,\,\,\, B, J \in GL(g, \ZZ_p[\lambda_i])~.
\eea
 In order for $Q$ to be an element of $GL(g, \ZZ_p)$, we require $\mathrm{det}(Q) = \mathrm{det}(B^{-1} J B) = \mathrm{det}(J) \in \ZZ_p^\times$. Thus even though the entries of $J$ may not in general be valued in $\ZZ_p^\times$, the product of the diagonal entries \textit{must} be valued in $\ZZ_p^\times$. 

The quantities $\lambda_i$ above are related to the modular transformation $F$ as follows. We begin by rewriting (\ref{eq:maineqclassS2}) as 
\bea
\label{eq:maineqclassS3}
F^T \widetilde{K}_{L} =  \widetilde{K}_L J~,\hspace{0.5 in}  \widetilde{K}_L :=  K_L B^{-1}~.
\eea
Writing $\widetilde{K}_L = (\widetilde v_1, \widetilde v_2)$ at genus 2, we then have 
\bea
F^T \,(\tilde v_1, \tilde v_2) = (\lambda_1  \tilde v_1, \lambda_2  \tilde v_2)~, \hspace{0.5 in} F^T \,( \tilde v_1, \tilde v_2) = (\lambda_1  \tilde v_1, \tilde v_1+ \lambda_1  \tilde v_2)~
\eea
for the two choices of $J$ in (\ref{eq:genus2Jposs}). 
Either way, this tells us that all $\lambda_i$ must be  eigenvalues of $F$. 

A further simplifying feature is that $F$ is always the generator of a finite group, meaning that $F^k = \mathds{1}$ for appropriate $k$. In the case that $J$ is diagonal, this tells us that 
\bea
(\tilde v_1, \tilde v_2) = (F^T)^k \,(\tilde v_1, \tilde v_2) = (\lambda_1^k  \tilde v_1, \lambda_2^k  \tilde v_2)
\eea
and hence that the $\lambda_i$ must be $k$-th roots of unity. On the other hand, if $J$ is a non-trivial Jordan block, this tells us that 
\bea
(\tilde v_1, \tilde v_2) = (F^T)^k \,(\tilde v_1, \tilde v_2) = (\lambda_1^k  \tilde v_1,\, \lambda_1^k  \tilde v_2 + k \lambda_1^{k-1} v_1)
\eea
The first equation in this pair would imply $\lambda_1^k=1$, upon which the second would imply that $k\lambda_1^{k-1} v_1=0$ mod $p$. Since neither $\lambda_1$ or $v_1$ can be trivial, we conclude that having a non-trivial Jordan block is only possible if $p|k$. 

Once we have identified a $J$ satisfying the above properties, it is guaranteed that there will exist a matrix $K_L$ with coefficients in $\ZZ_p$ satisfying (\ref{eq:maineqclassS2})---this follows since the entries of $K_L$ satisfy a set of linear equations with coefficients in $\ZZ_p$, which always admit solutions in $\ZZ_p$. However, it is not clear that the matrices obtained in this way will have columns spanning an isotropic sublattice. For $K_L$ to correspond to a legitimate global form, we must verify this. Here it is useful to note two facts: 
\begin{enumerate}
    \item If we write $K_L = (v_1, v_2)$ and $\widetilde K_L = (\tilde v_1, \tilde v_2)$, then the vectors $v_1, v_2$ span an isotropic subspace if and only if $\tilde v_1, \tilde v_2$ do. To prove e.g. the forward direction, we assume that $K_L$ is isotropic, i.e.
    \bea
    K_L^T \mathfrak{I} K_L = 0_{2\times 2}~,
    \eea
    which then implies that 
    \bea
    \widetilde K_L^T \mathfrak{I} \widetilde K_L = (K_L B^{-1})^T \mathfrak{I} K_L B^{-1} = (B^{-1})^T K_L^T \mathfrak{I} K_L B^{-1} = 0_{2\times 2}~.
    \eea
    The reverse direction follows analogously. 
    \item Let $v_1, v_2$ be two eigenvectors of $F$ with eigenvalues $\lambda_1, \lambda_2$. Since $F$ is symplectic  we have
\bea
  \label{eq:37}
 v_2^{T} \mathfrak{I} v_1 =  v_2^{T} F^{T} \mathfrak{I} F v_1 =  (F v_2)^{T} \mathfrak{I} F v_1  = \lambda_{1} \lambda_{2}  v_2^{T} \mathfrak{I} v_1~,
\eea
or in other words $(\lambda_2 \lambda_1 - 1) v_2^{T} \mathfrak{I} v_1=0$ mod $p$. This means that the span of $v_{1}$ and $v_{2}$ is necessarily isotropic unless $\lambda_{1} \lambda_{2} =1$ mod $p$.
\end{enumerate}
Using these two facts, we will be able to identify the solutions $K_L$ corresponding to legitimate global forms.

To summarize the discussion above, in order to identify invariant global forms we begin by identifying the eigenvalues of $F$, which is done by solving the characteristic equation. 
Not all of the eigenvalues will be elements of $\ZZ_p^\times$. We choose pairs of eigenvalues which are conjugates (i.e. such that $\lambda_1 \lambda_2 \in \ZZ_p^\times$ or $\lambda_1^2 \in \ZZ_p^\times$) and construct all possible $J$, subject to the constraint that $J$ can only have non-trivial Jordan blocks if $p$ divides the order of $F$. Each such $J$ will give rise to a generalized eigenvalue equation as in (\ref{eq:maineqclassS2}).  There will always exist a matrix $K_L$ with coefficients in $\ZZ_p$ satisfying this equation, but to ensure that this matrix corresponds to a legitimate global form we must check that its columns span an isotropic sublattices. This may equivalently be done by checking isotropicity of the solutions $\widetilde K_L$ to (\ref{eq:maineqclassS3}).

We now apply this procedure for each of the symmetries mentioned above. 

\paragraph{$\ZZ_{10}$:} We begin by asking if there exist global forms invariant under the generator $\phi$ of the $\ZZ_{10}$ symmetry. An explicit matrix form for $\phi$ is given in Appendix \ref{app:matrices}, from which we obtain the characteristic equation, 
\bea
\label{eq:phieigenvalueeq}
\mathrm{det} (\phi- \lambda \mathds{1}_{4 \times 4}) = \lambda^4 - \lambda^3 + \lambda^2 - \lambda + 1 = 0 \,\,\,\mathrm{mod}\,\,p~. 
\eea 
We begin by giving two representative examples before presenting the general solution. 

We first consider the case of $p=5$, for which the equation above can be rewritten as 
\bea
\lambda^4 + 4 \lambda^3 + 6 \lambda^2 + 4 \lambda + 1 = (\lambda+1)^4 = 0\,\,\,\mathrm{mod}\,\,p~. 
\eea
Thus there exists a single solution $\lambda = -1$ mod $5$ of multiplicity 4. Since $5$ divides the order  $|\phi| = 10$, we are allowed to have a non-trivial Jordan block, and hence the possible choices for $J$ are 
\bea
J = \left( \begin{matrix}- 1 & 0 \\ 0 & -1 \end{matrix} \right)~, \left(\begin{matrix} -1 & 1 \\ 0 & -1 \end{matrix} \right)~. 
\eea
The determinant of $J$ is automatically in $\ZZ_5^\times$ since the diagonal entries are. The most general matrices $\widetilde K_L$ satisfying (\ref{eq:maineqclassS3}) for these $J$ are then found to be (up to linear transformations on the columns),
\bea
\label{eq:g2p5Ksols}
\widetilde K_L = \left(\begin{matrix} 3 & 3 \\ 1 & 1 \\ 1 & 1 \\ 1 & 1 \end{matrix} \right)~, \,\, \left(\begin{matrix} 3 & 3 \\ 1 & 1 \\ 1 & 0 \\ 1 & 4 \end{matrix} \right)~,
\eea
respectively.
In the first case, we see that the two columns are identical, and will not span a maximal sublattice. Hence this does not correspond to a legitimate label for a global form. In the second case, the two columns are linearly independent, but we must check that they span an isotropic sublattice, i.e. $\widetilde K_L \mathfrak{I} \widetilde K_L = 0$. One easily confirms that this is the case. We thus conclude that for $p=5$ there is a single global form with an invertible $\phi$ symmetry, labelled by the second matrix in (\ref{eq:g2p5Ksols}).

We next consider $p=11$. In this case all four solutions of (\ref{eq:phieigenvalueeq}) are in $\ZZ_{11}$, and are given by 
\bea
\lambda = 2,\, 6,\, 7,\, 8~.
\eea
Because $11 \nmid 10$, no non-trivial Jordan blocks are allowed, so there are 6 possible $J$,\footnote{We neglect cases where $J$ is proportional to the identity, since in such cases the columns of $\widetilde{K}_L$ will not span an isotropic sublattice.} \bea
J = \mathrm{diag}(2,6), \, \mathrm{diag}(2,7),\,\mathrm{diag}(2,8),\,\mathrm{diag}(6,7),\,\mathrm{diag}(6,8),\,\mathrm{diag}(7,8)~.
\eea
The determinant of $J$ is automatically in $\ZZ_{11}^\times$ since the diagonal entries are. The corresponding matrices $\widetilde K_L$ satisfying (\ref{eq:maineqclassS3}) are found to be
\bea
\widetilde K_L = \left(\begin{matrix} 1 & 5 \\ 3 & 4 \\ 5 & 9 \\ 1 & 1 \end{matrix} \right)~,\,\, \left(\begin{matrix} 1 & 6 \\ 3 & 9 \\ 5 & 3 \\ 1 & 1 \end{matrix} \right)~,\,\,\left(\begin{matrix} 1 & 7 \\ 3 & 5 \\ 5 & 4 \\ 1 & 1 \end{matrix} \right)~,\,\,\left(\begin{matrix} 5 & 6 \\ 4 & 9 \\ 9 & 3 \\ 1 & 1 \end{matrix} \right)~,\,\,\left(\begin{matrix} 5 & 7 \\ 4 & 5 \\ 9 & 4 \\ 1 & 1 \end{matrix} \right)~,\,\,\left(\begin{matrix} 6 & 7 \\ 9 & 5 \\ 3 & 4 \\ 1 & 1 \end{matrix} \right)~. 
\eea
We must now check which of these has columns spanning an isotropic sublattice, i.e. $\widetilde K_L^T \mathfrak{I} \widetilde K_L = 0$.
As mentioned above, isotropicity is automatic if the eigenvalues of $J$ are not inverses, and in the current case we have $6 = 2^{-1}$ mod $11$ and $8 = 7^{-1}$ mod $11$. So we need only check the first and last matrices above, in which case we find $\widetilde K_L^T \mathfrak{I} \widetilde K_L = 6,4$ mod $11$ respectively. We see that these cases are not isotropic, and do not correspond to legitimate global forms. We conclude that for $p = 11$ there are four global forms with an invertible $\phi$ symmetry.

Having given two explicit examples, we now turn to generic $p$. We first look for the number of solutions in $\ZZ_p^\times$ to the  characteristic equation. The number of such solutions turns out to be 
\bea
\#\,\,\mathrm{solutions}= \left\{ \begin{matrix} 1 & & p = 5 \\ 4 & & p \in 10 \NN + 1 \\ 0 & & \mathrm{otherwise}\end{matrix} \right.~.
\eea
For generic $p \in 10 \NN+1$, we see that all four solutions $\{\lambda_1, \dots, \lambda_4\}$ are in $\ZZ_p^\times$. One can furthermore show that these come in inverse pairs, say $\lambda_2 = \lambda_1^{-1}$ and $\lambda_4 = \lambda_3^{-1}$.  Because $(10 \NN+1) \nmid 10$, it is not possible to have non-trivial Jordan blocks, so there are six choices for $J$, namely 
\bea
J = \mathrm{diag}(\lambda_i, \lambda_j)~, \hspace{0.5 in} 1 \leq i < j\leq 4~,
\eea
all of which have determinant in $\ZZ_p^\times$ since the diagonal entries are in $\ZZ_p^\times$. The corresponding matrices $\widetilde{K}_L$ satisfying (\ref{eq:maineqclassS3}) are 
\bea
\widetilde K_L = \left(\begin{matrix} p + \lambda_i - 1&p + \lambda_j - 1 \\ (p-1)\lambda_i^3 &(p-1)\lambda_j^3 \\ p-1+\lambda_i + (p-1)\lambda_i^2 + \lambda_i^3 &p-1+\lambda_j + (p-1)\lambda_j^2 + \lambda_j^3 \\ 1 & 1 \end{matrix} \right)~. 
\eea
What remains is to check that the columns span an isotropic sublattice. This is automatically the case if $\lambda_i$ and $\lambda_j$ are not modular inverses (which is four of the six cases), but must be checked otherwise. Taking $\lambda_j = \lambda_i^{-1}$, we find 
\bea
\tilde v_i^T \mathfrak{I} \tilde v_j = \lambda_i ( 2 \lambda_i - 1) - \lambda_j (2 \lambda_j - 1)\,\,\,\mathrm{mod}\,\,p~. 
\eea
It is easy to show that the right-hand side can never vanish for $\lambda_j \neq \lambda_i$, and hence these cases do not correspond to legitimate global forms. The remaining four cases do correspond to legitimate global forms though, and so we conclude that for any $p \in 10 \ZZ+1$ there exist four global forms with invertible $\phi$ symmetry. 

We finally turn to the case of $p>5$ and $p \not \in 10 \NN+1$. In this case there do not exist any solutions to the characteristic equation in $\ZZ_p^\times$. Furthermore, it can be shown that the product of any two elements which are not inverses of each other does not return an element of $\ZZ_p^\times$. Thus there are no $J$ that we can write down with the requisite properties, and hence no global forms with an invertible $\phi$ symmetry.  The results so far are summarized in the first line of Table \ref{table:genus2intrinsic}. 

 \begin{table}[!tp]
\begin{center}
\begin{tabular}{c|cccccccccc}
$p$ & 2 & 3 & 5 & 7 & 11 & 13 & 17 & 19 & 23 & 29
\\\hline
$\phi$ intrinsic? & \green{\cmark} & \green{\cmark} & \red{\xmark} & \green{\cmark} &  \red{\xmark} &\green{\cmark} & \green{\cmark} &\green{\cmark} &\green{\cmark} &\green{\cmark}
\\
$M_1$ intrinsic?  &  \red{\xmark} & \red{\xmark} &  \red{\xmark} & \red{\xmark} &  \red{\xmark} & \red{\xmark} &  \red{\xmark} &  \red{\xmark}  & \red{\xmark} & \red{\xmark}
\\
$M_2$ intrinsic?  &  \red{\xmark} & \red{\xmark} &  \red{\xmark} & \red{\xmark} &  \red{\xmark} & \red{\xmark} &  \red{\xmark} &  \red{\xmark}  & \red{\xmark} & \red{\xmark}
\\
$M_3$ intrinsic?  &  \red{\xmark} & \red{\xmark} &  \red{\xmark} & \red{\xmark} &  \red{\xmark} & \red{\xmark} &  \red{\xmark} &  \red{\xmark}  & \red{\xmark} & \red{\xmark}
\\
$M_7$ intrinsic?  &  \red{\xmark} & \red{\xmark} &  \red{\xmark} & \red{\xmark} &  \red{\xmark} & \red{\xmark} &  \red{\xmark} &  \red{\xmark}  & \red{\xmark} & \red{\xmark}
\\
$M_8$ intrinsic?  & \red{\xmark}& \red{\xmark} &  \red{\xmark} &\green{\cmark} & \red{\xmark} & \red{\xmark} &  \red{\xmark}& \red{\xmark} &\green{\cmark}& \red{\xmark}
\\
$N_2$ intrinsic?  & \red{\xmark}& \red{\xmark} &  \red{\xmark} & \red{\xmark} & \red{\xmark} & \red{\xmark} &  \red{\xmark}& \red{\xmark} & \red{\xmark} & \red{\xmark}
\\
$N_5$ intrinsic?   & \red{\xmark}& \red{\xmark} &  \red{\xmark} & \red{\xmark} & \red{\xmark} & \red{\xmark} &  \red{\xmark}& \red{\xmark} & \red{\xmark} & \red{\xmark}
\\
$N_6$ intrinsic?  & \red{\xmark}& \red{\xmark} &  \red{\xmark} & \red{\xmark} & \red{\xmark} & \red{\xmark} &  \red{\xmark}& \red{\xmark} & \red{\xmark} & \red{\xmark}
\end{tabular}
\end{center}
\caption{For any $p$, there are global variants such that the modular transformations in Tables \ref{tab:g2singmodtab1}, \ref{tab:g2singmodtab2}, \ref{tab:g2singmodtab3} give rise to non-invertible symmetries (upon appropriate dressing with topological manipulations $G$). We may then ask if these non-invertible symmetries are intrinsic or not. This is answered by looking for global forms left invariant by the modular transformations. The results are shown for the first few primes. }
\label{table:genus2intrinsic}
\end{table}%

 \paragraph{$GL(2,3)$:} We next consider the case of $GL(2,3)$. This group is generated by matrices $M_7$ and $M_8$, which are of respective order 2 and 8. The explicit form of the matrices is given in Appendix \ref{app:matrices}, which allows us to compute the relevant characteristic equations 
 \bea
 M_7: \hspace{0.2 in} \lambda^4 - 2 \lambda^2 + 1 = 0 \,\,\,\mathrm{mod}\,\,p~, \hspace{0.5 in} M_8: \hspace{0.2 in} \lambda^4 + 1 = 0\,\,\,\mathrm{mod}\,\,p ~. 
 \eea
 These admit the following number of solutions in $\ZZ_p^\times$, 
 \bea
 M_7 : \hspace{0.2 in} \#\,\,\mathrm{solutions}&=& \left\{ \begin{matrix} 1 & & p = 2 \\ 2 & & \mathrm{otherwise}\end{matrix} \right.
 \no\\
 M_8: \hspace{0.2 in} \#\,\,\mathrm{solutions}&=& \left\{ \begin{matrix} 1 && p=2\\  4 & & p \in 8 \NN+1  \\ 0 && \mathrm{otherwise}\end{matrix} \right.
 \eea
 We begin by analyzing $M_7$. Starting with $p=2$, the characterstic equation simplifies to $(\lambda +1)^4 = 0$ mod $2$ and the only solution is $\lambda = 1$. The possibilities for $J$ are $J=\left(\begin{smallmatrix} 1 & 0 \\ 0 & 1 \end{smallmatrix} \right),\, \left(\begin{smallmatrix} 1 & 1 \\ 0 & 1 \end{smallmatrix} \right)$; in the first case we have $\widetilde K_L = \left(\begin{smallmatrix} 1 & 1 & 1 & 1 \\ 0 & 0 & 1 &1  \end{smallmatrix} \right)^T$, while in the second case we can have either $\widetilde K_L = \left(\begin{smallmatrix} 1 & 1 & 1 & 1 \\ 0 & 1 & 0 &1  \end{smallmatrix} \right)^T,\,\left(\begin{smallmatrix} 1 & 1 & 0 & 0 \\ 0 & 1 & 0 &0  \end{smallmatrix} \right)^T$. It is straightforward to check that $\widetilde K_L^T \mathfrak{I} \widetilde K_L = 0$ mod $2$ in all cases, and hence all correspond to legitimate global forms. Thus for $p=2$ there are three global forms with an invertible $M_7$ symmetry. 
 
 Moving on to more general $p$, there are two solutions $\lambda = \pm 1$ mod $p$ in $\ZZ_p^\times$. The product of the remaining two solutions does not give an element in $\ZZ_p^\times$, so we need not consider them. Since in general $p \nmid 2$, there can be no non-trivial Jordan blocks, and  the only possible $J$ are 
 \bea
 J = \mathrm{diag}(1,1),\,\mathrm{diag}(1,-1),\, \mathrm{diag}(-1,1),\, \mathrm{diag}(-1,-1)~.
 \eea
The corresponding $\widetilde K_L$ take the form 
\bea\label{M7lattices}
\widetilde K_L = \left(\begin{matrix} a & b \\ a & b \\ c & d \\ c & d \end{matrix} \right)~, \hspace{0.5 in} a,b,c,d \in \ZZ_p~.
\eea
Since the eigenvalues are not modular inverses, all such $\widetilde K_L$ are isotropic and label legitimate global forms. In total there are $(p+1)^2$ distinct such $\widetilde K_L$, 
and hence for generic $p$ there are a total of $(p+1)^2$ global forms with an invertible $M_7$ symmetry.

We next turn to $M_8$. The exercise here proceeds in much the same way as before, and one begins by showing that there is a single global form with invariant $M_8$ symmetry for $p=2$ and four global forms with invariant $M_8$ symmetry for $p \in 8 \NN+1$. However, we now encounter a feature that we have not seen in previous examples. In particular, though there do not exist solutions to the characteristic equation in $\ZZ_p^\times$ for $p \not\in 8 \NN+1$, products of two of them can give elements of $\ZZ_p^\times$, and we may use these conjugate pairs to construct $J$. Indeed, over $\CC$ the solutions to the characteristic polynomial are simply $\lambda_n = e^{2 \pi i (2n-1) \over 8}$  for $n=1,\dots,4$, and we see in particular that $\lambda_1 \lambda_4 = \lambda_2 \lambda_3 = 1$, which is an element in $\ZZ_p^\times$. We must thus allow for the following matrices 
\bea
J = \mathrm{diag}( e^{2 \pi i \over 8},\, e^{14 \pi i  \over 8})~,\,\, \mathrm{diag}( e^{6 \pi i \over 8},\, e^{10 \pi i \over 8})~. 
\eea
Focusing on the first case for instance, we find that the corresponding $\widetilde{K}_L$ is of the form
\bea
\widetilde K_L = \left(\begin{matrix} 1 & 1 \\ 1 + \sqrt{2} & 1 + \sqrt{2} \\ -{1 \over \sqrt{2}} (1-i) & -{1 \over \sqrt{2}} (1+i)\\ {1\over \sqrt{2}}(1 + i(1+ \sqrt{2})) &  {1\over \sqrt{2}}(1 - i(1+ \sqrt{2}))\end{matrix} \right)~,
\eea
whose columns satisfy $\widetilde v_1^T \mathfrak{I} \widetilde v_2 = 4 i (1+ \sqrt{2}) $. We would like this quantity to be zero mod $p$ so that the columns span an isotropic sublattice, but before that we must first make sense of what $i$ and $i \sqrt{2}$ even mean in $\ZZ_p$. There are various inequivalent ways to make sense of these quantities: for example, we may demand the existence of elements $x$ and $y$ in $\ZZ_p$ satisfying either
\bea
x^2 = - 1\,\,\,\mathrm{mod}\,\,p~, \hspace{0.5 in} y^2 = 2\,\,\,\mathrm{mod}\,\,p~,
\eea
or 
\bea
x^2 = - 1\,\,\,\mathrm{mod}\,\,p~, \hspace{0.5 in} y^2 = -2\,\,\,\mathrm{mod}\,\,p~.
\eea
Considering all of the possibilities and demanding that $\widetilde v_1^T \mathfrak{I} \widetilde v_2 =0$ mod $p$, we find that there are no invariant global forms if and only if $p \in 8 \NN+7$. These results are collected in Table \ref{table:genus2intrinsic}.

Because the full $GL(2,3)$ symmetry is generated by both $M_7$ and $M_8$, we can now ask a somewhat more refined question: are there global forms which are left invariant under the \textit{full} $GL(2,3)$, i.e. under both $M_7$ and $M_8$? We will call the full non-Abelian symmetry $GL(2,3)$ intrinsic if there is no global form such that both $M_7$ and $M_8$ are invertible, even if there are global forms such that one or the other is individually invertible. In order to understand, for a given $p$, whether $GL(2,3)$ is intrinsic or not, we must know more than just the characteristic equation. Here we turn to a computerized search, which gives the results in Table \ref{tab:GL23intrinsic}. We see that for e.g. $p=2,3$ there exists a global variant such that the full $GL(2,3)$ becomes invertible,\footnote{More concretely, for $p=2$ there is one global variant fixed under both $M_7$ and $M_8$, 
\bea
K_{\cL} = \left(\begin{matrix}1 & 1 & 0 & 1 \\ 0 & 0 & 1 & 1 \end{matrix} \right)^T~, 
\eea
while for $p=3$ there are two global forms fixed by both,
\bea
K_{\cL} = \left(\begin{matrix}1 & 0 & 0 & 1 \\ 0 & 1 & 1 & 0  \end{matrix} \right)^T~, \hspace{0.5 in}K_{\cL} = \left(\begin{matrix}1 & 0 & 1 & 1 \\ 0 & 1 & 1 & 1\end{matrix} \right)^T~. 
\eea}
 whereas for $p=5,7$ no such global form exists.

\begin{table}[!tp]
\begin{center}
\begin{tabular}{c|cccccccccc}
$p$ & 2 & 3 & 5 & 7 & 11 & 13 & 17 & 19 & 23 & 29
\\\hline
${GL(2,3)}$ intrinsic? & \red{\xmark} & \red{\xmark} &  \green{\cmark}  & \green{\cmark} &  \red{\xmark} &\green{\cmark}&  \red{\xmark} & \red{\xmark} & \green{\cmark} & \green{\cmark}
\\
$(\ZZ_2 \times \ZZ_6)\rtimes \ZZ_2$ intrinsic? & \red{\xmark} & \red{\xmark} &  \green{\cmark}  & \red{\xmark}  & \green{\cmark} & \red{\xmark} & \green{\cmark} & \red{\xmark} &  \green{\cmark} & \green{\cmark}
\\
$D_8$ intrinsic? & \red{\xmark} & \red{\xmark} &  \red{\xmark} & \red{\xmark}  & \red{\xmark} & \red{\xmark} &  \red{\xmark} & \red{\xmark} & \red{\xmark} & \red{\xmark}
\\
$D_{12}$ intrinsic? & \red{\xmark} & \red{\xmark} &  \red{\xmark} & \red{\xmark}  &  \red{\xmark} & \red{\xmark} &  \red{\xmark} & \red{\xmark} &  \red{\xmark} &  \red{\xmark}
\\
$\ZZ_2\times \ZZ_2$ intrinsic? & \red{\xmark} & \red{\xmark} &  \red{\xmark}  & \red{\xmark}  &  \red{\xmark} & \red{\xmark} &  \red{\xmark} & \red{\xmark} &  \red{\xmark} &  \red{\xmark}
\end{tabular}
\end{center}
\caption{We call a rank $r>1$ group intrinsic if there is no global form such that \textit{all generators}  are simultaneously invertible. We do allow some of the generators to become invertible, though.}
\label{tab:GL23intrinsic}
\end{table}%

 \paragraph{$(\ZZ_2 \times \ZZ_6)\rtimes \ZZ_2$:} We finally mention the case of $(\ZZ_2 \times \ZZ_6)\rtimes \ZZ_2$, which is generated by the matrices $M_1, M_2,$ and $M_3$. The characteristic equations of $M_1$ and $M_2$ are identical, and the characteristic equation for $M_3$ coincides with that for $M_7$ above, 
 \bea
 M_1, M_2: \hspace{0.5 in}    \lambda^4 + 2 \lambda^2 + 1 =0 \,\,\,\mathrm{mod}\,\,p~, 
 \\
 M_3: \hspace{0.5 in}\lambda^4 - 2 \lambda^2 + 1 = 0 \,\,\,\mathrm{mod}\,\,p~. 
 \eea 
 The number of invariant global forms are to be the following
 \bea
 M_1, M_2:\hspace{0.2 in} \#\,\,\mathrm{invariant\,\,forms}&=& \left\{ \begin{matrix}  p+3 & & p \in 4\NN+1 \\p+1 & & \mathrm{otherwise} \end{matrix} \right. 
 \\
 M_3: \hspace{0.5 in} \#\,\,\mathrm{invariant\,\,forms}&=& \left\{ \begin{matrix} 3 & & p = 2 \\ (p+1)^2 & & \mathrm{other}\,\,p\end{matrix} \right. 
 \eea
 By an analysis similar to the previous ones, one finds that for all primes $p$ the symmetries $M_1$, $M_2$, and $M_3$ are non-intrinsic. However, the full group $(\ZZ_2 \times \ZZ_6)\rtimes \ZZ_2$ can still be intrinsic: that is, there exist some primes for which there does not exist a global form simultaneously invariant under all three of $M_1$, $M_2$, and $M_3$. This is summarized in Table \ref{tab:GL23intrinsic}. 

\paragraph{\textbf{Invertible subspaces, extended loci cases.}} Having exhaustively analyzed the isolated fixed loci, we now turn to the extended loci. Since the analysis is similar to the one above, we will be brief.

 \paragraph{$D_8$:} We begin with the case of $D_8$ generated by the matrices $M_7$ and $N_2$, given in Appendix \ref{app:matrices}. The generator $M_7$ was analyzed previously in the context of the $GL(2,3)$ symmetric point, with the conclusion that it is never intrinsic. As for $N_2$, the characteristic polynomial is given by
 \begin{equation}
    N_2: \hspace{0.5 in}    \lambda^4 + 2\lambda^2 + 1 =0 \,\,\,\mathrm{mod}\,\,p~,  
 \end{equation}
which is the same as for $M_1$ and $M_2$. 

The number of invariant global forms is found to be
\begin{equation}
   N_2:\hspace{0.2 in} \#\,\,\mathrm{invariant\,\,forms}= \left\{ \begin{matrix}  p+3 & & p \in 4\NN+1 \\p+1 & & \mathrm{otherwise} \end{matrix} \right.  
\end{equation}
This is obtained roughly as follows. For $p \in 4\NN+1$ the element $-1$ is a quadratic residue, meaning that there is an element $a\in \mathbb{Z}_p$ such that $a^2=-1$. The matrix $N_2$ then has two distinct eigenvalues $\lambda_{1,2}=\pm a$. Let the $a$ eigenspace be spanned by $v_1,\,v_2$ and the $-a$ eigenspace be spanned by $u_1,\,u_2$. It is guaranteed that $ v_2^{T} \mathfrak{I} v_1 = u_2^{T} \mathfrak{I} u_1  = 0$, and furthermore that the bases can be chosen to satisfy $ v_1^{T} \mathfrak{I} u_2 = v_2^{T} \mathfrak{I} u_1 = 0$. We then have the following isotropic lattices: $\langle v_1, v_2\rangle$, $\langle u_1, u_2\rangle$, $\langle v_1, u_2\rangle$, $\langle v_2, u_1\rangle$ and $\langle v_1 + k v_2, u_1 + s u_2 \rangle$, where $k\in\mathbb{Z}_p^\times$ and $s=-k^{-1}(v_2^T \mathfrak{J} u_2)/(v_1^T \mathfrak{J} u_1)$. There are thus a total of $4 + (p-1) = p+3$ invariant global forms.  A similar analysis can be performed for $p\not \in 4\NN+1$.

We can also ask if there are isotropic sublattices invariant under both $M_7$ and $N_2$. 
The simplest way to check this is to start with the sublattices invariant under $M_7$ described around (\ref{M7lattices}), and then to check which (if any) of them are invariant under $N_2$. A simple computation demonstrates that for $p>2$, out of the $(p+1)^2$ global forms invariant under $M_7$, there are $p-2$ global forms invariant under the whole $D_8$, while for $p=2$ there are two such forms. Thus the full $D_8$ is never intrinsic.

 \paragraph{$D_{12}$:} We next consider the case of $D_{12}$, generated by $N_5$ and $N_6$. Both generators have the same characteristic polynomial:
 \begin{equation}
    N_5,\,N_6: \hspace{0.5 in}    \lambda^2 - 2\lambda^2 + 1 =0 \,\,\,\mathrm{mod}\,\,p~,  
 \end{equation}
 which has one root for $p=2$ ($\lambda=1$) and two different roots for prime $p>2$ ($\lambda=\pm1$).  The number of invariant isotropic sublattices (which is also the same for both $N_5$ and $N_6$) is

\begin{equation}
   N_5,\,N_6:\hspace{0.2 in} \#\,\,\mathrm{invariant\,\,forms}= \left\{ \begin{matrix}  3 & & p=2 \\(p+1)^2 & & \mathrm{otherwise} \end{matrix} \right.  
\end{equation}

We can again ask if there are global forms invariant with respect to $N_5$ and $N_6$ simultaneously. We find the following

\begin{equation}
   \#\,\,\mathrm{simult.\, \,invariant\,\,forms}= \left\{ \begin{matrix}  2 & & p=2 \\ 7 & & p=3  \\p+1 & & \mathrm{otherwise} \end{matrix} \right.  
\end{equation}
from which we conclude that $D_{12}$ is never intrinsically non-invertible.

 \paragraph{$\mathbb{Z}_2\times\mathbb{Z}_2$:} Finally we consider the case of the $\ZZ_2 \times \ZZ_2$ symmetry spanned by $\{C, M_7\}$. The analysis in this case is particularly simple: we have already seen that $M_7$ fail to be intrinsic for every value of $p$, while the generator $C$ is obviously never intrinsic, since it leaves invariant every isotropic sublattice. The combined symmetry is also obviously non-intrinsic.

\section{Intrinsic versus non-intrinsic at higher genus}
\label{sec:non-invert-symm}
In the previous section we understood in detail the spectrum of (non-)invertible symmetries for class $\cS$ theories of genus $2$. We now turn to higher genus theories. 
The special points in moduli space have been classified up to genus 5 \cite{kuribayashi1990automorphism1,kuribayashi1990automorphism2}, and
many examples are known at higher genera. Our discussion here will not be as exhaustive as before, and will instead focus on cyclic subgroups of enhanced symmetries.

In the references \cite{kuribayashi1990automorphism1,kuribayashi1990automorphism2}, the stabilizer subgroups at the special points of moduli space are realized as subgroups of $GL(g,\mathbb{C})$ acting on holomorphic one-forms.
Given such a matrix $U = A + iB$ with $A$ and $B$ real, we can construct a $2g \times 2g$ matrix
\begin{align}
  \label{eq:3.1}
  M = \begin{pmatrix}
        A & -B \\
        B & A
      \end{pmatrix} ~.
\end{align}
For stabilizer subgroups this matrix is real symplectic, and moreover is similar to an integer symplectic matrix $F$. It is this integer matrix which we will analyze.

We warn the reader that the content of this section is rather technical, and the arithmetically-averse reader may wish to skip directly to Section \ref{sec:topops}. For such readers, let us simply note here that the content of this section translates to an answer to the following physical question. Given a $(\ZZ_p^{(2)})^{\otimes 2g}$ gauge theory in $(4+1)$d, there is an $Sp(2g, \ZZ_p)$ zero-form symmetry acting via outer automorphisms, and we may gauge certain subgroups of this symmetry. In general, the result will \textit{not} be another gauge theory, even for a higher-group symmetry. The results of this section tell us, for $g\leq 5$, when the result \textit{is} a gauge theory: namely, this happens whenever there is a non-zero entry in Tables \ref{tab:genus3table}, \ref{tab:genus4table}, or \ref{tab:genus5table}.

\subsection{Characteristic polynomials of modular matrices}
\label{sec:char-poly-duality}
We start by introducing some relevant machinery. Consider elements $F \in Sp(2g, \ZZ)$ leaving some point in moduli space fixed. Such elements are always of finite order. This implies that as complex matrices the $F$ are diagonalizable, and moreover that their eigenvalues are $k$-th roots of unity. These two facts together mean that the characteristic polynomials
$P_{F}(x)$ are products of cyclotomic polynomials $\phi_i(x)$ with total degree $2g$; see Appendices \ref{sec:roots-unity-over} and \ref{sec:integ-matr-finite} for a detailed discussion. Since cyclotomic polynomials of a given degree are limited, there are relatively few characteristic polynomials at a given genus. 

As
an example, consider the case of genus 1, for which we are interested in products of cyclotomic polynomials of total degree 2. There are two possible reducible characteristic polynomials:
\begin{align}
  \label{eq:3.2}
   \phi_{1}(x)^{2} = (x-1)^{2} = P_{\mathbbm{1}}(x)
  && \phi_{2}(x)^{2} = (x+1)^{2} =  P_{-\mathbbm{1}}(x) ~,
\end{align}
which are clearly the characteristic polynomials of $\mathbbm{1}$ and $-\mathbbm{1}$.  There are also two possible irreducible characteristic polynomials,\footnote{
Note that we could have also considered  $\phi_{3}(-x)$, but in general for a $2g \times 2g$ diagonalizable matrix $M$ we have $P_{-M}(x) = P_{M}(-x)$, and hence when
considering the action of stabilizer subgroups the matrices with characteristic polynomials $P(x)$ and $P(-x)$ are identified in ${Sp}(2g,\mathbb{Z}) / \mathbb{Z}_{2}$.}
\begin{align}
  \label{eq:3.3}
   \phi_{3}(x) = x^{2} + x + 1 = P_{\sfS\sfT}(x)~, && \phi_{4}(x)= x^{2} + 1 = P_{\sfS}(x) ~, 
\end{align}
which we recognize as the characteristic polynomials of the $\sfS\sfT$ and $\sfS$ elements of $SL(2, \ZZ)$.

Of special interest to us are the cases in which $P_{F}(x)$ is a single cyclotomic polynomial $\phi_{n}(x)$, as opposed to a product thereof. We have just seen that $F = \sfS$ and $\sfS\sfT$ are of this type, for example. Such matrices $F$ have order $n$, and as will be explained later serve as the building blocks for all other cases.
Note that the degree of the $n$-th cyclotomic polynomial $\phi_{n}(x)$ is Euler's totient function $\varphi(n)$, i.e the number of integers less than $n$ which are coprime to $n$. However, not all $\phi_{n}(x)$ such that $\varphi(n) = 2g$ need to be considered: indeed, a matrix with characteristic polynomial $\phi_{n}(x)$ generates a cyclic subgroup of order $n$, and since the maximum order of cyclic isometries is known to be $4g+2$ by the ``$(4g+2)$ theorem'' \cite{Farb2013APO}, only $\phi_{n}(x)$ with $n \leq 4g+2$ are relevant.

\subsubsection{Cyclotomic polynomials over finite fields}
\label{sec:cycl-polyn-over}
Over integers, the cyclotomic polynomials $\phi_n(x)$ are by definition irreducible, i.e. they do not factorize.  On the other hand, over complex numbers they factorize completely, with the roots being primitive $n$-th roots of unity $\exp(2 \pi i \frac{m}{n})$ with $m$ and $n$ coprime.  The situation over finite fields is somewhat intermediate, in that $\phi_{n}(x)$ can factorize partially, but not always completely. We are ultimately interested in prime fields $\mathbb{F}_{p}$, but it is
convenient to start with an appropriate extension of $\mathbb{F}_{p}$ in which $\phi_{n}(x)$ does split completely. 
 Over this field $\phi_{n}(x)$ has $\varphi(n)$ roots. These roots are not necessarily distinct: if $p^{k}$ is the highest power of $p$ that divides
$n$, the number of distinct roots is $\varphi\left(\frac{n}{p^{k}}\right)$,\footnote{Another way of saying this is that the group of $n$-th roots of unity in the splitting field
  of $\phi_{n}(x)$ over $\mathbb{F}_{p}$ has order $\varphi\left(\frac{n}{p^{k}}\right)$.} and each of these roots occurs with multiplicity $\varphi(p^{k})$, i.e. we have
\begin{align}
  \label{eq:3.4}
  \phi_{n}(x) &= \left(\phi_{\frac{n}{p^{k}}}(x)\right)^{\varphi(p^{k})} ~;
\end{align}
see Appendix \ref{sec:roots-unity-over} for more details.
As we will see below, this fact essentially fixes the Jordan normal form of all $F$ of interest. 

With the above factorization in hand, we now restrict to $\phi_{n}(x)$
with $\gcd(n,p) = 1$ and consider how $\phi_{n}(x)$ can factorize over $\mathbb{F}_{p}$. The simplest case is when $n$ divides $p-1$. In this case
$\mathbb{F}_{p}$ contains a multiplicative element of order $n$, which we call $\omega$, and $\phi_{n}(x)$ splits completely,
\begin{align}
  \label{eq:3.5}
  \phi_{n}(x) = \prod_{\substack{m < n\\\gcd(m,n)=1}} (x - \omega^{m}) ~.
\end{align}
This is not the only possible factorization that can happen. Indeed, since $\gcd(n,p) = 1$, the element $p$ is an invertible element in $\mathbb{Z}_{n}$. If $m$ is the order of $p$ in $\mathbb{Z}_{n}$, then
$\phi_{n}(x)$ can also split into $\frac{\varphi(n)}{m}$ factors over $\mathbb{F}_{p}$, all of which have the same degree $m$.

\subsection{Invariant subspaces of modular matrices}
\label{sec:invar-subsp-dual}
We have seen how the characteristic polynomials of $F$ are related to cyclotomic polynomials. We now explain how to use this fact to identify invariant global forms, i.e. maximal isotropic sublattices invariant under $F$. To begin, let us temporarily forget the constraints of maximality and isotropicity and just identify invariant subpaces in general.  
It is again convenient to start by considering $F$ such that the characteristic polynomial (over integers) of $P_{F}(x)$ is a single cyclotomic polynomial $\phi_{n}(x)$.

Let us first consider the generic case, in which we are working over a finite field $\mathbb{F}_{p}$ such that $p$ is not a factor of $n$. As explained above, in this case the eigenvalues of $F$ in a suitable extension of $\mathbb{F}_{p}$ are all pairwise distinct and
given by the roots of $\phi_{n}(x)$ as in \eqref{eq:3.5}. As a result, over this field the homology of $\Sigma_{g,n}$ splits into eigenspaces, each of which is
one-dimensional, and any invariant subspace can be written as a sum of these eigenspaces. From this fact we can derive a sufficient condition and a
necessary condition for the existence of non-trivial invariant subspaces.

First for the sufficient condition, when $p - 1 \in n \mathbb{N}$ all the roots of $\phi_{n}(x)$ are in $\mathbb{F}_{p}$ itself, and hence all the eigenspaces described above give a decomposition of
$H_{1}(\Sigma_{g,0},\mathbb{F}_{p})$. There are $\binom{2g}{d}$ invariant subspaces of dimension $d$.

The necessary condition comes from the fact that all invariant subspaces are annihilated by some factor of the characteristic polynomial. All factors of
$\phi_{n}(x)$ over $\mathbb{F}_{p}$ have degree $m$ where $m$ is the multiplicative order of $p$ in $\mathbb{Z}_{n}$. Hence invariant subspaces of dimension $d$ exist only if $d$ is a multiple of $m$, and in that case there are $\binom{\varphi(n) / m}{d / m}$ such subspaces.

\subsubsection{Invariant isotropic subspaces}
\label{sec:invar-isotr-subsp}
The considerations above allow us to understand which subspaces are invariant under the action of an enhanced symmetry $F$. But to identify invariant global forms we need to identify which of those subspaces have dimension $g$ and are isotropic. Naively there is no reason to expect that the Jordan decomposition of a matrix $F$ can tell us which of its invariant subspaces are isotropic, since all similar matrices have the same Jordan normal form and there is no reason to expect that
similarity transformations preserve isotropy. However, the fact that $F$ preserves the intersection form, together with the restrictions we
have derived using the finite order of $F$, actually suffices to fix which of the invariant subspaces are isotropic. 

To see this, we begin not in $\mathbb{F}_{p}$ but in the field extension which splits the characteristic polynomial $P_{F}(x) = \phi_{n}(x)$. We now consider two cases:

First, we consider the generic case in which $p$ is not a factor of $n$ and we take $n > 2$ since $n=1,2$ are trivial. Then all of the eigenvalues of $\phi_{n}(x)$ are
distinct. We saw around (\ref{eq:37}) that two eigenvectors $v_{1}$ and $v_{2}$ with eigenvalues $\lambda_{1}$ and $\lambda_{2}$ can intersect only if $\lambda_{1}\lambda_{2} = 1$ mod $p$, i.e. the two eigenvalues are
modular inverses of each other. From \eqref{eq:3.5}, it can be seen that if $\lambda$ is an eigenvalue then $\lambda^{-1}$ is also an eigenvalue. Since the eigenvector $v$ with
eigenvalue $\lambda$ does not intersect the complement of eigenvector $v^{\prime}$ with eigenvalue $\lambda^{-1}$, the eigenvectors $v$ and $v^{\prime}$ must have non-zero intersection, since the intersection form $\mathfrak{I}$ of $\Sigma_{g,0}$ is non-degenerate. As a result, if an invariant global form is to exist the polynomial $\phi_{n}(x)$ must factorize over $\mathbb{F}_{p}$
as
\begin{align}
  \label{eq:3.6}
  \phi_{n}(x) = P(x)Q(x) ~,
\end{align}
where both $P(x)$ and $Q(x)$ have degree $\frac{1}{2}\varphi(n)$. Over the splitting field of $\phi_{n}(x)$, both $P(x)$ and $Q(x)$ split as well,  and isotropy requires
that if $\lambda$ is a root of $P(x)$, then $\lambda^{-1}$ is a root of $Q(x)$. As a result, there are far fewer invariant isotropic subspaces than invariant
subspaces in general. For example, in the case when $p \in n \mathbb{N} + 1$ where $\phi_{n}(x)$ splits over $\mathbb{F}_{p}$, there are $\binom{2g}{g}$ invariant subspaces of dimension
$g$, but only $2^{g}$ of them are isotropic.

Next, if $p$ is a factor of $n$, then for each eigenvalue $\lambda$ there is a single Jordan block of size $\varphi(p^{k})$, where $p^{k}$ is the highest power of $p$ dividing $n$. The
eigenvalues themselves are roots of $\phi_{\frac{n}{p^{k}}}(x)$. This Jordan decomposition means that for each eigenvalue $\lambda$, we can find $\varphi(p^{k})$ generalized eigenvectors 
$v^{(a)}_{\lambda}$ such that,
\begin{align}
  \label{eq:3.7}
  Dv^{(0)}_{\lambda} &= \lambda v^{(0)}_{\lambda} ~, \nonumber \\
  Dv^{(a)}_{\lambda} &= \lambda v^{(a)}_{\lambda} + v^{(a-1)}_{\lambda} ~.
\end{align}
It can be shown that $\ev{v^{(a)}_{\lambda_{1}} , v^{(b)}_{\lambda_{2}}}$ is zero unless $\lambda_{1} \lambda_{2} = 1$ mod $p$ and $a+b \ge \varphi(p^{k})$; see Appendix \ref{sec:inters-jord-blocks}. Unpacking this a bit, if $n = p^{k}$ or
$n=2p^{k}$, then we obtain a single Jordan block of size $\varphi(n)$ and there is a single invariant isotropic subspace of dimension $d$ if $d \le \frac{\varphi(n)}{2} = g$.
This subspace is the span of $v^{(0)}_{\pm 1} , v^{(1)}_{\pm 1} , \dots , v^{(d-1)}_{\pm 1}$ where $1$ is the eigenvalue when $n = p^{k}$ and $-1$ is the eigenvalue
when $n=2p^{k}$. There are no invariant isotropic subspaces of dimension greater than $g = \frac{\varphi(n)}{2}$.

If $\frac{n}{p^{k}} > 2$, each Jordan block for a given eigenvalue is isotropic and the vector $v_{\lambda}^{(a)}$ can intersect only with $v_{\lambda^{-1}}^{(b)}$ if
$a+b \ge g$.

\subsection{Invariant global forms}
\label{sec:invar-glob-forms}
After the rather technical discussion above, we now determine the global forms left invariant by matrices $F$ generating isometries of Riemann surfaces. 
In this subsection we discuss all possible examples at genus 3, 4, 5 such that the characteristic polynomial is a cyclotomic polynomial. 
See Appendix \ref{sec:composite-char-pol} for a discussion of the more general case in which the characteristic polynomial is a product of cyclotomic polynomials.

\subsubsection{Genus 3}
\label{sec:genus-3}
We start with the case of genus 3, for which we are interested in cyclotomic polynomials of degree $2g=6$. There are 4 of them, namely $\phi_{7}(x), \phi_{9}(x), \phi_{14}(x),$ and $\phi_{18}(x)$. Since $18$ is larger than the
upper limit $4g+2 = 14$ for the order of cyclic subgroups of $Sp(2g, \ZZ)$, no $F$ can have $\phi_{18}(x)$ as its characteristic polynomial. On the other hand, if a matrix $F$ has characteristic
polynomial $\phi_{14}(x)$, then the matrix $-F$ has characteristic polynomial $\phi_{7}(x)$, and the invariant subspaces of $F$ and $-F$ are the same. Hence we may focus only on the cases of $\phi_{7}(x)$ and $\phi_{9}(x)$.

\paragraph{$\phi_{7}(x)$ :}
According to \cite{genus-3-isolated-fixed-points}, there are two fixed points at genus $3$ for which the stabilizer subgroup has order $7$. Defining the matrices 
\bea
A = \left(\begin{matrix} 0 & 1 & 0 \\ 1 & 0 & 1 \\ 0 & 1 & -1 \end{matrix} \right)~,\hspace{0.3 in} B = \left(\begin{matrix} {\alpha + \alpha^2 \over 1 + 3 \alpha} &  {\beta + \beta^2 \over 1 + 3 \beta} &  {\gamma + \gamma^2 \over 1 + 3 \gamma} \\ {1 + 2 \alpha \over 1 + 3 \alpha}& {1 + 2 \beta \over 1 + 3 \beta} & {1 + 2 \gamma \over 1 + 3 \gamma} \\ {\alpha^2 \over 1 + 3 \alpha} & {\beta^2 \over 1 + 3 \beta} & {\gamma^2 \over 1 + 3 \gamma}\end{matrix} \right)~, \hspace{0.3 in} C = \mathrm{diag}\left(1,1,-1 \right) 
\eea
for $\alpha = 2 \cos {2 \pi \over 7}$, $\beta = 2 \cos {4\pi \over 7}$, and $\gamma = 2 \cos {6 \pi \over 7}$, the two isolated fixed points are given by 
\beaa
\Omega_{7,1} &:=& -{1\over 2} A + i B\left(\mathds{1} - {1\over 4} B^TA^2 B \right)^{1\over 2}B^T~, 
\no\\
\Omega_{7,2} &:=& -{1\over 2} A(A+\mathds{1}) + i B\left[ \left(\mathds{1}- {1\over 4} B^T A^2 B \right)^{1\over 2} B^T(A+\mathds{1})BC\right] B^T
\eeaa
 All of the non-identity elements for
the symmetry groups at these points have $\phi_{7}(x)$ as their characteristic polynomials.

We begin by considering the generic case for which $p$ is not a factor of 7. This means that we consider $p \in 7 \NN+k$ for $k=1, \dots, 6$. Beginning with $p \in 7\mathbb{N}+1$, by the sufficient condition at the end of Section \ref{sec:invar-subsp-dual} we see that $\mathbb{F}_{p}$ is the splitting field of $\phi_{7}(x)$ and hence that 8 invariant subspaces exist. The non-invertible symmetries are non-intrinsic.

Moving on to  $p \in 7\mathbb{N} + 2$ or $7\mathbb{N} + 4$, we note that the elements $2$ and $4$ have degree $3$ in $\mathbb{Z}^{\times}_{7}$, so the polynomial $\phi_{7}(x)$ splits into two factors of degree $3$, c.f. (\ref{eq:3.4}). The only possible factorization that can make the corresponding kernels isotropic is $\phi_{7}(x) = P(x)Q(x)$ with
\begin{align}
  \label{eq:3.8}
  P(x) = (x - \lambda)(x-\lambda^{2})(x-\lambda^{4}) ~, && Q(x) = (x-\lambda^{3})(x-\lambda^{5})(x-\lambda^{6}) ~.
\end{align}
Here $\lambda$ is a primitive seventh root of unity that exists in the splitting field. We want to determine if $P(x)$ and $Q(x)$ are polynomials over $\mathbb{F}_{p}$.
This is the case if $\lambda + \lambda^{2} + \lambda^{4} \in \mathbb{F}_{p}$, which can be shown to happen if there exists a square root of $-7$ mod $p$. This is
indeed the case for $p \in 7\mathbb{N} + 2$ or $7\mathbb{N} + 4$. Hence in these cases there are two invariant global forms given by $\ker(P(F))$ and $\ker(Q(F))$.

In the cases of $p \in 7\NN+3$ and $7\NN+5$ we note that the elements $3$ and $5$ in $\mathbb{Z}_{7}$ both generate $\mathbb{Z}^{\times}_{7}$, and hence there are no invariant global forms and the symmetry is intrinsically non-invertible. Finally for the case of $p \in 7\NN+6$ we note that the element $6$ has order $2$ in $\mathbb{Z}_{7}^{\times}$, so the polynomial $\phi_{7}(x)$ splits into three factors of degree 2. In this case all invariant subspaces are even-dimensional, and in particular not equal to $g=3$. Hence there is no maximal invariant subspace, and hence no invariant global form.

Having discussed the generic case of $p$ not dividing 7, we now discuss the exceptional case of $p = 7$. This is part of the family discussed in \cite{Bashmakov:2022jtl}, where it was seen to have a single invariant global form. This can be recovered from our analysis in Section \ref{sec:invar-isotr-subsp}, which shows that there is a single Jordan block with eigenvalue $1$  in this case. The unique invariant global form is the span of first three vectors in the Jordan block.

The results so far are summarized in the first line of Table \ref{tab:genus3table}.

\begin{table}[!tp]
\begin{center}
\begin{tabular}{c|cccccccccc}
$p$ & 2 & 3 & 5 & 7 & 11 & 13 & 17 & 19 & 23 & 29
\\\hline
$\phi_7(x)$ & 2 & 0 &  0  & 1 &2   &0 &  0&0  & 2 & 8
\\
$\phi_9(x)$ & 0&1  &0  & 2 &0 &2  &0  &8  &0  & 0
\end{tabular}
\end{center}
\caption{The number of invariant global forms for enhanced symmetries $F$ with characteristic polynomials given by $\phi_7(x)$ and $\phi_9(x)$, relevant for genus 3. The $F$ give rise to intrinsically non-invertible symmetries if and only if the entry above is $0$.}
\label{tab:genus3table}
\end{table}%

\paragraph{$\phi_{9}(x)$ :} The reference \cite{genus-3-isolated-fixed-points} also contains one fixed point with stabilizer subgroup of order 9. Defining the matrices
\beaa
A &=& \left( \begin{matrix} 0 & 1 & 0 \\ 1 & -1 & 1 \\ 0 & 1 & 1 \end{matrix} \right)~, \hspace{0.1 in}B = {1\over 3} \left(\begin{matrix}-3 + \alpha+\alpha^2 & -3 + \beta + \beta^2 & - 3 + \gamma + \gamma^2 \\ -1+\alpha^2 & -1 + \beta^2 & -1 + \gamma^2 \\ 1 + \alpha & 1 + \beta & 1 + \gamma \end{matrix}  \right)
\eeaa
for $\alpha = 2 \cos {2 \pi \over 9}$, $\beta + 2 \cos {4 \pi \over 9}$, and $\gamma = 2 \cos {8 \pi \over 9}$, the isolated fixed point occurs at the following value of the period matrix, 
\bea
\Omega_9 := -{1 \over 2} A + i B \left(\mathds{1}-{1\over 4} B^T A^2 B\right)^{1\over 2} B^T~.
\eea
 Any generator of the subgroup preserving this value has $\phi_{9}(x)$ as its characteristic polynomial.

We begin with the generic case of $p$ not dividing 9. We would naively have to consider the cases of $p \in 9 \NN + k$ for $k=1, \dots, 8$, but since $p$ is prime we may skip the cases of $k=3,6$. Using the same arguments as for $\phi_{7}(x)$ above, we find that the symmetry is intrinsically non-invertible unless $p$ has order $3$ in $\mathbb{Z}_{9}^{\times}$. The latter occurs whenever $p \in 9\mathbb{N} + 1, 9\mathbb{N} + 4,$ or $9\mathbb{N} + 7$, i.e. when $p \in 3\mathbb{N} + 1$. Let $\lambda$ be the primitive $9$-th root of unity in the splitting field. For $p \in 3\mathbb{N} + 1$, a primitive third root of
unity exists in $\mathbb{F}_{p}$ and must be $\lambda^{3}$. It can be shown that
\beaa
  \label{eq:3.9}
  P(x) &=& (x - \lambda)(x-\lambda^{4})(x-\lambda^{7}) = x^{3} - \lambda^{3} ~, 
 \no\\
   Q(x) &=& (x - \lambda^{2})(x - \lambda^{5})(x - \lambda^{7}) = x^{3} - \lambda^{6} ~.
\eeaa
Hence $P(x)$ and $Q(x)$ are both polynomials with coefficients in $\mathbb{F}_{p}$, and they satisfy the criterion for the corresponding kernels to be isotropic.
If $p \in 9\mathbb{N} + 4$ or $ 9\mathbb{N} + 7$, then these are the only two invariant global forms. If $p \in 9\mathbb{N} + 1$, then $\lambda \in \mathbb{F}_{p}$ and there are $8$ invariant global forms.

We finally discuss the exceptional case of $p=3$. In this case we obtain a single Jordan block of size 6 and eigenvalue $1$, and hence there is a
single invariant global form consisting of the span of the first three vectors in the Jordan block.

The results of the above discussion are given in the second line of Table \ref{tab:genus3table}.

\subsubsection{Genus 4}
\label{sec:genus-4}
For genus 4, we are interested in cyclotomic polynomials $\phi_{n}(x)$ with degree $2g=8$ and $n \le 4g+2=18$. There are again only two physically distinct possibilities, namely  $\phi_{15}(x)$ and $\phi_{16}(x)$. The stabilizer subgroups
at genus 4 were classified in \cite{kuribayashi1990automorphism1} and contain cyclic groups of order $15$ and $16$. The characteristic polynomial of any
generator $F$ of these groups was indeed found to be $\phi_{15}(x)$ and $\phi_{16}(x)$. 

\paragraph{$\phi_{15}(x): $} We begin with the case of $\phi_{15}(x)$. One can again split the discussion into two cases:  $p$ not being a factor of 15, and $p$ being a factor of 15. We will begin by analyzing the latter here, i.e. the cases of $p=3$ and $5$.

For $p=3$, the cyclotomic polynomial partially factorizes as 
\begin{align}
  \label{eq:3.10}
  \phi_{15}(x) = \phi_{5}(x)^{2} = (x^{4} + x^{3} + x^{2} + x +1)^{2} ~.
\end{align}
Since $\phi_{5}(x)$ does not split over $\mathbb{F}_{3}$, the only maximally invariant subspace (i.e. invariant subspace of dimension 4) is $\ker(\phi_{5}(F))$. This subspace is  isotropic. This can be seen by passing to the splitting field of $\phi_{5}(x)$ and using the discussion in Section \ref{sec:invar-isotr-subsp}, which tells us that the non-zero intersections are
\begin{align}
  \label{eq:3.11}
  \ev{v_{\lambda}^{(0)} , v_{\lambda^{4}}^{(1)}} &= \ev{v_{\lambda}^{(1)} , v_{\lambda^{4}}^{(0)}} = \ev{v_{\lambda^{2}}^{(0)} , v_{\lambda^{3}}^{(1)}} = \ev{v_{\lambda^{2}}^{(1)} , v_{\lambda^{2}}^{(0)}} = 1 ~
\end{align}
where $\lambda$ is any the root of $\phi_{5}(x)$. On the other hand $\ker{(\phi_{5}(M))} = \mbox{span}\{v_{\lambda}^{(0)} , v_{\lambda^{2}}^{(0)} , v_{\lambda^{3}}^{(0)} , v_{\lambda^{4}}^{(0)}\}$ and is hence isotropic. Thus there is a single invariant global form.

Next we consider $p=5$, for which
\begin{align}
  \label{eq:3.12}
  \phi_{15}(x) = (\phi_{3}(x))^{4} = (x^{2} + x +1)^{4} ~.
\end{align}
Once again, there is a single candidate for an invariant global form, namely $\ker(\phi_{3}(F)^{2})$. It can be seen to be isotropic using the same argument as for $p=3$.

The generic case when $p \ne 3,5$, requires some more involved algebra, which we relegate to Appendix \ref{sec:fact-over-finite}. Here we only summarize the results:
\begin{itemize}
  \item The only case with intrinsic non-invertible symmetry is $p \in 15 \mathbb{N} + 14$.
  \item If $p \in 15 \mathbb{N} + 2, 15 \mathbb{N} + 7, 15 \mathbb{N} + 8$ or $15 \mathbb{N} + 13$ there are two invariant global forms.
  \item If $p \in 15 \mathbb{N} + 4 $ or $15 \mathbb{N} + 11$ there are four invariant global forms.
  \item If $p \in 15 \mathbb{N} + 1$ there are eight invariant global forms.
\end{itemize}

\begin{table}[!tp]
\begin{center}
\begin{tabular}{c|cccccccccc}
$p$ & 2 & 3 & 5 & 7 & 11 & 13 & 17 & 19 & 23 & 29
\\\hline
$\phi_{15}(x)$ & 2 & 1 &  1  & 2 &4   &2 &2  &4 & 2 & 0
\\
$\phi_{16}(x)$ & 1 &0  & 0 &  0& 0& 0 &16  &0  &0  & 0
\end{tabular}
\end{center}
\caption{The number of invariant global forms for symmetry generators with characteristic polynomials given by $\phi_{15}(x)$ and $\phi_{16}(x)$, relevant for genus 4. There exist intrinsically non-invertible symmetries if and only if the entry above is $0$.}
\label{tab:genus4table}
\end{table}%

\paragraph{$\phi_{16}(x)$ : } We next consider the case of $\phi_{16}(x)$. The discussion again splits into the generic case for which $p$ is not a factor of 16, and the exceptional case when $p=2$. In the latter case there is a single Jordan block with eigenvalue $1$, and hence a single invariant subspace spanned by
$v_{1}^{(0)} , v_{1}^{(1)} , v_{1}^{(2)}$, and $v_{1}^{(3)}$.

For $p>2$, the cyclotomic polynomial always factorizes to some degree. To carry out the analysis, we start in the splitting field which has a primitive sixteenth root of unity $\lambda$. In the splitting field we may write the cyclotomic polynomial as 
\begin{align}
  \label{eq:3.13}
  \phi_{16}(x) = (x-\lambda)(x-\lambda^{3})(x-\lambda^{5})(x-\lambda^{7})(x-\lambda^{9})(x-\lambda^{11})(x-\lambda^{13})(x-\lambda^{15}) ~.
\end{align}
When $p \in 16\mathbb{N}+1$, the root $\lambda$ is in $\mathbb{F}_{p}$ itself, and there are 16 invariant global forms, c.f. the necessary condition in Section \ref{sec:invar-subsp-dual}.

When $p \in 16 \mathbb{N} +9$, there exists a primitive eighth root of unity $\sigma$ in $\mathbb{F}_{p}$ with $\sigma = \lambda^{2}$ and $\sigma^{4} = -1$. Hence the product of two
factors $(x-\lambda^{2i+1})(x-\lambda^{2j+1})$ is a polynomial over $\mathbb{F}_{p}$ if
\begin{align}
  \label{eq:3.14}
  \lambda^{2i+1} + \lambda^{2j+1} \in \mathbb{F}_{p} \hspace{0.5 in} \Rightarrow \hspace{0.5 in} \lambda(\sigma^{i}+\sigma^{j}) \in \mathbb{F}_{p}~.
\end{align}
Since $\lambda \notin \mathbb{F}_{p}$, this happens whenever $\sigma^{i} = -\sigma^{j}$, which in turn happens whenever $i = j+4$. Hence
\begin{align}
  \label{eq:3.15}
  \phi_{16}(x) = (x^{2}-\sigma)(x^{2}+\sigma)(x^{2} - \sigma^{3})(x^{2}+\sigma^{3}) ~
\end{align}
and there are $4$ basic invariant subspaces of dimension 2. All of them are isotropic.

When $p \in 8 \mathbb{N} + 5$ a primitive fourth root of unity $\kappa$ exists and $\kappa = \lambda^{4}$. The cyclotomic polynomial $\phi_{16}(x)$ then factorizes over $\mathbb{F}_{p}$ into two factors of degree 4, i.e $\phi_{16}(x) = P(x)Q(x)$.
Isotropy requires that if $\lambda$ is a root of $P(x)$, then $\lambda^{-1}$ is a root of $Q(x)$. Imposing this together with requiring that the coefficient of the constant term is
a power of $\kappa$ leaves four possibilities,
\begin{align}
  \label{eq:3.16}
  P(x) &= (x-\lambda)(x-\lambda^{3})(x-\lambda^{5})(x-\lambda^{7})~, &&& Q(x) &= (x-\lambda^{9})(x-\lambda^{11})(x-\lambda^{13})(x-\lambda^{15}) ~, \nonumber \\
  P(x) &= (x-\lambda)(x-\lambda^{3})(x-\lambda^{9})(x-\lambda^{11})~, &&& Q(x) &= (x-\lambda^{5})(x-\lambda^{7})(x-\lambda^{13})(x-\lambda^{15}) ~, \nonumber \\
  P(x) &= (x-\lambda)(x-\lambda^{5})(x-\lambda^{7})(x-\lambda^{11})~, &&& Q(x) &= (x-\lambda^{3})(x-\lambda^{9})(x-\lambda^{13})(x-\lambda^{15}) ~, \nonumber \\
  P(x) &= (x-\lambda)(x-\lambda^{7})(x-\lambda^{11})(x-\lambda^{13})~, &&& Q(x) &= (x-\lambda^{3})(x-\lambda^{5})(x-\lambda^{9})(x-\lambda^{15}) ~.
\end{align}
In all four cases the requirement that the coefficient of $x^{2}$ is an element of $\mathbb{F}_{p}$ leads to the requirement that $w^{2} \in \mathbb{F}_{p}$, which
is impossible for $p \in 8 \mathbb{N} + 5$. Hence there are no invariant global forms in this case. 

For the remaining case of $p \in 4\mathbb{N} + 3$, a similar analysis shows that there are again no invariant global forms. The results are summarized in Table \ref{tab:genus4table}.

\subsubsection{Genus 5}
\label{sec:genus-5}
Finally for genus $5$ the relevant cyclotomic polynomials are $\phi_{11}(x)$ and $\phi_{22}(x)$. If a matrix $F$ has characteristic polynomial $\phi_{22}(x)$ then $-F$ has
characteristic polynomial $\phi_{11}(x)$ and both have the same invariant subspaces, and hence we deal only with $\phi_{11}(x)$ here.

\begin{table}[!tp]
\begin{center}
\begin{tabular}{c|cccccccccc}
$p$ & 2 & 3 & 5 & 7 & 11 & 13 & 17 & 19 & 23 & 29
\\\hline
$\phi_{11}(x)$ & 0 & 0 &  2  & 2 & 1  &0 &0  &2 & 32 & 2
\end{tabular}
\end{center}
\caption{The number of invariant global forms for symmetry generators with characteristic polynomial given by $\phi_{11}(x)$, relevant for genus 5. There exist intrinsically non-invertible symmetries if and only if the entry above is $0$.}
\label{tab:genus5table}
\end{table}%

For the generic case in which $p$ does not divide 11, the only possibility for the factorization of $\phi_{11}(x) = P(x)Q(x)$ satisfying the isotropy criterion is
\beaa
  \label{eq:3.17}
  P(x) &=& (x-\lambda)(x-\lambda^{3})(x-\lambda^{4})(x-\lambda^{5})(x-\lambda^{9})~, 
  \no\\
  Q(x) &=& (x-\lambda^{2})(x-\lambda^{6})(x-\lambda^{7})(x-\lambda^{8})(x-\lambda^{10}) ~
\eeaa
with $\lambda$ a primitive eleventh root of unity.
It can be shown that $P(x)$ and $Q(x)$ are polynomials over $\mathbb{F}_{p}$ if $\eta:= \lambda + \lambda^{3} + \lambda^{4} + \lambda^{5} + \lambda^{9} \in \mathbb{F}_{p}$. Note that $\eta$ satisfies a quadratic equation $\eta^{2} + \eta+ 3 = 0$ which has a solution in $\mathbb{F}_{p}$ if $-11$ has a square root in $\mathbb{F}_{p}$. This happens for
$p \in 11 \mathbb{N} + k$ for $k=1,5,7,8,$ and $9$. For the last four cases there is no further factorization and there are only
two invariant global forms. For $p \in 11 \mathbb{N} + 1$ the polynomial $\phi_{11}(x)$ splits completely and there are $32$ invariant global forms.

For the remaining cases of $p \in 11 \mathbb{N} + k$ for $k=2,3,4,6,$ and $10$ the symmetry is intrinsically non-invertible. Finally for the exceptional case of $p =11$  there is a single invariant global form. These results are summarized in Table \ref{tab:genus5table}.

\section{A higher-dimensional perspective}
\label{sec:topops}
 
In this final section we give a higher-dimensional perspective on the non-invertible symmetries discussed above. To make the discussion self-contained, we begin with a review of the relative nature of the 6d (2,0) theory. For most of this section we will work the (2,0) theory of type $\mathfrak{a}_{N-1}$ without assuming that $N$ is prime, though this assumption will reappear towards the end.\footnote{We thank Kantaro Ohmori, Gabi Zafrir, and Yunqin Zheng for numerous crucial discussions about the content of this section.} 

\subsection{The 6d (2,0) theory as a relative theory}
\label{sec:6d20review}

As was already mentioned in the introduction and Section \ref{sec:classSintro}, for generic $N$ the type $\mathfrak{a}_{N-1}$ 6d (2,0)  theory is a relative theory, which means that it is best thought of as living on the boundary of a non-trivial 7d TQFT. This situation was shown in Figure \ref{fig:6drel}. In order to specify the state $|\mathfrak{a}_{N-1}\rangle$ corresponding to the 6d (2,0) theory, we must first specify a basis for the Hilbert space of the 7d TQFT. To do so, we might naively aim to  specify boundary conditions for all fields in the bulk 7-dimensional theory, in particular the dynamical three-form field $c$. 
However, due to a term proportional to $\int_{W_7} c\wedge d c$ in the action of the bulk TQFT, the field $c$ becomes its own canonically conjugate momentum, and thus there are non-trivial commutation relations between $c$ on distinct 3-cycles. This means that it is not actually possible to specify the boundary value of $c$ on all 3-cycles of $X_6$. 

Let us be more concrete. Our starting point is the 7d Abelian CS theory\footnote{For simplicity, we will assume the presence of a Wu 4-structure (so that the CS term is well-defined for arbitrary $N$) and take $c$ to be topologically trivial (so that we do not need to work with differential cohomology). }
\bea
\label{eq:7dCStheory}
S_{7d}={N \over 4 \pi}\int_{W_7}c\wedge d c~
\eea
which we write in differential form notation $c \in H^3(W_7, U(1))$. This theory has the following Wilson surfaces,
\bea
\Phi_q(M_3) := e^{i q \oint_{M_3} c}~, \hspace{0.6 in}q \in \ZZ_N~, \hspace{0.2 in} M_3 \in H_3(W_7, \ZZ)~.
\eea

Let us consider a configuration of two Wilson surfaces $\Phi_q(M_3)$ and $\Phi_{q'}(M'_3)$, with $M_3$ and $M'_3$ forming a Hopf link in $W_7$. To evaluate this configuration via the path integral, we consider the action with insertions (see e.g. \cite{Kaidi:2022cpf})
\beaa
S_{\mathrm{CS}}&=&{N \over 4 \pi}\int_{W_7} c\wedge d c +  q \int_{M_3} c + q' \int_{M_3'} c 
\\
&=& {N \over 4 \pi}\int_{W_7} c \wedge d c + \int_{W_7} (q \omega_{M_3} + q' \omega_{M_3'} )\wedge c~,
\eeaa
where $\omega_M$ is the Poincar{\'e} dual of $M$. Integrating out $c$ then imposes 
\bea
d c = -{2 \pi \over N} (q \omega_{M_3} + q' \omega_{M_3'})~.  
\eea
If we define a manifold $V_4$ such that 
\bea
\partial V_4 =\left( q M_3 + q' M_3'\right)~, 
\eea
then we have $c = -{2 \pi \over N} \mathrm{PD}(V_4)$. Plugging this back into the action gives a term 
\beaa
{\pi \over N} \int_{M_7} (q \omega_{M_3}+ q' \omega_{M_3'}) \wedge \mathrm{PD}(V_4) &=& { \pi \over N} \int_{M_7}\mathrm{PD}\left((qM_3 + q' M_3') \cap V_4 \right) 
\no\\
&=& {\pi \over N}\int_{M_7} \left[ q \mathrm{PD}(M_3 \cap V_4) + q' \mathrm{PD}(M_3' \cap V_4) \right]
\no\\
&=& {2 \pi \over N} q q' \mathrm{link}(M_3, M_3')~,
\eeaa
where $\mathrm{link}(M_3, M_3')$ is the linking number of $M_3$ and $M_3'$ in $W_7$. 
This linking comes about since $V_4$ is the Seifert surface for $q M_3 + q' M_3'$, and hence each time $M_3$
 pierces $V_4$ it links $M_3'$ once. 
 
 To summarize, we have found that the Hopf link of Wilson surfaces supported on $M_3$ and $M'_3$ in $W_7$ evaluates to\footnote{This is the direct analog of the S-matrix element $S_{ab}$ between charge $a$ and $b$ anyons in $U(1)_k$ CS theory in 3d, which is given by $S_{ab} = e^{{2 \pi i \over k} a b} $.}
 \bea
 \label{eq:linkingofflux}
 \langle \Phi_q(M_3) \Phi_{q'}(M_3') \dots \rangle = e^{{2 \pi i\over N} q q' \mathrm{link}(M_3, M_3')} \langle \dots \rangle~. 
 \eea
 We may now push the configuration above to a 6d plane $X_6 \subset W_7$. This gives a statement about the equal-time commutation relation 
\bea
\label{eq:Heisenbergalg}
\Phi_q(M_3) \Phi_{q'}(M_3') = e^{{2 \pi i \over N} q q' \langle M_3, M_3' \rangle}\Phi_{q'}(M_3')\Phi_q(M_3)~. 
\eea
We see that the linking number in 7d has reduced to the (signed) intersection number in 6d. One should be careful to note that the parity properties of the two bilinear pairings are opposite: whereas the linking pairing was symmetric, the intersection pairing is anti-symmetric.

 The Wilson surfaces also satisfy the quantum torus algebra on a 6d slice,
\bea
\label{eq:quanttorusalg}
\Phi_q(M_3) \Phi_{q}(M_3') = e^{{2 \pi i \over N} {q^2 \over 2} \langle M_3, M_3' \rangle}\Phi_{q}(M_3+M_3')~.
\eea
Note that it is not obvious that the factor $e^{{2 \pi i \over N} {q^2 \over 2}}$ appearing in the quantum torus algebra is well-defined. This is because the charge $q$ was only defined modulo $N$. Under $q\rightarrow q+N$ the factor  $e^{{2 \pi i \over N} {q^2}}$ is single-valued, but this can fail to be true when we take the square root.

For $N$ odd, there is no issue: noting that $e^{{2 \pi i \over N} {q^2} }= e^{{2 \pi i \over N} (N+1){q^2}}$, we can simply define 
\bea
e^{{2 \pi i \over N} {q^2 \over 2} }: = e^{{2 \pi i \over N}(N+1) {q^2 \over 2}}
\eea
which is manifestly well-defined since $N+1 \over 2$ is an integer. A simpler way of saying this is that $2$ is invertible in $\ZZ_N$ for $N$ odd. 

For $N$ even, the discussion is more subtle. To begin, let us return to $e^{{2 \pi i \over N} {q^2}}$, which as we said is well-defined. In fact, even this is slightly non-obvious, since one has to check that $q^2$ is well-defined modulo $N$ when $q \rightarrow q+N$. This is indeed the case, since 
\bea
(q+N)^2 - q^2 = N(2q+N)
\eea
and hence $q^2 = (q+N)^2$ mod $N$. But now we see something more: when $N$ is even, then we in fact have 
\bea
(q+N)^2 - q^2 = 2N \left(q+{N\over 2}\right)
\eea
and hence $q^2 = (q+N)^2$ mod $2N$. This means that we can promote $q^2$ to an element in $\ZZ_{2N}$, as opposed to an element in $\ZZ_N$ like $q$ was. We denote this promoted quantity by $\cP(q) \in \ZZ_{2N}$, which is known as a quadratic refinement of the original pairing in $\ZZ_N$. We may now define 
\bea
e^{{2 \pi i \over N} {q^2 \over 2} }: =e^{{2 \pi i \over N} {\cP(q) \over 2} }~,
\eea
which is well-defined since we only need single-valuedness under $\cP(q) \rightarrow \cP(q) + 2N$.

We now return to the specification of the basis of the Hilbert space of the 7d TQFT. Noting that the factor of $q$ can be absorbed by replacing $M_3$ with $qM_3$, we may without loss of generality restrict to $\Phi_1(M_3)$, which for simplicity we will denote by $\Phi(M_3)$, and $M_3 \in H_3(X_6, \ZZ_N)$. Because of the non-commutativity of $\Phi(M_3)$ for generic $M_3\in H_3(X_6, \ZZ_N)$, it is not possible to write a simultaneous eigenvector for all $\Phi(M_3)$---in other words, it is not possible to specify the boundary value of $c$ on all three-cycles of $X_6$. Instead, the best one can do is to choose a \textit{maximal isotropic sublattice} $\cL \in H_3(X_6, \ZZ_N)$ and to specify all of the boundary values associated with $\cL$. Here isotropicity means that we have
\bea
\langle M_3, M_3' \rangle = 0 ~, \hspace{0.8 in}\forall \, M_3, M_3' \in \cL~. 
\eea
In terms of this isotropic sublattice we may define a state $|\cL, 0 \rangle$ such that 
\bea
\label{eq:DBC7d}
\Phi(M_3) | \cL , 0 \rangle = | \cL , 0 \rangle ~, \hspace{0.5 in} \forall\,\,M_3 \in \cL~. 
\eea

On the other hand, applying $\Phi(M_3')$ for $M_3'$ in the complement $\cL^\perp := H^3(X_6, \ZZ_N) / \cL$ gives a new state, 
\bea
|\cL, M_3'\rangle : = \Phi(M_3')|\cL, 0\rangle 
\eea
which by the commutation relations (\ref{eq:Heisenbergalg}) satisfies 
\bea
\Phi(M_3) | \cL , M_3' \rangle = e^{ {2\pi i\over N} \langle M_3, M_3'\rangle} | \cL , M_3' \rangle ~, \hspace{0.5 in} \forall\,\, M_3 \in \cL~. 
\eea
These states form an irreducible representation of the algebra (\ref{eq:Heisenbergalg}).

Physically, we may interpret the state $| \cL , 0 \rangle$ as imposing Dirichlet boundary conditions $c|_\p = 0$ on all $M_3 \in \cL$. This means the bulk Wilson surface $\Phi(M_3)=e^{i\oint_{M_3}c}$ trivializes when moved to the boundary, reproducing the statement in (\ref{eq:DBC7d}).  By the canonical commutation relations,  Dirichlet boundary conditions on $M_3 \in \cL$ require Neumann boundary conditions on $M_3' \in \cL^\perp$, and hence $\Phi(M_3')$ for $M_3'\in \cL^\perp$ gives a new state, which is $| \cL , 0 \rangle$ with the insertion of a Wilson surface on $M_3'$. This is the state $|\cL, M_3'\rangle$.\footnote{These are the analogues of the non-identity characters $\chi_a$ in 2d RCFT. The Wilson surfaces discussed here are the analogues of the Verlinde lines in that context.}

In terms of the basis of states $|\cL, M_3'\rangle$, we may then write the state $|\mathfrak{a}_{N-1}\rangle$ corresponding to the relative 6d (2,0) theory as 
\bea
\label{eq:classSstate}
|\mathfrak{a}_{N-1} \rangle = \sum_{M_3' \in \cL^\perp} Z_{\cL}[M_3'] |\cL, M_3'\rangle~
\eea
for some appropriate coefficients $Z_{\cL}[M_3']$.
This vector is sometimes referred to as the ``partition vector'' of the 6d theory.  We should be careful to note that the states $|\cL, \cB\rangle$ do not define topological boundary conditions for the bulk 7d TQFT; indeed, there do not in general  exist topological boundary conditions for this theory. Thus when one refers to the 6d (2,0) ``theory,'' what one is really referring to is the above vector.

\subsection{Compactification on $\Sigma_{g,n}$}

We now take $X_6 = \Sigma_{g,n} \times X_4$, as for the class $\cS$ construction.  The 7d bulk theory is likewise put on $W_7= \Sigma_{g,n} \times W_5$, where $\p M_5 = X_4$.\footnote{Despite our notation, the product here is not necessarily trivial, as we will discuss shortly.}
Because the four-dimensional spacetime $X_4$ is attached to a bulk five-dimensional spacetime $W_5$, we may expect that the theory is again only relative. However, we will now show that the theory in $W_5$ can be made invertible, and hence that this procedure defines an absolute theory in 4d. 

 Given the 6d spacetime $X_6 = \Sigma_{g,n} \times X_4$, there is a particular class of 7-manifolds $W_7$ obtained by taking $W_7 = V_{g,n} \times X_4$ with $V_{g,n}$ a three-manifold with boundary $\Sigma_{g,n}$. For a given Riemann surface $\Sigma_{g,0}$, there are many inequivalent three-manifolds with boundary $\Sigma_{g,0}$. A particularly simple class of such manifolds are the so-called handlebodies, whose basic properties are reviewed in Appendix \ref{app:handlebodies}. 
As discussed there, in order to construct a handlebody, one must choose a set of $g$ meridians, i.e. a set of $g$ generators of $H_1(\Sigma_{g,0},\ZZ)$ which become trivialized as elements of $H_1(V_{g,0},\ZZ)$. The remaining $g$ generators of $H_1(\Sigma_{g,0},\ZZ)$  lift to generators of  $H_1(V_{g,0},\ZZ)$ and are referred to as ``longitudes.'' We will denote the meridians by $\mu_i$ and the longitudes by $\lambda_i$ for $i=1,\dots,g$. Note that not every choice of $g$ generators of $H_1(\Sigma_{g,0},\ZZ)$ gives a legitimate set of meridians. Indeed, as explained in Appendix \ref{app:handlebodies}, each legitimate choice of meridians corresponds to a maximal isotropic sublattice $L \subset H_1(\Sigma_{g,0},\ZZ)$. We will denote the handlebody specified by the choice of meridians $L$ by $V_{g,0}^L$. 

Given a handlebody $V_{g,0}^L$, there are multiple choices of longitudes, which differ by shifts by meridians. Indeed, given one choice of longitudes $\lambda_i$, the choice of longitudes 
\bea
\label{eq:alternatelongitudes}
\lambda_i' = \lambda_i + \sum_{j=1}^g k_{ij} \,\mu_j~, \hspace{0.5 in} k_{ij}\in \ZZ
\eea
is another legitimate choice. We may also allow for linear combinations of the $\lambda_i$, but we will fix a basis such that the intersection pairing of $\lambda_i$ with the meridians $\mu_i$ is 
\bea
\label{eq:easyintpar}
\langle \mu_i, \lambda_j \rangle=-\langle \lambda_j , \mu_i\rangle = \delta_{ij}~. 
\eea

We now consider the 7d CS theory (\ref{eq:7dCStheory}) on this handlebody geometry. First we define the following fields 
\bea
b_i = \oint_{\mu_i} c~, \hspace{0.5 in} \widehat b_i = \oint_{\lambda_i} c~, \hspace{0.5 in} i = 1, \dots, g~. 
\eea
Because the meridians $\mu_i$ are naively contractible, we might assume that all of the $b_i$ are trivial. This however is not necessarily true. The origin of the discrepancy is that we can have Wilson surfaces supported on longitudes $\lambda_i$ which link with $\mu_i$, and hence effectively make $\mu_i$ non-contractible. 

To understand this, let us denote the Wilson surfaces wrapping meridians and longitudes by 
\bea
\Phi_i(M_2) &:=& \Phi(M_2 \times \mu_i) =  e^{{2\pi i\over N} \oint_{M_2} b_i}~, 
\no\\
\widehat \Phi_i(M_2) &:=& \Phi(M_2 \times \lambda_i)= e^{{2\pi i\over N} \oint_{M_2}\widehat  b_i}~,
\eea
respectively, where we have converted to discrete cocycle notation $(b_i, \widehat b_i) \rightarrow {2\pi i \over N}(b_i, \widehat b_i)$.  By (\ref{eq:linkingofflux}), these surfaces may have non-trivial linking in 5d. Indeed, let us insert a Wilson 3-surface on the longitude $\lambda_j$ in 7d, with the remaining support on $M_2' \subset X_4$. This amounts to inserting $\widehat \Phi_j(M_2')$ in 5d.  
Equation (\ref{eq:linkingofflux}) then gives 
\bea
\langle \Phi_i(M_2) \widehat \Phi_j(M_2') \dots \rangle = e^{{2 \pi i \over N} \mathrm{link}(M_2 \times \mu_i, M_2' \times \lambda_j)} \langle \dots \rangle~. 
\eea
Clearly this means that we cannot set $\Phi_i(M_2)=1$, and likewise $b_i$ is not simply zero.

Let us now compactify the theory on $\Sigma_{g,0}$ in the presence of a series of Wilson 3-surfaces on $\lambda_j \times M_{2,j}'$. Denote the Poincar{\'e} duals of $M_{2,j}'$ by $B_j \in H^2(X_4, \ZZ_N)$. The linking equation above becomes an operator equation determining $\Phi_i(M_2)$, 
\bea
\label{eq:Wilsonsurfacefixed}
\Phi_i(M_2) = e^{{2\pi i \over N}  \sum_{j=1}^g\langle \mu_i, \lambda_j \rangle (M_2, M_{2,j}')}= e^{{2\pi i \over N}  \sum_{j=1}^g\langle \mu_i, \lambda_j \rangle \oint_{M_2} B_j}~. 
\eea
Another way to say this is that in any 4d computation involving $\Phi_i$ we may commute it to the right of all $\widehat \Phi_j$ using (\ref{eq:Heisenbergalg}), at which point it can be shrunk away. Hence  insertion of $\Phi_i$ is equivalent to insertion of the phase appearing in (\ref{eq:Heisenbergalg}).

We see that the insertion of the longitudinal Wilson surface $\widehat \Phi_j(M_{2,j}')$ effectively fixes
\bea
b_i = \sum_{j=1}^g\langle\mu_i,\lambda_j\rangle B_j~.
\eea
In particular, if we work with a basis of longitudes satisfying (\ref{eq:easyintpar}), we have simply $b_i =B_i$.  The insertion of Wilson lines on longitudes has allowed us to make $b_i$ non-zero, though it is still a constant, i.e. a background field.

\begin{figure}[t]
    \centering
    \hspace{-0.65in}
\begin{tikzpicture}
%%%%%%%%tori%%%%%%%%%

\begin{scope}[rotate=270,scale=0.8]

\begin{scope}[scale=.8]
\shade[left color=blue!30, right color=blue!10, distance=4in] (2,0) 
to [out=270, in=270](-2,0)
to [out = 0, in = 0] (2,0); 
\end{scope}

 %Torus
\draw[thick] (0,0) ellipse (1.6 and .9);
\shade[bottom color=blue!40, top color=blue!10] (0,0) ellipse (1.6 and .9);
%Hole
\begin{scope}[scale=.8]
\clip (0,1.3) circle (1.55);
\fill[white] (0,-1.27) circle (1.55);
\end{scope}
\begin{scope}[scale=0.8]
\path[rounded corners=24pt] (-.9,0)--(0,.6)--(.9,0) (-.9,0)--(0,-.56)--(.9,0);
%\draw[][rounded corners=28pt] (-1,.1)--(0,-.6)--(1,.1);
\draw[thick][rounded corners=24pt] (-0.9,0)--(0,.8)--(0.9,0);
\draw[thick][rounded corners=8pt] (-0.77,0)--(0,-0.4)--(0.77,0);

\draw[thick, distance=4in] (2,0) to [out=270, in=270](-2,0); 

\draw[thick,dgreen] (0,0) ellipse (1.4 and .7);

\end{scope}

\end{scope}

 \begin{scope}[xshift=1.4in, yshift=-1.2 in]
 \shade[top color=red!40, bottom color=red!10,rotate=90]  (0,2.5) -- (2,2.5) -- (2.6,3) -- (0.6,3)-- (0,2.5);
 \draw[thick,rotate=90]  (0,2.5) -- (2,2.5);
    \draw[thick,rotate=90] (2,2.5) -- (2.6,3);
     \draw[thick,rotate=90] (2.6,3) -- (0.6,3);
       \draw[thick,rotate=90] (0.6,3)-- (0,2.5);
\end{scope}

\draw[thick,dashed,stealth-](-4.2,-1.7)--(0.8,-1.7);

\node[right] at (1,-3) {$X_4$};

\draw [thick, decorate,
    decoration = {brace}] (1,-3.5) --  (-4.4,-3.5);

\node[right] at (-2.1,-4) {$W_7$};

\node[right] at (0.25,0.8) {$\small{\widehat \Phi_i(M_{2,i})}$};
\node[right] at (0.7,0.1) {$\Sigma_{g,0}$};
\node[right] at (-2.3,0.1) {$V^L_{g,0}$};

\draw[thick, snake it, <->] (2.6,-1) -- (4.4,-1);
\node[above] at (3.5,-1) {Shrink\,\,$\Sigma_{g,0}$};

\begin{scope}[xshift=3.5 in, yshift=-1.2in]
 \shade[top color=red!40, bottom color=red!10,rotate=90]  (0,-0.7) -- (2,-0.7) -- (2.6,-0.2) -- (0.6,-0.2)-- (0,-0.7);

 \draw[thick,rotate=90] (0,-0.7) -- (2,-0.7);
\draw[thick,rotate=90] (0,-0.7) -- (0.6,-0.2);
\draw[thick,rotate=90]  (0.6,-0.2)--(2.6,-0.2);
\draw[thick,rotate=90]  (2.6,-0.2)-- (2,-0.7);
\draw[thick,rotate=90]  (2.6,-0.2)-- (2.6,3);
\draw[thick,rotate=90,dashed]  (0.6,-0.2)-- (0.6,3);
\draw[thick,rotate=90]  (0,-0.7)-- (0,2.5);
\draw[thick,rotate=90,]  (2,-0.7)-- (2,2.5);
\node[below] at (0.5, 3.5) {$\cT_L^{N,g,0}[\Omega,B_i]$};

\node[below] at (0.5, 0){$X_4$};

\node[below] at (-1.2,1.5) {$W_5$};
\end{scope}

\end{tikzpicture}

\vspace{-1.2 in}
    \caption{We begin with a handlebody configuration in 7d specified by $L$, together with a set of Wilson surfaces $\widehat \Phi_i(M_{2,i})$ on the longitudes. This configuration has a single boundary, located on the right. One then compactifies on $\Sigma_{g,n}$ to get a 5d bulk with a single 4d boundary. The 5d theory can be shown to be invertible, and hence the boundary theory is a well-defined 4d theory labelled by $L$, the longitudes $\lambda_i$ (capturing stacking with SPT phases), and the background fields $B_i = \mathrm{PD}(M_{2,i})$.}
    \label{fig:7dconfig}
\end{figure}
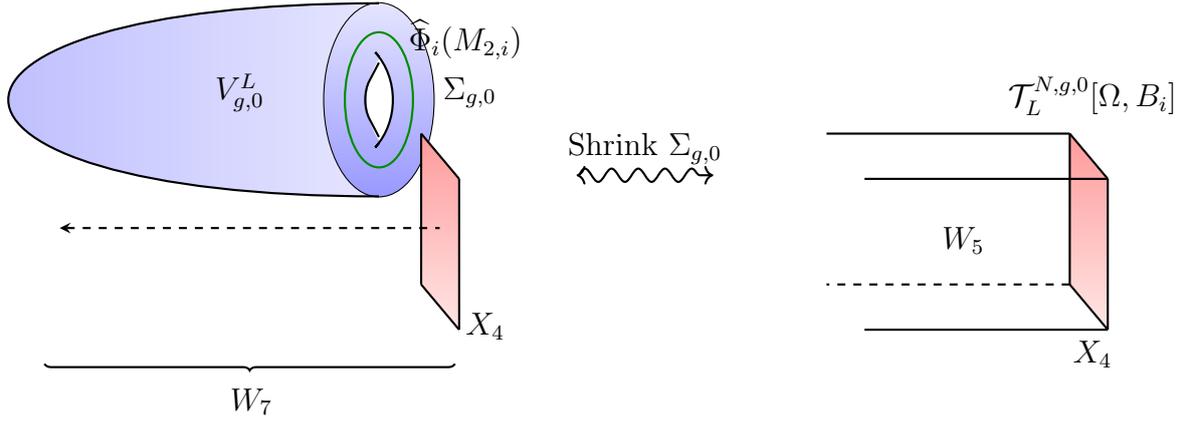

Upon compactification on $\Sigma_{g,0}$, the bulk CS theory of (\ref{eq:7dCStheory}) then reduces to,
\bea
\label{eq:7dto5daction1}
S_{7d} = {N \over 4 \pi} \int_{W_7} c\wedge d c \,\,\,\xrightarrow{\text{compactify\,\,on\,\,}\Sigma_{g,0}}\,\,\, S_{5d} &=&{2 \pi \over N}\sum_{i=1}^{g} \int_{W_5} B_i \cup \delta\widehat b_i~.
\eea
Unlike usual BF theory, only one of the fields appearing above is dynamical---in other words, this is the theory of $g$ free $\ZZ_N$ background gauge fields $B_i$. This theory is obviously invertible, which means that the boundary four-dimensional theory is well-defined as a standalone theory. This is how one obtains an absolute 4d theory from the relative 6d $(2,0)$ theory. 

Of course, the set of Wilson surfaces $\widehat \Phi_j(M_{2,i}')$ that we inserted in 7d was just one choice. We could have modified this choice by placing the Wilson surfaces on different 2-cycles $M_{2,i}'$. From the four-dimensional perspective, this changes the background fields $B_i$ to which the one-form symmetries are coupled. 

Alternatively, we could have changed the longitudes $\lambda_i$ on which the Wilson surfaces were inserted. Say that we change longitudes to $\lambda_i'$ as defined in (\ref{eq:alternatelongitudes}), with the corresponding Wilson surfaces denoted by $\widehat \Phi'_i(M_{2,i})$. Using the commutation relations (\ref{eq:Heisenbergalg}) and the quantum torus algebra (\ref{eq:quanttorusalg}), we find that
\beaa
\prod_i\widehat{\Phi}_i'(M_{2,i}) &=& \prod_i \left(e^{ {2 \pi i \over N} \sum_{j=1}^g k_{ij} \langle \mu_i, \lambda_j \rangle \int_{X_4} {B_i \cup B_i \over 2}}\,\widehat \Phi_i(M_{2,i}) \prod_{j=1}^g\Phi_j^{k_{ij}}(M_{2,j}) \right)
\\
&=& \mathrm{exp}\left[{2\pi i \over N}\left( \sum_i k_{ii}  \int_{X_4}{B_i \cup B_i \over 2} + \sum_{i<j} k_{ij} \int_{X_4} B_i \cup B_j\right)\right] \prod_i\widehat \Phi_i(M_{2,i})
\no
\eeaa
where in the second line we have used (\ref{eq:easyintpar}), and after commuting all $\Phi_j$ operators to the right of $\widehat \Phi_i$ operators we have shrunk them away.
The factor of ${B_i \cup B_i \over 2}$ appearing above should be understood as the appropriate quadratic refinement when applicable. We see that changing to a different longitude stacks the 4d theory with an SPT phase.

The final picture is thus as in Figure \ref{fig:7dconfig}. We begin in 7d by specifying a maximal isotropic sublattice $L$, which specifies a way to fill in the Riemann surface $\Sigma_{g,0}$ to get a handlebody $V_{g,0}^L$.  We may also choose to insert some set of Wilson surfaces $\widehat \Phi_i(M_{2,i})$ along the longitudes $\lambda_i$. Upon compactification on $\Sigma_{g,0}$, we obtain a well-defined 4d theory coupled to an invertible 5d bulk. The choice of meridians $L$ in 7d determines the charge lattice of the 4d theory. On the other hand, the choice of longitudinal Wilson surfaces determines both the background fields $B_i$ for the  1-form symmetry (coming from the choice of $M_{2,i}$) together with the possible invertible phases that can be stacked with the 4d theory (coming from the choice of $\lambda_i$). 

\subsection{From the Anomaly TFT to the Symmetry TFT}
\label{sec:condensationdefects}

We have now seen how the (trivial) Anomaly TFT of the 4d theory can be obtained by reducing the 7d CS term on an appropriate handlebody. It is also enlightening to see how the Symmetry TFT (SymTFT) \cite{Freed:2012bs,Freed:2018cec,Gaiotto:2020iye,Apruzzi:2021nmk,Apruzzi:2022dlm, Burbano:2021loy, Freed:2022qnc,Kaidi:2022cpf}  emerges in this picture. 

At any point $y$ along the handlebody direction $\RR_+$, we may choose to split the handlebody into an interior portion $V_{g,0}^{L,\text{in}}$ with $y\geq y_*$ and an exterior portion $V_{g,0}^{\text{ext}}:=\Sigma_{g,0}\times [0,y_*]$. If we focus on the latter and forget about boundary conditions, the reduction on the Riemann surface gives 
\bea
\label{eq:7dto5daction3}
S_{7d} = {N \over 4 \pi} \int_{W_7} c\wedge d c \,\,\,\xrightarrow{\text{compactify\,\,on\,\,}\Sigma_{g,0}}\,\,\, S_{5d} &=&{2 \pi \over N}\sum_{i=1}^{g} \int_{X_4 \times [0,y_*]} b_i \cup  \delta\widehat b_i~,
\eea
where now \textit{both} $b_i$ and $\widehat b_i$ are dynamical---indeed, both the longitudes $\lambda_i$ and meridians $\mu_i$ are non-trivial in $V_{g,0}^{\text{ext}}$. 

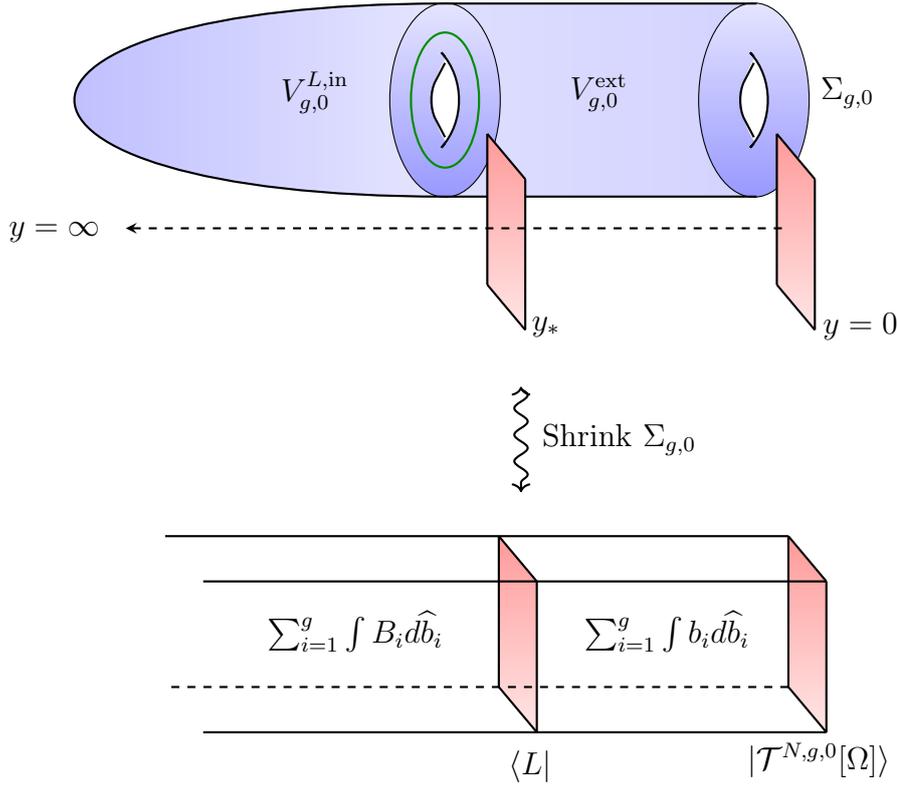
\begin{figure}[t]
    \centering
    \vspace{-1.3 in}
    \hspace{-0.65in}
\begin{tikzpicture}
%%%%%%%%tori%%%%%%%%%

\begin{scope}[rotate=270,scale=0.8]

\begin{scope}[scale=.8]
\shade[left color=blue!30, right color=blue!10, distance=4in] (2,0) 
to [out=270, in=270](-2,0)
to [out = 0, in = 0] (2,0); 
\end{scope}

\begin{scope}[scale=.8]
\shade[left color=blue!10, right color=blue!20, distance=4in] (0-2,0+0) 
to [out=90, in=270](0-2,6.2+0)
to [out=0, in=180](4-2,6.2+0)
to [out = 270, in = 90] (4-2,0+0);
to [out = 180, in = 0] (0-2,0+0);  
\end{scope}

\begin{scope}[yshift=2in]
\draw[thick] (0,0) ellipse (1.6 and .9);
\shade[bottom color=blue!40, top color=blue!10] (0,0) ellipse (1.6 and .9);
\begin{scope}[scale=.8]
\clip (0,1.3) circle (1.55);
\fill[white] (0,-1.27) circle (1.55);
\end{scope}
\begin{scope}[scale=0.8]
\path[rounded corners=24pt] (-.9,0)--(0,.6)--(.9,0) (-.9,0)--(0,-.56)--(.9,0);
%\draw[][rounded corners=28pt] (-1,.1)--(0,-.6)--(1,.1);
\draw[thick][rounded corners=24pt] (-0.9,0)--(0,.8)--(0.9,0);
\draw[thick][rounded corners=8pt] (-0.77,0)--(0,-0.4)--(0.77,0);

\end{scope}

\end{scope}

 %Torus
\draw[thick] (0,0) ellipse (1.6 and .9);
\shade[bottom color=blue!40, top color=blue!10] (0,0) ellipse (1.6 and .9);
%Hole
\begin{scope}[scale=.8]
\clip (0,1.3) circle (1.55);
\fill[white] (0,-1.27) circle (1.55);
\end{scope}
\begin{scope}[scale=0.8]
\path[rounded corners=24pt] (-.9,0)--(0,.6)--(.9,0) (-.9,0)--(0,-.56)--(.9,0);
%\draw[][rounded corners=28pt] (-1,.1)--(0,-.6)--(1,.1);
\draw[thick][rounded corners=24pt] (-0.9,0)--(0,.8)--(0.9,0);
\draw[thick][rounded corners=8pt] (-0.77,0)--(0,-0.4)--(0.77,0);

\draw[thick, distance=4in] (2,0) to [out=270, in=270](-2,0); 

\draw[thick,dgreen] (0,0) ellipse (1.4 and .7);

\end{scope}

\end{scope}

\draw[thick](-0,-1.28)--(4.1,-1.28);
\draw[thick](-0,1.28)--(4.1,1.28);

 \begin{scope}[xshift=1.4in, yshift=-1.2 in]
 \shade[top color=red!40, bottom color=red!10,rotate=90]  (0,2.5) -- (2,2.5) -- (2.6,3) -- (0.6,3)-- (0,2.5);
 \draw[thick,rotate=90]  (0,2.5) -- (2,2.5);
    \draw[thick,rotate=90] (2,2.5) -- (2.6,3);
     \draw[thick,rotate=90] (2.6,3) -- (0.6,3);
       \draw[thick,rotate=90] (0.6,3)-- (0,2.5);
\end{scope}

 \begin{scope}[xshift=2.9in, yshift=-1.2 in]
 \shade[top color=red!40, bottom color=red!10,rotate=90]  (0,2.5) -- (2,2.5) -- (2.6,3) -- (0.6,3)-- (0,2.5);
 \draw[thick,rotate=90]  (0,2.5) -- (2,2.5);
    \draw[thick,rotate=90] (2,2.5) -- (2.6,3);
     \draw[thick,rotate=90] (2.6,3) -- (0.6,3);
       \draw[thick,rotate=90] (0.6,3)-- (0,2.5);
\end{scope}

\draw[thick,dashed,stealth-](-4.2,-1.7)--(4.5,-1.7);

\node[left] at (-4.4,-1.75) {$y=\infty$};
\node[right] at (1,-3) {$y_*$};
\node[left] at (6.1,-3) {$y=0$};

\node[right] at (4.8,0.1) {$\Sigma_{g,0}$};
\node[right] at (1.5,0.1) {$V^{\text{ext}}_{g,0}$};
\node[right] at (-2.3,0.1) {$V^{L,\text{in}}_{g,0}$};

\draw[thick, snake it, <->] (1,-3.8) -- (1,-5.2);
\node[right] at (1,-4.5) {\,\,Shrink\,\,$\Sigma_{g,0}$};

\begin{scope}[xshift=1.7 in, yshift=-3.3in]
 \shade[top color=red!40, bottom color=red!10,rotate=90]  (0,-0.7) -- (2,-0.7) -- (2.6,-0.2) -- (0.6,-0.2)-- (0,-0.7);

  \begin{scope}[xshift=-1.5in]
 \shade[top color=red!40, bottom color=red!10,rotate=90]  (0,-0.7) -- (2,-0.7) -- (2.6,-0.2) -- (0.6,-0.2)-- (0,-0.7);
 \draw[thick,rotate=90] (0,-0.7) -- (2,-0.7);
\draw[thick,rotate=90] (0,-0.7) -- (0.6,-0.2);
\draw[thick,rotate=90]  (0.6,-0.2)--(2.6,-0.2);
\draw[thick,rotate=90]  (2.6,-0.2)-- (2,-0.7);
\end{scope}

 \draw[thick,rotate=90] (0,-0.7) -- (2,-0.7);
\draw[thick,rotate=90] (0,-0.7) -- (0.6,-0.2);
\draw[thick,rotate=90]  (0.6,-0.2)--(2.6,-0.2);
\draw[thick,rotate=90]  (2.6,-0.2)-- (2,-0.7);
\draw[thick,rotate=90]  (2.6,-0.2)-- (2.6,8);
\draw[thick,rotate=90,dashed]  (0.6,-0.2)-- (0.6,8);
\draw[thick,rotate=90]  (0,-0.7)-- (0,7.5);
\draw[thick,rotate=90,]  (2,-0.7)-- (2,7.5);

\node[below] at (0.6, 0){$|\cT^{N,g,0}[\Omega]\rangle$};
\node[below] at (-3.2, -0.1){$\langle L|$};

\node[below] at (-1.4,1.75) {$\sum_{i=1}^g\int b_i d \widehat b_i$};
\node[below] at (-5.5,1.75) {$\sum_{i=1}^g\int B_i d \widehat b_i$};
\end{scope}

\end{tikzpicture}

    \caption{Splitting the handlebody into an inner handlebody $V^{L,\text{in}}_{g,0}$ and an outer shell $V^{\text{ext}}_{g,0}$ separates the (trivial) Anomaly TFT from the Symmetry TFT. The location $y_*$ of this splitting is arbitrary in 7d, and hence the boundary $\langle L|$ separating the Anomaly and Symmetry TFTs is topological.}
    \label{fig:7dconfig2}
\end{figure}

This gives us a picture in which a dynamical BF theory inhabits $ [0,y_*]$, with a  boundary condition at $y=y_*$ specified by $L$. In particular, $L$ specifies which directions of the Riemann surface become trivial in the handlebody (i.e. meridians), and hence which of the dynamical fields of the BF theory become background. In other words, each choice of $L$ is a choice of Dirichlet boundary conditions for different sets of fields of the BF theory. The setup is shown in Figure \ref{fig:7dconfig2}. Because the 7d geometry is completely independent of the location of $y_*$, the 5d picture should likewise be independent of  $y_*$, meaning that the interface between the BF theory and the anomaly theory is topological. 

While it is tempting to refer to the dynamical BF theory we have obtained as the SymTFT for the boundary 4d theories, one should be careful with this language. In general, the SymTFT is defined such that it captures the \textit{full} symmetry (higher-)category of the boundary theory, and it is not always true that the bulk BF theory does so.\footnote{In fact, if we also account for zero-form symmetries such as the $\ZZ_{2N}$ chiral symmetry of the boundary, then the BF theory is \textit{never} the full SymTFT; at best it only captures the one-form symmetry sector. We will focus only on this one-form symmetry sector here, defining the ``full SymTFT'' as such. } Indeed, the BF theory above is a simple example of a Dijkgraaf-Witten (DW) theory, and as explained in \cite{Kaidi:2022cpf} the SymTFT is a DW theory if and only if the boundary fusion category has no intrinsically non-invertible symmetries. Hence straightforward reduction of the 7d CS theory on a Riemann surface can only capture the full SymTFT in cases in which all potential non-invertible symmetries are non-intrinsic. To obtain the SymTFT for intrinisically non-invertible symmetries, it turns out that the 7d CS theory must be coupled to topological gravity, as we will now explain.\footnote{Of course, it is not inconsistent to have the 4d theory coupled to the 5d BF theory alone, so the sum over geometries is ultimately optional. It is only necessary if we want to obtain the 4d theory coupled to the full SymTFT for the one-form symmetry sector.  } This gives a physical relevance for the notion of intrinsic non-invertibility.

To begin our discussion, we define two families of operations on the handlebody configuration.  The first are modular transformations of the Riemann surface, which act throughout the entire handlebody and generically change the period matrix of the Riemann surface at each cross section. Note that because the theory in the bulk of the handlebody is topological, the only sensitivity to the change in the period matrix is on the boundary Riemann surface, where the 6d $(2,0)$ theory lives. 

The second family of operations involve excising from the handlebody an inner handlebody, like in the case of the splitting into $V_{g,0}^{L, \text{in}}$ and $V_{g,0}^{ \text{ext}}$ from before, but now gluing them back together with a non-trvial element of the modular group $Sp(2g, \ZZ_N)$---in other words, we perform ``surgery'' on the three-manifold \cite{rolfsen2003knots}.  
Because this operation happens in the interior of the handlebody, it does not affect the boundary Riemann surface, and in particular it does not change the period matrix in the region in which the theory is sensitive to it. This operation does however change the global form $L$, because it changes which cycles in the handlebody are contractible (fixing the conventions for the cycles on the boundary Riemann surface). 

The handlebody configuration we have been describing thus far has two ``boundaries'': one is the obvious boundary where the 6d (2,0) theory  lives, while the other is off at infinity, where the longitudinal Wilson surfaces fixing the background fields live.
We can now perform surgery by an element $F$ of the modular group, as described above. Since this happens in the interior of the handlebody, it does not affect the boundary where the 6d (2,0) theory lives, but it does change the ``inner boundary'', i.e. it changes the global form. This operation may then be combined with a modular transformation $F$ acting on the entire handlebody. The modular transformation reverts the interior of the handlebody back to its original form, but changes the value of the period matrix of the boundary Riemann surface.  If we fix ourselves to a value of the period matrix that is invariant under the action of $F$, then the geometry with the combined surgery and modular transformations has \textit{the same boundaries} as the original geometry. Importantly though, the full geometry is not identical to the original one, due to the internal twist. 

In quantum gravity (and quantum field theory in general), one is instructed to sum over all configurations with the same asymptotic boundary conditions. If we were to treat the 7d CS theory in a (topological) quantum gravity context, then we would likewise sum over handlebody configurations with all such internal twists. Furthermore, we should allow for different twists at each point of $X_4$, captured by a discrete metric connection.\footnote{Without the sum over geometries, this setup is similar to that of theories of ``class $\mathcal{F}$ '' \cite{Lawrie:2018jut}, though in the current case we still take the boundary $X_6 = \Sigma_{g,n} \times X_4$ to be a trivial product. }

How do we interpret this topological gravity computation from the point of view of the compactified 5d theory? As we will explain in the next subsection, the surgery defects described above descend to condensation defects implementing outer automorphism symmetries of the 5d BF theory. Promoting background gravity to dynamical gravity in 7d then corresponds to gauging of these automorphisms, with the role of the discrete gauge field being played by the dimensional reduction of the discrete metric connection.  The result is not a BF theory in general. 

We may write the action for the resulting 5d theory in terms of twisted cocycles as follows \cite{Kaidi:2022cpf}. First, we write (\ref{eq:7dto5daction3}) in the form of a $K$-matrix theory, 
 \bea
 S_{5d} = {2\pi \over 2N} \int_{W_5} \mathbf{b}^T \cup K \delta\mathbf{b}~, \hspace{0.5 in} K = \left(\begin{matrix} \widetilde{K} &  \mathds{1}_{g\times g}   \\ -\mathds{1}_{g\times g}  & 0\end{matrix}\right)~,
 \eea
 where we have defined $\mathbf{b} = (b_1, \dots, b_{g}, \widehat b_1, \dots, \widehat b_g)$. The matrix $\widetilde K$ has elements $(\widetilde K)_{ij} = k_{ij}$, with $k_{ij}$ as defined in (\ref{eq:alternatelongitudes}).
 This theory has an $Sp(2g, \ZZ_N)$ outer automorphism symmetry which acts on $\mathbf{b}$ in the usual matrix representation, and is a zero-form symmetry.
 
Let us assume that the enhanced symmetry is cyclic and is generated by a single element $F \in Sp(2g, \ZZ_N)$. Gauging this symmetry amounts to promoting $\mathbf{b}$ to a twisted cocycle. If $F$ is of order $|F|$, the action after gauging is given by 
 \bea
 \label{eq:BFpluseta}
 S_{5d} = {2\pi \over 2 N} \int_{W_5}  \mathbf{b}^T \cup_\eta K \delta_\eta\mathbf{b} + {2 \pi \over |F|}\int_{W_5} x \cup \delta \eta 
 \eea
 where $\eta$ is the background gauge field for $F$, descending in an obvious manner from the discrete metric connection in 7d, and $x$ is a Lagrange multiplier field enforcing that $\eta$ is a $\ZZ_{|F|}$-valued cocycle. The twisted cup products and differentials are defined by 
 \bea
 (\mathbf{a} \cup \mathbf{b})_{ijk\ell m}:=  \mathbf{a}_{ijk} F^{\eta_{ik}} \mathbf{b}_{k \ell m}~, \hspace{0.3 in} (\delta_\eta \mathbf{b})_{ijk\ell}:= F^{\eta_{ij}} \mathbf{b}_{jk\ell}-\mathbf{b}_{ik\ell}+\mathbf{b}_{ij\ell}-\mathbf{b}_{ijk}~, 
 \eea
so that the term appearing in the integrand in components is 
\bea
(\mathbf{b}^T \cup_\eta K \delta_\eta \mathbf{b})_{i j k \ell p q} = \mathbf{b}^T_{ijk} F^{\eta_{ik}}K\left(F^{\eta_{k\ell}}\mathbf{b}_{\ell p q} - \mathbf{b}_{kpq}+\mathbf{b}_{k \ell q} - \mathbf{b}_{k\ell p} \right)~.
\eea
Here we take $F$ to be given in the appropriate matrix representation. Note that similar expressions can be written down in the non-cyclic (and even non-Abelian) cases, though the expressions are more complicated and we do not write them down here. This gives a complete description of the SymTFT for the 4d theory.
\newline

What we have just described is the generic scenario. However, in some cases something special happens: it can sometimes be the case that there is a global form  $L$ (i.e. a choice of handlebody) which is left invariant by modular transformations, and hence also by surgery.\footnote{For the current discussion we set the background gauge fields in 4d to zero---i.e. we do not insert longitudinal Wilson lines inside the handlebody.} In this case, the geometry with the combined surgery and modular transformation not only has the same boundary as the original geometry, but is exactly identical to it (at fixed points of the period matrix). In this case there are not distinct geometries to sum over, and upon compactification we obtain (\ref{eq:BFpluseta}) with the field $\eta$ now only a background field. The same holds for global forms which are related to an invariant global form by a surgery operation.  This is exactly as expected: when there is an invariant global form $L$, the symmetries are non-intriniscally non-invertible, and this is only possible if the bulk theory is Dijkgraaf-Witten (and in this case BF).

Incidentally, note that this gives an alternative, though obviously equivalent, approach to searching for invariant global forms: we ask when an isometry of the Riemann surface $\Sigma_{g,n}$ can be extended to an isometry of $V_{g,n}^L$, with coefficients valued in $\ZZ_N$. It is known that this is possible if and only if the isometry maps meridians of $V_{g,n}^L$ to meridians \cite{rolfsen2003knots} modulo $\ZZ_N$. It is easy to convince oneself that for $N=p$ prime this gives the same results as identified using 4d techniques before. We may also understand the transformation law given in (\ref{eq:maintransformationlaw}) from the current perspective: in the current language, $F$ correspond to modular transformations while $G$ corresponds to surgery operations. Whereas the former acts by changing the basis in which the global form $M_{L,\cB}$ is expressed, the latter fixes the basis but changes the ``charges''  labelling the global variant.

Finally, let us note that for $(g,n)=(1,0)$ and $F=\mathsf{S}$, the theory obtained by compactifying the 7d CS + topological gravity theory on $V_{1,0}^L$ is precisely the SymTFT for $(3+1)$d theories with duality defects, studied in detail in \cite{Kaidi:2022cpf}. One question which was left unanswered in that work was when the SymTFT can be recast as a Dijkgraaf-Witten theory. Our current results allow us to answer this question when $N=p$ is prime, as was already discussed in the introduction in Theorems \ref{thm:1} and \ref{thm:2}. We may use the results in Sections \ref{sec:genus2} and \ref{sec:non-invert-symm} to make analogous statements about discrete gaugings of $(\ZZ_p^{(2)})^{\otimes g}$ gauge theory in $(4+1)$d, with $g \leq 5$.

\subsection{Surgery defects}
\label{sec:condensationdefects}

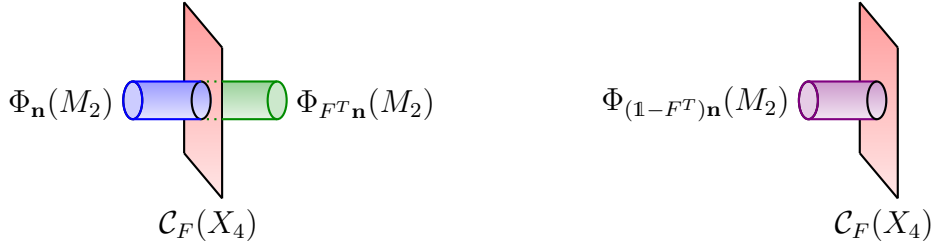
\begin{figure}[t]
    \centering

\begin{tikzpicture}

\begin{scope}[xshift=0 in, yshift=-1.2in]

   \shade[top color=red!40, bottom color=red!10,xshift=-0.6in, rotate=90]  (0,-0.7) -- (2,-0.7) -- (2.6,-0.2) -- (0.6,-0.2)-- (0,-0.7);

\draw[thick,xshift=-0.6in,rotate=90] (0,-0.7) -- (2,-0.7);
\draw[thick,xshift=-0.6in,rotate=90] (0,-0.7) -- (0.6,-0.2);
\draw[thick,xshift=-0.6in,rotate=90]  (0.6,-0.2)--(2.6,-0.2);
\draw[thick,xshift=-0.6in,rotate=90]  (2.6,-0.2)-- (2,-0.7);

 \shade[top color=blue!40, bottom color=blue!10]  (-2,1.55) -- (-1.1,1.55) 
 to[out=-10,in=10] (-1.1,1.05)  -- (-2,1.05)-- (-2,1.55);
  \shade[top color=dgreen!40, bottom color=dgreen!10] (-0.1,1.05) -- (-0.8,1.05)-- (-0.8,1.55) -- (-0.1,1.55)-- (-0.1,1.05) ;
    \draw[thick] (-1.1,1.3) ellipse (0.125cm and 0.25cm);
     \draw[thick,fill=blue!20] (-2,1.3) ellipse (0.125cm and 0.25cm);
         \draw[thick,dgreen,fill=dgreen!20] (-0.1,1.3) ellipse (0.125cm and 0.25cm);

\node[below] at (-1, 0){$\cC_{F}(X_4)$};

 \draw[thick,blue] (-2,1.3) ellipse (0.125cm and 0.25cm);
 \draw[thick,blue] (-2,1.55)--(-1.1,1.55);
  \draw[thick,blue] (-2,1.05)--(-1.1,1.05);

     \draw[thick,dgreen] (-0.1,1.55)--(-0.8,1.55);
  \draw[thick,dgreen] (-0.1,1.05)--(-0.8,1.05);
  \draw[thick,dotted,dgreen] (-0.8,1.55)--(-1.1,1.55);
  \draw[thick,dotted,dgreen] (-0.8,1.05)--(-1.1,1.05);
    
    \node[left] at (-2.1,1.25) {$\Phi_{\mathbf{n}}(M_2)$};
    
     \node[right] at (0,1.25) {$\Phi_{F^T\mathbf{n}}(M_2)$};

\end{scope}

\begin{scope}[xshift=3.5 in, yshift=-1.2in]
  
   \shade[top color=red!40, bottom color=red!10,xshift=-0.6in, rotate=90]  (0,-0.7) -- (2,-0.7) -- (2.6,-0.2) -- (0.6,-0.2)-- (0,-0.7);

\draw[thick,xshift=-0.6in,rotate=90] (0,-0.7) -- (2,-0.7);
\draw[thick,xshift=-0.6in,rotate=90] (0,-0.7) -- (0.6,-0.2);
\draw[thick,xshift=-0.6in,rotate=90]  (0.6,-0.2)--(2.6,-0.2);
\draw[thick,xshift=-0.6in,rotate=90]  (2.6,-0.2)-- (2,-0.7);

 \shade[top color=violet!40, bottom color=violet!10]  (-2,1.55) -- (-1.1,1.55) 
 to[out=-10,in=10] (-1.1,1.05)  -- (-2,1.05)-- (-2,1.55);

     \draw[thick,fill=violet!20] (-2,1.3) ellipse (0.125cm and 0.25cm);

\node[below] at (-1, 0){$\cC_{F}(X_4)$};

 \draw[thick,violet] (-2,1.3) ellipse (0.125cm and 0.25cm);
 \draw[thick,violet] (-2,1.55)--(-1.1,1.55);
  \draw[thick,violet] (-2,1.05)--(-1.1,1.05);
  
   \draw[thick] (-1.1,1.3) ellipse (0.125cm and 0.25cm);
   
       \node[left] at (-2.1,1.25) {$\Phi_{(\mathds{1}-F^T)\mathbf{n}}(M_2)$};

\end{scope}

\end{tikzpicture}

    \caption{The defects $\cC_F(X_4)$ can absorb Wilson surfaces $\Phi_{(\mathds{1}-F^T)\mathbf{n}}(M_2)$, and hence should be condensates of them.}
    \label{fig:defectpics}
\end{figure}

We close this section by providing more detail on the surgery defects discussed above. In particular, let us see how they descend to condensation defects in 5d. 
First we define the following surface operators,
\bea
\label{eq:general5dWilson}
\Phi_{\mathbf{n}}(M_2)= e^{{2\pi i \over N} {1\over 2}e \cdot m (M_2, M_2)}\prod_{i=1}^g \widehat \Phi_i^{e_i}(M_2) \prod_{i=1}^g \Phi_i^{m_i}(M_2)~, \hspace{0.5 in} \mathbf{n}=(e_1,\dots, e_g; m_1, \dots, m_g)~,
\eea
where $e \cdot m := \sum_{j=1}^g e_j m_j$. These are the most general Wilson surfaces that can be constructed in the 5d TQFT. Each of these surface operators descends from a Wilson three-surface in 7d, and is associated with a cycle $\gamma_\mathbf{n}$ on $\Sigma_{g,n}$ via
\bea
\Phi_{\mathbf{n}}(M_2) = \Phi(M_2 \times \gamma_\mathbf{n})~, \hspace{0.5 in} \gamma_\mathbf{n}:= \sum_{j=1}^g e_j \lambda_j + \sum_{j=1}^g m_j \mu_j~.
\eea
The surgery defects in 7d correspond to loci across which $F$ acts on $\gamma_\mathbf{n}$ to produce $F(\gamma_\mathbf{n})=\gamma_{F^T \mathbf{n}}$. We thus expect that surgery defects in 7d should reduce to defects $\cC_F(X_4)$ in 5d implementing 
\bea
\cC_F(X_4): \,\,\,\Phi_{\mathbf{n}}(M_2) \mapsto  \Phi_{F^T\mathbf{n}}(M_2)~\hspace{0.8 in} M_2 \subset X_4~. 
\eea
It is easy to see that such defects are condensation defects. Indeed, consider Figure \ref{fig:defectpics}. On the left-hand side of the figure we have a configuration in which $\Phi_{\mathbf{n}}(M_2)$ enters $\cC_F(X_4)$ from the left and $\Phi_{F^T \mathbf{n}}(M_2)$ exits from the right. We may now fold  $\Phi_{F^T\mathbf{n}}(M_2)$ to the left to obtain the figure on the right. This configuration tells us that $\cC_F(X_4)$ must be able to absorb $\Phi_{(\mathds{1}-F^T) \mathbf{n}}(M_2)$, for any $\mathbf{n}$.  Let us assume that $\mathbf{n} = (1,0, \dots, 0)$. Then $\Phi_{(\mathds{1}-F^T) \mathbf{n}} = \Phi_{(\mathds{1}-F^T)_1}$, where we use the notation that ${M_i}$ represents the $i$-th column of a matrix $M$. We conclude that $\cC_F(X_4)$ can absorb $\Phi_{(\mathds{1}-F^T)_1}$, and hence $\cC_F(X_4)$ should be a condensate of $\Phi_{(\mathds{1}-F^T)_1}$. The same argument can be used for $\mathbf{n} = (0,\dots, 0,1,0, \dots, 0)$ with the $1$ in the $i$-th slot, which tells us that $\cC_F(X_4)$ must be a simultaneous condensate of all $\Phi_{(\mathds{1}-F^T)_i}$ for $i=1, \dots, 2g$.

One subtlety is that, for certain values of $N$, two separate vectors $(\mathds{1}-F^T)_i$ and $(\mathds{1}-F^T)_j$ may be equal modulo $N$. In such cases the full one-form symmetry is not quite $\ZZ_N^{2g}$, but a quotient thereof. This happens for example in the case of $F = \sfS$ at genus 1, where the two vectors $(1-\sfS)_ i$ are $(1,-1)$ and $(1,1)$, which are equivalent for $N=2$; a treatment of the subtlety in this case can be found in \cite{Kaidi:2022cpf}. More generally, some of the  $(\mathds{1}-F^T)_i$ or multiples thereof may be linear combinations of the others. For simplicity here we will only focus on the case for which this does not happen, i.e. $\mathrm{ker}(\mathds{1}-F^T)= 0$ mod $N$, and will furthermore assume that $N=p$ is prime. The more general case is discussed in Appendix \ref{app:altCF}.

With this motivation, we may now define the following condensation defect
\bea
\cC_F(X_4) :={|H^0(X_4, \ZZ_p)|^{2g} \over |H^1(X_4, \ZZ_p)|^{2g} } \sum_{M_2^1, \dots, M_2^{2g} \in H_2(X_4, \ZZ_p)} e^{ {2\pi i \over p}{1\over 2} \sum_{i,j=1}^{2g}R^F_{ij} (M_2^i, M_2^j)} \prod_{i=1}^{2g} \Phi_{(\mathds{1}-F^T)_i}(M_2^i)~.
\eea
The $R^F_{ij}$ are $F$-dependent coefficients, whose explicit form will be given shortly.  In practice, it is actually more useful to rewrite the operators $\Phi_{(\mathds{1}-F^T)_i}(M_2)$ appearing above in terms of products of the elementary operators $\widehat \Phi_j(M_2)$ and $ \Phi_j(M_2)$ introduced previously. To do so, we write
\bea
\Phi_{(\mathds{1}-F^T)_i}(M_2^i) =\Phi(M_2^i\times \Gamma_i)~, \hspace{0.5 in} \Gamma_i :=\sum_{j=1}^g \alpha^i_j \lambda_j + \sum_{j=1}^g\beta_j^i \mu_j~.
\eea
The coefficients $\alpha^i_j$ and $\beta^i_j$ are fixed in terms of the matrix elements of $F$ straightforwardly,
\bea
\label{eq:Falphabetarel}
F = \mathds{1}-\left(\begin{matrix} \alpha^1_1 & \dots & \alpha_1^{2g}
\\
\vdots&&\vdots
\\
\alpha^1_g & \dots & \alpha_g^{2g}
\\
\beta^1_1 & \dots & \beta_1^{2g}
\\
\vdots&&\vdots
\\
\beta^1_g & \dots & \beta_g^{2g}
\end{matrix} \right)~.
\eea
Using (\ref{eq:general5dWilson}) and the commutation relations (\ref{eq:Heisenbergalg}), we may then expand 
\beaa
\prod_{i=1}^{2g} \Phi_{(\mathds{1}-F^T)_i}(M_2^i) &=& \mathrm{exp}\left\{ {2\pi i \over p} \left[ {1\over 2} \sum_{i=1}^{2g} \alpha^i \cdot \beta^i (M_2^i, M_2^i) + \sum_{i<j}^{2g} \beta^i \cdot \alpha^j (M_2^i, M_2^j) \right] \right\}
\no\\
&\vphantom{,}& \hspace{0.5 in} \times\prod_{j=1}^g \widehat \Phi_j\left(\sum_{i=1}^{2g}\alpha_j^i M_2^i\right)\prod_{j=1}^g \Phi_j\left(\sum_{i=1}^{2g}\beta_j^i M_2^i\right)
\eeaa
with $\alpha^i \cdot \beta^i := \sum_{j=1}^{2g} \alpha^i_j \beta^i_j$, whence the surgery defect may be rewritten as
\beaa
\label{eq:CFfinalans}
\cC_F(X_4) &=& {|H^0(X_4, \ZZ_p)|^{2g} \over |H^1(X_4, \ZZ_p)|^{2g} } \sum_{M_2^1, \dots, M_2^{2g} \in H_2(X_4, \ZZ_p)}  e^{{2 \pi i \over p}{1\over 2} \sum_{i,j=1}^{2g} (\beta^i \cdot \alpha^j+R^F_{ij}) (M_2^i, M_2^j)} 
\no\\
&\vphantom{.}&\hspace{0.8 in} \times\,  \prod_{j=1}^g \widehat \Phi_j\left(\sum_{i=1}^{2g}\alpha_j^i M_2^i\right)\prod_{j=1}^g \Phi_j\left(\sum_{i=1}^{2g}\beta_j^i M_2^i\right)~.
\eeaa
Having introduced this notation, we may now give the expression for the coefficients $R^F_{ij}$ in terms of $\alpha_i^j$ and $\beta_i^j$, 
\bea
\label{eq:RFij}
R^F_{ij} = \left\{ \begin{matrix} - \beta^i_j & & j \leq g \\ \alpha_{j-g}^i && j > g 
\end{matrix}\right. 
\eea
which completes the definition of the surgery defects $\cC_F(X_4)$. 

Let us also note that this defect can be written in a more streamlined fashion by defining it in terms of a sum over a single manifold $M_2$ in $H_2(X_4, \ZZ_p^{2g})$, as opposed to $2g$ manifolds $M_2^i$ in  $H_2(X_4, \ZZ_p)$, 
\begin{align}
  \label{eq:B.7.3}
  \cC_{F}(X_4) = {|H^0(X_4, \ZZ_p^{2g})| \over |H^1(X_4, \ZZ_p^{2g})| }\sum_{M_{2} \in H_{2}(X_{4} , \mathbb{Z}_{p}^{2g})} \exp(\frac{2\pi i}{2p}\ev{F M_{2},M_{2} }) \Phi((\mathds{1}-F)M_{2}) ~.
\end{align}
The notation here is such that $F$ now acts on the coefficient system of the manifold, instead of on the manifolds themselves. We discuss this expression in more detail in Appendix \ref{app:altCF}.

We may similarly define the conjugate defect $\overline{\cC_F}(X_4) := \chi(X_4, \ZZ_p)^{-2g}\cC_F(X_4)^\dagger$, where we have used the freedom to normalize this operator with an Euler counterterm. The dagger acts as complex conjugation, and also transposes the order of operators. With these definitions, one finds the following fusion rules,
\beaa
\cC_F(X_4) \times \overline{\cC_F}(X_4) &=& \overline{\cC_F}(X_4) \times \cC_F(X_4)  \,\, = \,\, \mathds{1}~, 
\no\\
\Phi_{\mathbf{n}}(M_2) \times \cC_F(X_4)  &=& \cC_F(X_4)  \times \Phi_{F^T\mathbf{n}}(M_2)~, 
\no\\
\Phi_{F^T\mathbf{n}}(M_2) \times \overline{\cC_F}(X_4)  &=& \overline{\cC_F}(X_4)  \times \Phi_{\mathbf{n}}(M_2)~.
\eeaa
In particular, we see that the fusions are generically non-Abelian. 

An important point is that the fusion rules above are valid only subject to certain constraints on the matrix $F$. In particular, explicit computation leads one to conclude that the following constraints on $\alpha^i_j$, $\beta^i_j$ must hold, 
\beaa
\label{eq:symplecticconstraints}
 \alpha^{i} \cdot \beta^{j} - \beta^{i} \cdot \alpha^{j} &=&  \beta_{i}^{j} - \beta_{j}^{i} \nonumber \\
  \alpha^{i} \cdot \beta^{j+g} - \beta^{i} \cdot \alpha^{j+g} &=& \beta_{i}^{j+g} + \alpha_{j}^{i} \nonumber \\
  \alpha^{i+g} \cdot \beta^{j+g} - \beta^{i+g} \cdot \alpha^{j+g} &=& \alpha_{j}^{i+g} -\alpha_{i}^{j+g} 
\eeaa
Remarkably, these conditions are satisfied exactly when $F$ is symplectic. Indeed, if $F$ is symplectic then $F^T \mathfrak{I} F = \mathfrak{I}$, and hence 
\bea
(\mathds{1}-F)^T \mathfrak{I} (\mathds{1}-F) = \mathfrak{I} (\mathds{1}-F) + (\mathds{1}-F)^T \mathfrak{I} ~. 
\eea
This gives a series of equations which are quadratic in $\alpha^i_j$, $\beta^i_j$ on the left and linear in $\alpha^i_j$, $\beta^i_j$ on the right, which are precisely the equations in (\ref{eq:symplecticconstraints}).

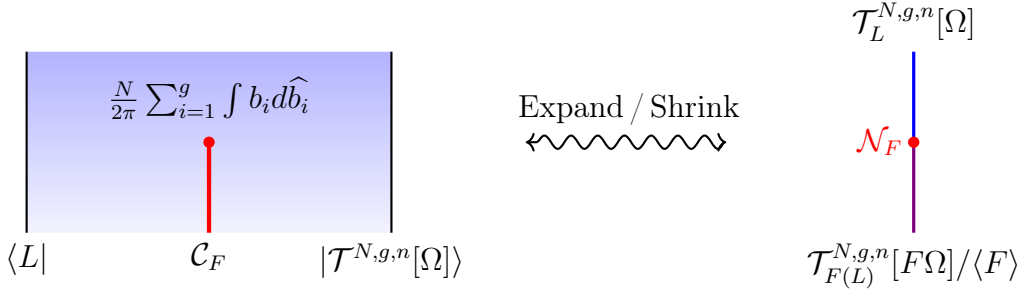
\begin{figure}[!tbp]
	\centering
	\begin{tikzpicture}[scale=0.8]

	\begin{scope}[xshift=-5.2 in]
	\shade[line width=2pt, top color=blue!30, bottom color=blue!5] 
	(0,0) to [out=90, in=-90]  (0,3)
	to [out=0,in=180] (6,3)
	to [out = -90, in =90] (6,0)
	to [out=180, in =0]  (0,0);
	
	\draw[thick] (0,0) -- (0,3);
	\draw[thick] (6,0) -- (6,3);
	\draw[ultra thick,red] (3,0) -- (3,1.5);
	\node[above] at (3,1.7) {${N \over 2 \pi} \sum_{i=1}^g \int b_i d \widehat{b_i}$};
	\node at (3,1.5) [circle,fill,red, inner sep=1.5pt]{};
	\node[below] at (0,0) {$\langle L| $};
	\node[below] at (6,0) {$|\cT^{N,g,n}[\Omega]\rangle $}; 
	\node[below] at (3,0) {$\cC_F$};
	\end{scope}

	\draw[thick, snake it, <->] (-1.7,1.5) -- (-5, 1.5);
	\node[above] at (-3.3,1.6) {Expand\,/\,Shrink};

	\begin{scope}[xshift=3.3 in]
	\draw[very thick, violet] (-7,0) -- (-7,1.5);
	\draw[very thick, blue] (-7,1.5) -- (-7,3);
	\node[above] at (-7,3) {$\cT^{N,g,n}_L[\Omega]$};
	\node[below] at (-7,0) {$\cT^{N,g,n}_{F(L)}[F\Omega]/\langle F\rangle$};
	\node at (-7,1.5) [circle,fill,red, inner sep=1.5pt]{};
	\node[left,red] at (-7,1.5) {$\cN_F$};
	\end{scope}

	\end{tikzpicture}
	
	\caption{When the surgery defect $\cC_F$ is given a topological boundary, one obtains a twist defect. Shrinking the slab, this twist defect becomes an $|F|$-ality interface $\cN_F$ in $(3+1)$d. 
	}
	\label{fig:twistdefshrink}
\end{figure}

For every element of $F \in Sp(2g, \ZZ_p)$, we have thus constructed codimension-1 defects implementing the symmetry in the bulk 5d TQFT. These condensation defects can now be used to construct $|F|$-ality defects in the boundary 4d gauge theories. Indeed, each of the condensation defects above admits at least one topological boundary condition (namely, Dirichlet boundary conditions for the condensate) which allows us to terminate them in the bulk. The result can be thought of as a non-genuine 3-manifold defect attached to a 4-manifold. Such non-genuine operators were referred to as ``twist defects'' in \cite{2015arXiv150306812T, Barkeshli:2014cna,Kaidi:2022cpf}, and we denote them by $T_F(M_3, M_4)$. Concretely, we have
\bea
T_F(M_3, M_4):={|H^0(M_4, M_3,  \ZZ_p)|^{2g} \over |H^1(M_4, M_3, \ZZ_p)|^{2g} }\sum_{M_{2} \in H_{2}(M_{4} , \mathbb{Z}_{p}^{2g})} \exp(\frac{2\pi i}{2p}\ev{F M_{2},M_{2} }) \Phi((1-F)M_{2}) ~,
\eea
where the only difference with the expression for the condensation defect $\cC_F(M_4)$ is in the overall normalization factor, which is now in relative cohomology.\footnote{Note however that the sum is still over \textit{absolute} homology, since the Lefschetz dual of relative cohomology is absolute homology.}
The fusion rules follow simply from those above.
By moving the twist defects to the boundary as shown in Figure \ref{fig:twistdefshrink}, one obtains an interface between the four-dimensional theory, 
\bea
\cN_F(M_3) := T_F(M_3, M_4)|_{y\rightarrow 0}~, 
\eea
which fuse as 
\bea
\cN_F(M_3) \times \overline{\cN_F}(M_3) = {1\over |H^0(M_3, \ZZ_p)|^{2g}} \sum_{M_2^1, \dots, M_2^{2g} \in H_2(M_3, \ZZ_p)}  \prod_{j=1}^g \widehat \Phi_j\left(\sum_{i=1}^{2g}\alpha_j^i M_2^i\right)\prod_{j=1}^g \Phi_j\left(\sum_{i=1}^{2g}\beta_j^i M_2^i\right)~.
\eea
One can further obtain the fusion rules of $\cN_F(M_3)$ with $\cN_F(M_3)$  or $\cN_G(M_3)$ for generic $G\in Sp(2g,\ZZ_p)$, thereby proving that they are $|F|$-ality defects as claimed. Since these fusion rules will appear in an upcoming work \cite{SISSA}  we do not expand on them here.

\section*{Acknowledgements}

We would like to thank Christian Copetti, 
Luis Diogo, Alice Hedenlund, Kantaro Ohmori, Jian Qiu, Gabi Zafrir, and Yunqin Zheng for many useful conversations. We also thank Yunqin Zheng for a careful reading of the manuscript. The work of MDZ and AH is supported by the European Research Council (ERC) under the European Union's Horizon 2020 research and innovation program
(grant agreement No. 851931). MDZ also acknowledges support from the Simons Foundation Grant $\#$888984 (Simons Collaboration on Global Categorical Symmetries). VB is supported from the Knut and Alice Wallenberg Foundation under grant KAW 2021.0170, VR grant 2018-04438 and the Olle Engkvists Stiftelse grant n. 2180108.

\newpage
\begin{appendix}
\section{Symmetry generators at genus $2$}
\label{app:matrices}

In this appendix we give explicit matrix representation of the generators of isometries for genus 2 Riemann surfaces. These matrices were also given in \cite{Nilles:2021glx}. %We use the notation $\rho := e^{2 \pi i /3}$ and $\varepsilon := e^{2 \pi i /5}$. 
First, the matrices appearing in Table \ref{tab:g2singmodtab1} are given by,

\beaa
\phi &=&  \left(\begin{matrix}0 & - 1 & - 1 & - 1 \\ 0 & 0 & -1 & 0 \\ 0 & 0 & 0 & -1 \\ 1 & 0 & 0 & 1 \end{matrix}\right)~, \hspace{0.5 in} M_1 = \left(\begin{matrix} 0 & 0 & 0 & 1 \\ 0 & 0 & 1 & 1 \\ 1 & -1 & 0 & 0 \\ -1 & 0 & 0 & 0 \end{matrix} \right)~, 
\no\\\no\\
M_2 &=&  \left(\begin{matrix} 0 & 0 & 1 & 1 \\ 0 & 0 & 1 & 0 \\ 0 & -1 & 0 & 0 \\ -1 & 1 & 0 & 0 \end{matrix}\right)~, \hspace{0.5 in} M_3=  \left(\begin{matrix}0 & 0 & 0 & 1 \\ 0 & 0 & -1& 0 \\ 0 & -1 &0 & 0 \\ 1 & 0 & 0 & 0 \end{matrix}\right)~, 
\no\\\no\\
M_4 &=&  \left(\begin{matrix}0 & 0 & 1 & 0 \\ 0 & 0 & 0 & 1 \\ -1& 0 & -1& 0 \\ 0 & -1& 0 & 0 \end{matrix}\right)~, \hspace{0.5 in} M_5=  \left(\begin{matrix} 0 & 0 & 1 & 0 \\ 0 & -1 & 0 & 0 \\ -1 & 0 & 0 & 0 \\ 0 & 0 & 0 & -1\end{matrix}\right)~, \no
\\\no\\
M_6 &=&  \left(\begin{matrix} 0 & 0 & -1 & 0 \\ 0 & 0 & 0 & 1 \\ 1 & 0 & 0 & 0 \\ 0 & -1 & 0 & 0 \end{matrix}\right)~, \hspace{0.5 in} M_7=  \left(\begin{matrix} 0 & 1 & 0 & 0 \\ 1 & 0 & 0 & 0 \\ 0 & 0 & 0 & 1 \\ 0 & 0 & 1 & 0\end{matrix}\right)~, 
\no\\\no\\
M_8 &=&  \left(\begin{matrix} -1& 1 & 1 & 0 \\ 1 &0 & 0 & 1 \\ -1 & 0 & 0 & 0 \\ 1 & -1 &0 & 1 \end{matrix}\right)~, \hspace{0.5 in} M_{9}=  \left(\begin{matrix}0 & 0 & 0 & -1 \\ 1 & 0 & 1 & 0 \\ 0 & 1 & 0 & 1 \\ -1& 0 & 0 & 0  \end{matrix}\right)~,
\no\\\no\\
M_{10}&=&  \left(\begin{matrix} 0 & 0 & 1 & 0 \\ 0 & 0 & 0 & -1 \\ -1& 0 & -1& 0 \\ 0 & 1 & 0 & 1\end{matrix}\right)~, \hspace{0.5 in} C=  \left(\begin{matrix} -1 & 0 &0& 0 \\ 0 & -1 & 0 & 0 \\ 0 & 0 & -1& 0 \\ 0 & 0 & 0 & -1\end{matrix}\right)~.
\eeaa

In addition, the remaining matrices appearing in Tables \ref{tab:g2singmodtab2} and \ref{tab:g2singmodtab3} are given by
\beaa
N_1 &=& \left(\begin{matrix}0 & 0 & 1 & 0 \\ 0 & 1 & 0 & 0 \\ -1 & 0 & 0 & 0 \\ 0 & 0 & 0 & 1 \end{matrix} \right)~, \hspace{0.5 in} N_2 = \left(\begin{matrix}0 & 1 & -1 & 0 \\ -1 & 0 & 0 & 1 \\ 0 & 0 & 0 & 1\\ 0 & 0 & -1& 0\end{matrix} \right)~, 
\no\\
N_3&=& \left(\begin{matrix} 0 & -1 & 0 & 0 \\ 1 & 0 & 0 & 0 \\ 0 & 0 & 0 & -1 \\ 0 & 0 & 1 & 0 \end{matrix} \right)~, \hspace{0.5 in} N_4= \left(\begin{matrix}1 & 0 & 1 & 0 \\ 0 & 1 & 0 & 0 \\ -1 & 0 & 0 & 0 \\ 0 & 0 & 0 & 1 \end{matrix}\right)~, 
\no\\
N_5 &=& \left(\begin{matrix} -1 & 1 & 0 & 0 \\ 0 & 1 & 0 & 0 \\ 0 & 0 & -1 & 0 \\ 0 & 0 & 1 & 1\end{matrix} \right)~,  \hspace{0.5 in}  N_6 = \left(\begin{matrix}1 & 0 & 0 & 0 \\ 1 & -1& 0 & 0 \\ 0 & 0 & 1 & 1\\ 0 & 0 & 0 & -1 \end{matrix}\right)~, 
\no\\
P_1&=& \left(\begin{matrix}1& 0 & 0 & 0 \\ 0 & -1& 0 & 0 \\0 & 0 & 1 & 0 \\0 & 0 & 0 & -1 \end{matrix}  \right)~. 
\eeaa

\section{Some algebra}
\label{app:arithmetic}

In this appendix we collect and give details on various arithmetic statements that are used in the main text.

\subsection{Roots of unity over finite fields}
\label{sec:roots-unity-over}
Over the complex numbers the polynomial $x^{n} - 1$ has $n$ roots, each with multiplicity one. We can ask if the same is true over an extension of $\mathbb{F}_{p}$ which splits
$x^{n}-1$.

Over any field, if a polynomial $P(x)$ has a root $\lambda$ with multiplicity greater than $1$, then $\lambda$ is also a root of $\dv{}{x}P(x)$. For $x^{n}-1$
this means that there are roots with multiplicity greater than $1$ if and only if $n\lambda^{n-1}$ = 0. Since $\lambda^{n-1}$ can't be zero, this is possible only if
$n = 0 \mod p$. Hence if $p$ is not a factor of $n$, there are $n$ distinct roots of unity.

Since the product of two $n$-th roots of unity is another $n$-th
root of unity, $n$-th roots of unity form a group. This group is always cyclic, just like the group of complex $n$-th roots of unity. To see this let us assume
the contrary, namely that there
% is an $n$-th root of unity which does not have order $n$. In this case we can find 
exist two $n$-th roots of unity $\sigma$ and $\kappa$ such that the intersection of
their orbits is the identity, and such that they generate a subgroup which is not cyclic. Let $k$ be the order of $\sigma$ and $l$ the order of $\kappa$. Non-cyclicity means that
$\gcd(k,l) = m > 1$. Hence, $\sigma^{\frac{k}{m}}$ and $\kappa^{\frac{l}{m}}$ both generate a subgroup of order $m$. Unique factorization of $x^{m}-1$ then demands that
$\sigma^{\frac{k}{m}} = \kappa^{\frac{l}{m}}$, which contradicts the assumption that the orbits do not intersect.

Hence if $p$ is not a factor of $n$, the group of $n$-th roots of unity is a cyclic group of order $n$.

Let us next consider the case where $p$ is a factor of $n$, namely $n = p^{k}m$ with $p$ not dividing $m$. Since
\begin{align}
  \label{eq:B.33}
  \binom{p^{k}}{\ell} = 0 \mod p \hspace{0.5 in} \mbox{ if }  \hspace{0.5 in}  \ell \ne 0,\,p^{k} ~
\end{align}
we have $(x^{m}-1)^{p^{k}} = \sum_{\ell=0}^{p^k} \binom{p^{k}}{\ell} (-1)^{p^k - \ell}x^{m \ell}= x^{n} - 1$ mod $p$. So the roots of $x^{n}-1$ are roots of $x^{m}-1$ with multiplicity $p$. Since $p$ is not a factor of $m$, the group of
$n = p^{k}m$ roots of unity is a cyclic group of order $m$.

Noting further that 
\begin{align}
  \label{eq:B.34}
  x^{n}-1 = \prod_{k|n} \phi_{k}(x) ~
\end{align}
with $\phi_k(x)$ the $k$-th cyclotomic polynomial, 
we can also determine the factorization of $\phi_{n}(x)$ from this. When $p$ is not a factor of $n$, the roots of $\phi_{n}(x)$ are primitive $n$-th roots of unity. More
explicitly, if $\lambda$ is the generator of $n$-th roots of unity then the roots of $\phi_{n}$ are $\lambda^{r}$ with $\gcd(r,n) = 1$.

When $n = p^{k}m$, \eqref{eq:B.34} along with an inductive argument show that
\begin{align}
  \label{eq:B.35}
  \phi_{n}(x) = (\phi_{m}(x))^{\varphi(p^{k})} ~.
\end{align}
In this case the roots of $\phi_{n}(x)$ are the primitive $m$-th roots of unity, each with multiplicity $\varphi(p^{k}) = p^{k} - p^{k-1}$.

\subsection{Integer matrices of finite order}
\label{sec:integ-matr-finite}
Let $F$ be a matrix of order $k$ with integer coefficients. Then as a complex matrix it has a Jordan normal form; see the discussion around (\ref{eq:Jordannormaldef}) for the definition of Jordan normal form. Each Jordan block corresponds to an eigenvalue $\lambda$
and contains an eigenvector $v$ with $\lambda$ as its eigenvalue. The condition that $F$ has order $k$ then translates to,
\begin{align}
  \label{eq:B.36}
  F^{k}v = \lambda^{k}v = v ~.
\end{align}
Hence $\lambda$ must be a $k$-th root of unity. Furthermore, every Jordan block must be of size $1$, i.e. $F$ is diagonalizable as a complex matrix. To see this, let us
suppose that a Jordan block with eigenvalue $\lambda$ is not of size 1. Then in addition to the eigenvector $v$, there is a vector $u$ such that
\begin{align}
  \label{eq:B.37}
  Fu = \lambda u + v ~.
\end{align}
Applying $F$ repeatedly, we deduce
\begin{align}
  \label{eq:B.38}
  F^{k}u = \lambda^{k}u + k\lambda^{k-1}v ~.
\end{align}
Since $F^{k} = \mathbbm{1}$, the above condition tells us that $k\lambda^{k-1}v = 0$, which is impossible. Thus all integers matrices of finite order are diagonalizable as complex matrices. Things will be different when we consider matrices over finite fields.

Since $F$ is an integer matrix, its characteristic polynomial $P_{F}(x)$ is an integer polynomial. Let $\lambda_{i}$ be the (complex) eigenvalues of $F$. Then by definition
\begin{align}
  \label{eq:B.39}
  P_{F}(x) = \prod_{i} (x - \lambda_{i}) ~.
\end{align}
As explained above, all eigenvalues of $F$ are roots of unity. Hence the only possible integer eigenvalues are $\pm 1$. All other eigenvalues are complex, and their
product must give an integer polynomial. Complex numbers which are roots of monic polynomials with integer coefficients are known as ``algebraic
integers'', and hence the eigenvalues are all algebraic integers.

 Every algebraic integer $\lambda$ has a corresponding minimal polynomial $\phi(x)$, which is the unique monic polynomial of the smallest degree such that $\lambda$ is a root of $\phi(x)$. If $P(x)$ is any other integer polynomial with root $\lambda$, then the minimal polynomial $\phi(x)$ of $\lambda$ must be a factor of $P(x)$. Roots of unity are algebraic integers since they are roots of $x^{n}-1$. The minimal polynomial for a primitive $k$-th root of unity is the cyclotomic polynomial $\phi_{k}(x)$. As a result, the
characteristic polynomial $P_{F}(x)$ of a finite order integer matrix $F$ must be a product of cyclotomic polynomials. Hence if one primitive $k$-th root of unity is a
root of $P_{F}(x)$,  then all other primitive $k$-th roots of unity must also be roots,  all with the same multiplicity.

\subsection{Intersections and the Jordan blocks}
\label{sec:inters-jord-blocks}
Recall that the action of a modular transformation on the cycles in $H_{1}(\Sigma_{g,0} , \mathbb{F}_{p})$ is represented by a matrix $F \in Sp(2g,\mathbb{F}_{p})$. This means that $F$ preserves the symplectic form $\mathfrak{I}$,
\begin{align}
  \label{eq:B.40}
  F^{T} \mathfrak{I} F = \mathfrak{I} ~.
\end{align}
If we work over the splitting field of the characteristic polynomial of $F$, the cycles organize themselves into Jordan blocks which are (partially) labelled by the
eigenvalues $\lambda$ of $F$. A Jordan block with eigenvalue $\lambda$ is spanned by vectors $v^{(i)}_{\lambda}$ such that
\begin{align}
  \label{eq:B.41}
  F v^{(0)} &= \lambda v^{(0)} ~, \nonumber \\
  F v^{(i)} &= \lambda v^{(i)} +  v^{(i-1)} ~.
\end{align}
In particular $v^{(0)}_{\lambda}$ is an eigenvector with eigenvalue $\lambda$. Since $F$ satisfies \eqref{eq:B.40}, we have 
\begin{gather}
  \label{eq:B.42}
  (v^{(0)}_{\lambda_{1}})^{T}\mathfrak{I} v^{(0)}_{\lambda_{2}} = (v^{(0)}_{\lambda_{1}})^{T}F^{T}\mathfrak{I}F v^{(0)}_{\lambda_{2}} = \lambda_{1}\lambda_{2}(v^{(0)}_{\lambda_{1}})^{T}\mathfrak{I} v^{(0)}_{\lambda_{2}} ~,\nonumber
\end{gather}
which then implies $ (\lambda_{1}\lambda_{2} - 1)(v^{(0)}_{\lambda_{1}})^{T}\mathfrak{I} v^{(0)}_{\lambda_{2}} = 0 $ mod $p$. 
Hence $\ev{v_{\lambda_{1}}^{(0)} , v_{\lambda_{2}}^{(0)}} = 0$ unless $\lambda_{2} = \lambda_{1}^{-1}$ mod $p$. Following the same steps iteratively, we can obtain that
$\ev{v_{\lambda_{1}}^{(i)} , v_{\lambda_{2}}^{(i)}} = 0$ unless $\lambda_{2} = \lambda_{1}^{-1}$ mod $p$ for any $i$. This means in particular that Jordan blocks themselves are isotropic unless the
eigenvalue $\lambda = \pm 1$.

Now let us apply this to integer matrices of finite order as above. As explained, the characteristic polynomial of such a matrix is a product of cyclotomic polynomials,
\begin{align}
  \label{eq:B.43}
  P_{F}(x) = (\phi_{n_{1}}(x))^{k_{1}} (\phi_{n_{2}}(x))^{k_{2}} \dots (\phi_{n_{l}}(x))^{k_{l}}
\end{align}
The module $\mathbb{Z}^{2g}$ on which $F$ acts splits as
\begin{align}
  \label{eq:B.44}
  \mathbb{Z}^{2g} \cong \Lambda_{1} \oplus \Lambda_{2} \oplus \dots \oplus \Lambda_{l} ~, && \Lambda_{i} := \ker(\Phi_{n_{i}}(F)) ~.
\end{align}
This splitting relies on the diagonalizability of $F$. The $\Lambda_{i}$ are invariant under the action of $F$, and using the intersections in diagonalized form we can
deduce that the vectors in $\Lambda_{i}$ do not intersect vectors in $\Lambda_{j}$ unless $i = j$. Furthermore, we can split each $\Lambda_{i}$ into $k_{i}$ parts,
\begin{align}
  \label{eq:B.45}
  \Lambda_{i} = \Lambda_{i}^{(1)} \oplus \Lambda_{i}^{(2)} \oplus \dots \oplus \Lambda_{i}^{(k_{i})} ~,
\end{align}
where each $\Lambda_{i}^{(1)}$ is of dimension $\varphi(n_{i})$ and is annihilated by $\phi_{n_{i}}(F)$. This means that regarded as a complex vector space, the space $\Lambda_{i}^{(a)}$ contains one eigenvector for each primitive $n$-th root of unity, and furthermore vectors in $\Lambda_{i}^{(a)}$ don't intersect those in $\Lambda_{i}^{(b)}$ unless $a=b$.

This remains true on passage from integers to finite fields. As a result, whenever $\phi_{n_{i}}(x)$ has $\varphi(n_{i})$ distinct roots over $\mathbb{F}_{p}$, each
$\Lambda_{i}^{(a)}$ will contain one eigenvector for each of these roots. On the other hand, if $p$ is a factor of $n$ there can be non-trivial
Jordan blocks. In fact, in this case each $\Lambda^{(a)}_{i}$ contains one Jordan block of size $\varphi(n)$ for each root of $\phi_{\frac{n_{i}}{p^{k}}}(x)$ where $p^{k}$ is the
highest power of $p$ that divides $n_{i}$. This can be seen as a statement about the order of $\tilde{F}$, defined to be $F$ restricted to $\Lambda_{i}^{(a)}$. Over
integers $\tilde{F}$ has order $n = m p^{k}$, while over $\mathbb{F}_{p}$ it has this order iff there is a single Jordan block for each root of $\phi_{m}(x)$. To see this, let us suppose that the order of $\tilde{F}$ is $mp^{\ell}$ with $\ell \leq k$. In that case we must have that over integers,
\begin{align}
  \label{eq:B.46}
  \tilde{F}^{mp^{\ell}} &= \mathbbm{1} + p N
\end{align}
where $N$ is some non-zero matrix. Let us recall that over its splitting field the matrix $\tilde{F}$ is diagonalizable, with eigenvalues that are
primitive $n$-th roots of unity. The eigenvalues of $\tilde{F}^{mp^{\ell}}$ are then primitive $p^{k-\ell}$-th roots of unity, each with multiplicity
$\frac{\varphi(n)}{\varphi(p^{k-\ell})}$. Hence we obtain
\begin{align}
  \label{eq:B.47}
  \det(x - M^{mp^{\ell}}) = \left(\phi_{p^{k-\ell}}(x)\right)^{\frac{\varphi(n)}{\varphi(p^{k-\ell})}} \Rightarrow \det(\mathbbm{1} - M^{mp^{\ell}}) = p^{\frac{\varphi(n)}{\varphi(p^{k-\ell})}}
\end{align}
where we have used the fact that $\phi_{p^{k}}(1) = p$ for $k > 1$ and $p > 2$. Now going back to \eqref{eq:B.46}, we see that this means,
\begin{align}
  \label{eq:B.48}
  p^{\varphi(n)}\det(N) &= p^{\frac{\varphi(n)}{\varphi(p^{k-\ell})}}
\end{align}
Hence $N$ cannot be an integer matrix unless $k = \ell$.

We can now ask about the intersection number of a vector $v$ in a Jordan block of eigenvalue $\lambda$ with another vector $u$ in a different Jordan block. %inside $\Lambda_{i}^{(a)}$
 First we note that since $\mathfrak{I}$ is non-degenerate there must exist at least one vector which intersects $v$, and from the discussion above it must
lie in a Jordan block of eigenvalue $\lambda^{-1}$. Now we can check that
\begin{align}
  \label{eq:B.49}
  (v^{(0)}_{\lambda^{-1}})^{T}F^{T}\mathfrak{I} Fv^{(1)}_{\lambda} =  (v^{(0)}_{\lambda^{-1}})^{T}\mathfrak{I} v^{(1)}_{\lambda}  \hspace{0.3 in}
  \Rightarrow \hspace{0.3 in}(v^{(0)}_{\lambda^{-1}})^{T} v^{(0)}_{\lambda} = 0 ~,
\end{align}
i.e. if there is a non-trivial trivial Jordan block for $\lambda$, then the eigenvector $v^{(0)}_{\lambda}$ cannot intersect $v^{(0)}_{\lambda^{-1}}$. The argument above obviously
does not work if $v^{(1)}_{\lambda}$ does not exist. We can then iterate this computation and deduce that $\ev{v_{\lambda}^{(0)} , v_{\lambda^{-1}}^{(a)}} = 0$, unless
$v_{\lambda^{-1}}^{(a+1)}$ doesn't exist. In that case we need to compute $(v^{(a+1)}_{\lambda^{-1}})^{T}F^{T}\mathfrak{I} Fv^{(0)}_{\lambda}$. Similarly, by an appropriate sequence of computations of
$(v^{(a)}_{\lambda^{-1}})^{T}F^{T}\mathfrak{I} Fv^{(b)}_{\lambda}$, we can conclude that $\ev{v^{(a)}_{\lambda} , v^{(b)}_{\lambda^{-1}}} = 0$ unless $a+b-1$ is greater than or equal to the size of the
Jordan blocks, i.e $\varphi(p^{k})$. In fact, this can be simplified even further: the $v^{(a)}_{\lambda}$ are not uniquely determined by \eqref{eq:B.41}. In fact we can change $v^{(a)}_{\lambda}$ by any arbitrary linear combination of $v^{(b)}_{\lambda}$ with $b < a$, and \eqref{eq:B.41} will still hold. By using this freedom we can choose $v^{(b)}_{\lambda}$ in such a way that the non-zero intersection numbers are $
  \ev{v_{\lambda}^{(a)} , v_{\lambda^{-1}}^{(\varphi(p^{k})-1-a)}}$.

\subsection{Factorization $\phi_{15}(x)$ of over finite fields}
\label{sec:fact-over-finite}

We now focus on the study of invariant subspaces of $\phi_{15}(x)$, which is relevant to the study of global forms of genus 4 class $\cS$ theories invariant under a certain order 15 symmetries, discussed in Section \ref{sec:genus-4}.

Passing to the splitting field of $\phi_{15}(x)$, we obtain a primitive fifteenth root of unity $\lambda$. The roots of $\phi_{15}(x)$ are then:
$\lambda,\lambda^{2},\lambda^{4},\lambda^{7},\lambda^{8},\lambda^{11},\lambda^{13},$ and $\lambda^{14}$. Since $\lambda^{3}$ is a primitive fifth root of unity and $\lambda^{5}$ is a primitive third root of unity, we obtain two
relations,
\begin{align}
  \label{eq:B.51}
  \phi_{3}(\lambda^{5}) &\,\,=\,\, \lambda^{10} + \lambda^{5} + 1 = 0~, \nonumber \\
  \phi_{5}(\lambda^{3}) &\,\,=\,\, \lambda^{12} + \lambda^{9} + \lambda^{6} + \lambda^{3} + 1 = 0~.
\end{align}

The multiplicative groups $\ZZ_{15}^{\times}$ is not cyclic, and as a result there is no element of order $8$ in it. Thus over any $\mathbb{F}_{p}$ we can write
$\phi_{15}(x) = P(x)Q(x)$ where $P(x),Q(x)$ are polynomials over $\mathbb{F}_{p}$ that can potentially factorize further. We want to determine when they determine isotropic
invariant global form. Here we need to divide into three cases,
\begin{enumerate}
  \item\label{item:1} We first assume that the constant term of $P(x)$ and $Q(x)$ is $1$, since that is only power of $\lambda$ that is in $\mathbb{F}_{p}$ for all $p$, and hence
  the only value allowed without further assumptions on $p$. The only possible factorization satisfying this and the isotropy criterion is:
  \beaa
    \label{eq:B.52}
    P(x) &=& (x-\lambda)(x-\lambda^{2})(x-\lambda^{4})(x-\lambda^{8}) ~,
    \no\\
    Q(x) &=& (x-\lambda^{7})(x-\lambda^{11})(x-\lambda^{13})(x-\lambda^{14}) ~.
  \eeaa
  These are polynomials over $\mathbb{F}_{p}$ if $\eta := \lambda + \lambda^{2} + \lambda^{4} + \lambda^{8} \in \mathbb{F}_{p}$. Using the relations \eqref{eq:B.51}, it can be checked that
  $\eta$ satisfies,
  \begin{align}
    \label{eq:B.53}
    \eta^{2} - \eta + 4 = 0 \hspace{0.3 in}\Rightarrow\hspace{0.3 in} \eta = \frac{1 \pm \sqrt{-15}}{2} ~.
  \end{align}
  Hence $P(x)$ and $Q(x)$ are polynomials over $\mathbb{F}_{p}$ whenever $-15$ has a square root in $\mathbb{F}_{p}$. Using the Legendre symbol, this can be seen to happen
  whenever $p \in 15 \mathbb{N} + k$ for $k=1,2,4,$ and $8$. Notice that when $k=1,4$ there are also other values available for the constant term of $P(x)$ and $Q(x)$. This means that $P(x)$ and $Q(x)$ can split further for these cases, and as we will see below that they do.
  For the other cases $k=2,8$ there are exactly 2 invariant global forms.

  \item\label{item:2} If $p \in 3 \mathbb{N} + 1$ then $\lambda^{5} \in \mathbb{F}_{p}$ and hence $P(x)$ and $Q(x)$ can have a constant term which is a power of $\lambda^{5}$. As a result, in addition to the possibility given in \eqref{eq:B.52}, there is another possibility,
  \beaa
    \label{eq:B.54}
    P(x) &=& (x-\lambda)(x-\lambda^{4})(x-\lambda^{7})(x-\lambda^{13}) ~,
    \no\\
     Q(x) &=& (x-\lambda^{2})(x-\lambda^{8})(x-\lambda^{11})(x-\lambda^{14}) ~.
  \eeaa
  Using \eqref{eq:B.51}, it can be verified that
  \begin{align}
    \label{eq:B.55}
    (-\lambda - \lambda^{4} - \lambda^{7} - \lambda^{13})^{3} = (-\lambda^{2} - \lambda^{8} - \lambda^{11} - \lambda^{14})^{3} = 1~.
  \end{align}
  This together with \eqref{eq:B.51} can then be used to show that $P(x)$ and $Q(x)$ are polynomials over $\mathbb{F}_{p}$. This means that when
  $p \in 15 \mathbb{N} + 7$ or $15 \mathbb{N} + 13$, there are 2 invariant global forms. For $p \in 15\mathbb{N} + 4$ there are 4.

  \item\label{item:3} Finally if $p \in 5\mathbb{N} + 1$, then $\lambda^{3} \in \mathbb{F}_{p}$. Moreover using \eqref{eq:B.51} we can show that,
  \begin{align}
    \label{eq:B.56}
    (- \lambda - \lambda^{11})^{5} = (-\lambda^{2} - \lambda^{7})^{5} = (-\lambda^{4} - \lambda^{14})^{5} = (-\lambda^{8}-\lambda^{13})^{5} = 1~.
  \end{align}
  Hence all of the four sums are fifth roots of unity, and thus in $\mathbb{F}_{p}$. This gives four factors of $\phi_{15}$ which are polynomials over
  $\mathbb{F}_{p}$, namely $(x-\lambda)(x-\lambda^{11})$, $(x-\lambda^{2})(x-\lambda^{7})$, $(x - \lambda^{4})(x-\lambda^{14})$, and $(x-\lambda^{8})(x - \lambda^{13})$. In all 4 cases, the coefficient of
  both the linear and constant term is a fifth root of unity. Hence there are $4$ invariant global forms when $p \in 15 \mathbb{N} + 6$ or $p \in 15 \mathbb{N} + 11$.
\end{enumerate}

When $p \in 15 \mathbb{N} + 1$, all 3 cases above coincide and there are 8 invariant global forms. In the remaining case, i.e. when $p \in 15 \mathbb{N} + 14$, the factorization gives non-isotropic invariant forms and hence in that case there are no invariant global forms.

\subsection{Composite characteristic polynomial}
\label{sec:composite-char-pol}
Finally, in this appendix we make some comments on the case in which the characteristic polynomial of $F$ is not a cyclotomic polynomial, but a product thereof.  Concretely, take  $P_{F}(x) = (\phi_{n}(x))^{k}$ with $n > 2$. In this case, over the complex numbers
each of the $n$ eigenvalues of $P_{F}(x)$ has multiplicity $k$. For the generic case when $p$ is not a factor of $n$, this fact carries over to the field extension of $\mathbb{F}_{p}$ which splits $P_{F}(x)$. Hence each
eigenvalue $\lambda$ has an eigenspace of degeneracy $k$. We can always choose a basis $v_{\lambda}^{(i)}, 1 \le i \le k$ of these spaces such that the non-zero intersection
numbers are,\footnote{The ordering in the expression is arbitrary and its only purpose is to fix the sign of the intersection pairing.}
\begin{align}
  \label{eq:3.18}
  \ev{v_{\lambda}^{(i)} , v_{\lambda^{-1}}^{(i)}} = 1~, &\hspace{0.5 in}\mbox{ if }\,\,\, \lambda > \lambda^{-1} ~.
\end{align}
This degeneracy means that often a large number of invariant subspaces exist.
%and examples of this can be seen at genus 2. 
In fact, if $k = 2l$ is even then $F$ has invariant spaces for every prime field $\mathbb{F}_{p}$. This is because any subspace spanned by $l$ eigenvectors of each eigenvalue is annihilated by
$(\phi_{n}(F))^{l}$ and hence is an invariant subspace, and because of the intersections given in \eqref{eq:3.26} many of these are isotropic.

As an illustration let us consider the case of $k=2$. If $\phi_{n}(x)$ is irreducible over $\mathbb{F}_{p}$, then an invariant global form consists of precisely
one vector from each of the eigenvalues $\lambda^{i}$, where $1 \le i < n, \gcd(i,n) = 1$ and $\lambda^{n} = 1$. There are $p+1$ ways of picking an eigenvector (up to scalar
multiplication) with the eigenvalue $\lambda^{i}$ from the 2 dimensional eigenspace. Once we pick an eigenvector $v_{\lambda}$ for $\lambda^{i}$, then isotropy forces us to pick the
unique (again up to scalar multiplication) eigenvector $u_{\lambda^{-1}}$ for $\lambda^{-i} = \lambda^{n-i}$ which does not intersect $u$. As a result, there are
$(p+1)^{\frac{\varphi(n)}{2}}$ invariant isotropic subspaces. If $\phi_{n}(x)$ factorizes then more invariant global forms exist and can be counted using similar
arguments, we will refrain from a general exposition here.

If $k$ is odd, then invariant global forms require factorization of $\phi_{n}(x)$ over $\mathbb{F}_{p}$. The counting of the number of invariant forms is tedious,
but is a straightforward application of the arguments above combined with those used in the preceding sections. The non-generic case when $p$ is a factor of $n$ is
yet more involved because there are multiple non-trivial Jordan blocks.

Finally we discuss the most general case,
\begin{align}
  \label{eq:3.19}
  P_{F}(x) = \prod_{n} (\phi_{n}(x))^{k_{n}} ~.
\end{align}
Over the complex numbers each $\ker(\phi_{n}(F)^{k_{n}})$ is an invariant subspace and, moreover, a vector in such a subspace only intersects with other vectors in
the same subspace.

For the generic prime fields $\mathbb{F}_{p}$\footnote{Generic means that if $k_{n} \ne 0$ then $p$ is not a factor of $n$. This is the generic case
  because for a given $F$, there are only finitely many primes not satisfying this.} this means that an invariant global form consists of invariant maximal isotropic
lattice from each of $\ker(\phi_{n}(F)^{k_{n}})$. As a result the number of invariant global forms is a product: If we denote by
$N^{(p)}_{n_{1} , k_{1} ; n_{2} , k_{2} ; \dots ; n_{a} , k_{a}}$, the number of global forms for,
\begin{align}
  \label{eq:3.20}
  P_{F}(x) = (\phi_{n_{1}}(x))^{k_{1}} \dots (\phi_{n_{a}}(x))^{k_{a}} ~,
\end{align}
then for generic $p$,
\begin{align}
  \label{eq:3.21}
  N^{(p)}_{n_{1} , k_{1} ; n_{2} , k_{2} ; \dots ; n_{a} , k_{a}} = N^{(p)}_{n_{1},k_{1}} N^{(p)}_{n_{2},k_{2}} \dots N^{(p)}_{n_{a},k_{a}} ~.
\end{align}
This factorization essentially holds for non-generic $p$ too, but multiplicities need to be adjusted to take into account the factorization \eqref{eq:3.4} and the
more complicated structure of the Jordan blocks.

\subsubsection{An all genus example}
\label{sec:an-all-genus}
As an example with composite characteristic polynomials, we consider an infinite family of fixed point indexed by genus $g$. This fixed point is a genus $g$ surface obtained by identifying the diagonally opposite sides of a regular $(4g+2)$-sided polygon in hyperbolic space. This surface has a $\mathbb{Z}_{4g+2}$
isometry corresponding to rotation by $\frac{2\pi}{4g+2}$. The non-zero entries for the matrix $F$ for the corresponding transformation take the form (in a basis with non-standard
intersection form, c.f. \cite{Bashmakov:2022jtl}),
\begin{align}
  \label{eq:3.22}
  F_{i+1,i} = 1 ~,\hspace{0.3in} F_{2g,i}  = (-1)^{i}~, \hspace{0.5 in}0 < i < 2g~.
\end{align}
The characteristic polynomial of $F$ can be evaluated using,
\begin{align}
  \label{eq:3.23}
  \det(X) = \sum_{\pi \in S_{n}} \sign(\pi) \prod_{i=1}^{n} X_{\pi(i),i} ~
\end{align}
where $X$ is an $n \times n$ matrix and $S_{n}$ is the group of permutations of $n$ objects. Since $F$ has few non-zero entries, it is possible to enumerate
all the permutations $\pi$ that contribute to $\det(x-F)$. We do so by identifying permutations contributing to a particular power of $x$. For $x^{2n}$, the
permutations that contribute to its coefficient are those that fix precisely $2n$ of the $2g$ objects. The rest of the $2n-2g$ objects are permuted among themselves.
For such a permutation to have a non-zero contribution we must have $\pi(i) = i+1$ for all those $2n-2g$ objects, except for $i = 2g$. Completing the cycle then requires $\pi(2g) = 2g-2n+1$. This is an odd permutation, and thus the coefficient is $1$.

Next  consider the odd powers of $x$. For $x^{2n-1}$, the only
permutation that contributes is the identity permutation since the diagonal element $F_{2g,2g} = 1$, and hence the coefficient is $-1$. For other odd powers
$x^{2n+1}$ the only permutation that contributes is again the one consisting of the cycle $\pi(i) = i+1$ for $i \ge 2g-2n$ and now $\pi(2g) = 2g-2n$. This is an even
permutation and the coefficient  is $-1$. Hence
\begin{align}
  \label{eq:3.24}
  P_{F}(x) = x^{2g} - x^{2g-1} + x^{2g-2} + \dots + x^{2} - x + 1 ~.
\end{align}
We can mod out by $\mathbb{Z}_{2}$ by considering $F^{2}$ as the generator of $\mathbb{Z}_{2g+1}$. Since we know the eigenvalues from the characteristic polynomial, it is
straightforward to obtain that,
\begin{align}
  \label{eq:3.25}
  P_{F^{2}}(x) = x^{2g} + x^{2g-1} + \dots + 1 = \frac{x^{2g+1} - 1}{x-1} ~.
\end{align}
Hence it is easy to obtain the expansion of $P_{F^{2}(x)}$ in terms of cyclotomic polynomials,
\begin{align}
  \label{eq:3.26}
  P_{F^{2}}(x) = \prod_{m \neq 1, m | 2g+1}\phi_{m}(x) ~.
\end{align}

We can now easily determine when the polynomial $P_{F^{2}(x)}$ splits completely over $\mathbb{F}_{p}$ when $p$ is not a divisor of $2g+1$. This requires that
for any divisor $m$ of $2g+1$, $p = 2m+1$. Requiring that this happens simultaneously for all the divisors means that $p \in (2g+1) \mathbb{N}+1$ and in this case there are
$2^{2g}$ invariant global forms. We can also easily find a sufficient criteria for the existence of intrinsic non-invertible symmetry. The group $\mathbb{Z}_{n}^{\times}$ is
cyclic if $n$ is a prime and there are $\varphi(n-1)$ generators in this group. Hence if $p \in (2g+1) \mathbb{N} + k$ where $k$ is one of the $\varphi(2g)$ generators of $\mathbb{Z}_{n}^{\times}$
then the symmetry generated by $M$ is intrinsically non-invertible. However whenever $p$ is not in one of these cases, i.e. when $\phi_{2g+1}(x)$ is not irreducible or
split over $\mathbb{F}_{p}$, then we need the details of how it factors to decide whether the symmetry is invertible or not.

The next simplest case is when $2g+1 = q^{k}$ with $q$ a prime. Here the sufficient condition for the invariant global forms to exist is that $p \in q^{k} \mathbb{N} + 1$.
Since $\phi_{q}(x)$ is a factor of $P_{F^{2}}(x)$ we have intrinsic non-invertible symmetry whenever $p \in (2g+1)\mathbb{N} + k$ where $k$ is one of $\varphi(q-1)$ generators of
$\mathbb{Z}_{q}^{\times}$.

For the most general case we can write down a similar pair of sufficient conditions for complete splitting and irreducibility of
$P_{F}(x^{2})$ when $2g+1 = q_{1}^{k_{1}} \dots q_{m}^{k_{m}}$. When $p = 2l (2g+1)+1$, then $2^{2g}$ invariant forms always exist. Similarly, if any of the
cyclotomic polynomials appearing in \eqref{eq:3.26} are irreducible over $\mathbb{F}_{p}$ then the symmetry is intrinsically non-invertible. So if
$p \in (2g+1)\mathbb{N} + k$ where $k$ generates one of $\mathbb{Z}_{q_{i}}^{\times}$, then the symmetry is intrinsically non-invertible.

\section{Basics of handlebodies}
\label{app:handlebodies}
In this appendix we review some basic facts about handlebodies which will be used in Section \ref{sec:topops}. These facts can be found in many standard math texts, such as \cite{rolfsen2003knots}. 

\subsection{Solid tori}

Handlebodies are a class of three manifolds obtained by ``filling in'' a genus $g$ Riemann surface (for simplicity we will assume no punctures in this appendix). The simplest case is filling in a torus to obtain a solid torus, homeomorphic to $S^1 \times D^2$.\footnote{Not every three-manifold with torus boundary is a solid torus. Indeed, for any knot $K$ we could consider the knot exterior in $S^3$. For any $K$ besides the unknot this will not be a solid torus, but rather a ``cube with knotted hole''.} More concretely, given a torus $\Sigma_{1,0}$, we may  construct a solid torus $V_{1,0}$ by choosing any curve $\mu\in H_1(\Sigma_{1,0},\ZZ)$ on $\Sigma_{1,0}$ to be the ``meridian'', i.e. the curve which becomes contractible in the handlebody. The remaining generator of $ H_1(\Sigma_{1,0},\ZZ)$, which remains as a non-trivial generator of $H_1(V_{1,0},\ZZ)$, is referred to as the ``longitude'', and will be denoted by $\lambda$. 

Meridians and longitudes are not on equal footing: all meridians of $V_{1,0}$ are ambiently isotopic, and hence it makes sense to discuss \textit{the} meridian of $V_{1,0}$, but there are infinitely many ambient isotopy classes of longitudes.\footnote{Any two longitudes are related by homeomorphism, though.} These statements can be better understood by first introducing the following non-trivial theorem:

\begin{theorem}
The set of isotopy classes $[\gamma]$ of non-trivial closed curves $\gamma$ on $\Sigma_{1,0}$ is in bijection with ordered pairs of coprime integers $(m,n)$, such that $[\gamma]$ corresponds to $(m,n)$ if and only if $\gamma$ is homologous to $m \mu + n \lambda$.
\end{theorem}

Since $\mu$ is trivial in $\pi_1(V_{1,0})$, whereas $\lambda$ generates $\pi_1(V_{1,0})$, a curve $\gamma$ is homotopically trivial in $V_{1,0}$ if and only if it is labelled by $(m,n) = (\pm 1 ,0)$. Thus all homotopically trivial curves in $V_{1,0}$, i.e. all meridians, are in a single isotopy class (modulo change of orientation). 

On the other hand, a curve $\gamma$ labelled by $(m,n)$ represents a generator of $H_1(V_{1,0})$, and hence a longitude, if and only if $n = \pm 1$. This means that any curve labelled by $(m,\pm 1)$ is a longitude. There are thus an infinite number of them, labelled by integers $m$. Note that each $(m,\pm 1)$ can be obtained from the curve $(m,\pm 1)$ and doing $m$ Dehn twists around the meridian. 

\subsection{Higher genus handlebodies}
We may now proceed to the case of more general handlebodies $V_{g,0}$. In order to specify the handlebody, we must again choose a set of $g$ generators of $H_1(\Sigma_{g,0},\ZZ)$ to be meridians, i.e. to be trivialized in $H_1(V_{g,0},\ZZ)$. Given a meridian $\mu_1$, note that any $\gamma$ intersecting $\mu_1$ at a single point \textit{must} be a longitude. Indeed, we have the following, 

\begin{proposition}
Given two curves $\gamma_1$ and $\gamma_2$ with labels $(m_1,n_1)$ and $(m_2, n_2)$, the algebraic intersection number is given by 
\bea
\langle \gamma_1, \gamma_2 \rangle = m_1 n_2 - n_1 m_2~. 
\eea
\end{proposition}

This proposition implies that if $\langle\gamma_1, \gamma_2 \rangle = \pm 1$, then we have $m_1 n_2 - n_1 m_2 = \pm 1$. Now if we assume that $\gamma_1$ is a meridian, we must have $(m_1, n_1) = (\pm 1,0)$, and hence we conclude that $n_2 = \pm 1$. This means that $\gamma_2$ is indeed a longitude. 

In the current context, what this tells us is that when we choose our meridians, we must choose them such that they have zero intersection pairing. In other words, the choice of meridians is given by a choice of maximal isotropic sublattice $L \subset H_1(\Sigma_{g,0},\ZZ)$. 

As we discussed above for solid tori, the set of meridians $L$ is a well-defined property of the handlebody. On the other hand, there are many choices for the set of longitudes. Each choice is related by shifts by the meridians. The invariant quantity associated to the handlebody is thus not the set of longitudes themselves, but the coset $L^\perp = H_1(\Sigma_{g,0},\ZZ)/L$.

\section{Alternative expression for $\cC_F(X_4)$}
\label{app:altCF}

In this appendix we give details on the more streamlined expression for $\cC_F(X_4)$ given in (\ref{eq:B.7.3}). The basic idea is to replace the $2g$ two-cycles $M_2^i$ valued in $\mathbb{Z}_{p}$ with a single two-cycle $M_2$ valued in $\mathbb{Z}_{p}^{2g}$. Another way of saying the same thing is we consider
a single cycle in $H_{2}(X_{4} , \mathbb{Z}_{p}^{2g})$ instead of $2g$ cycles in $H_{2}(X_{4} , \mathbb{Z}_{p})$. This also makes direct contact with the 6d picture since,\footnote{As in the main text, we assume $H_{1,3}(X_4, \ZZ) = 0$. }
\begin{align}
  \label{eq:B.7.1}
  H_{3}(\Sigma_{g,0} \times X_{4} , \mathbb{Z}_{p}) \cong H_{2}(X_{4} , H_{1}(\Sigma_{g,0} , \mathbb{Z}_{p})) \cong  H_{2}(X_{4} , \mathbb{Z}_{p}^{2g}) ~.
\end{align}
In this picture given $H_{2}(X_{4} , \mathbb{Z}_{p}^{2g})$ we can define the antisymmetric pairing $\ev{\cdot,\cdot}$ which is the combination of the symmetric intersection
pairing on homology with the antisymmetric Dirac paring on $\mathbb{Z}_{p}^{2g}$. In class $\mathcal{S}$ theories, this is the same as the intersection pairing on
$H_{3}(\Sigma_{g,0} , \mathbb{Z}_{p})$ by \eqref{eq:B.7.1}. As a result, the operators $\Phi(M_{2})$ with $M_{2} \in H_{2}(X_{4} , \mathbb{Z}_{p}^{2g})$ satisfy the algebra in (\ref{eq:Heisenbergalg}). There is also a symmetric pairing $(\cdot,\cdot)$ which is the combination of symmetric intersection pairing with the
canonical symmetric pairing on the group $\mathbb{Z}_{p}^{2g}$ of coefficients. The two are related by,
\begin{align}
  \label{eq:B.7.6}
  \ev{M_{2} , M'_{2}} = (M_{2},\mathfrak{I} M'_{2})~.
\end{align}

We would like to construct a defect $\cC_{F}(X_4)$ that acts on $\Phi(M_{2})$ with $M_{2} \in H_{2}(X_{4} , \mathbb{Z}_{p}^{2g})$ as
\begin{align}
  \label{eq:B.7.2}
  \cC_{F}(X_4):\,\,\, \Phi(M_{2}) \mapsto \Phi(FM_{2}) ~,\hspace{0.5 in} M_2 \subset X_4~.
\end{align}
This is equivalent to the fusion rule,
\begin{align}
  \label{eq:B.7.4}
  \Phi(M_{2}) \times \cC_{F}(X_4) = \cC_{F}(X_4) \times \Phi(F M_{2}) ~,\hspace{0.5 in} M_2 \subset X_4~.
\end{align}
In the equations above $F$ is a $2g \times 2g$ matrix with coefficients in $\mathbb{Z}_{p}$, and $FM_{2}$ represents an automorphism of the coefficient system of $H_{2}(X_{4},\mathbb{Z}_{p}^{2g})$.

The most general form $\cC_{F}(X_4)$ can take is
\begin{align}
  \label{eq:B.7.3}
  \cC_{F}(X_4) = {|H^0(X_4, \ZZ_p^{2g})| \over |H^1(X_4, \ZZ_p^{2g})| }\sum_{M_{2} \in H_{2}(X_{4} , \mathbb{Z}_{p}^{2g})} \exp(\frac{2\pi i}{2p}\ev{ M_{2},M_{2} }_{F}) \Phi((\mathds{1}-F)M_{2}) ~,
\end{align}
where $\ev{\cdot,\cdot}_{F}$ is a bilinear form which we will determine by demanding that the fusion rules given by \eqref{eq:B.7.4} are satisfied. From \eqref{eq:B.7.3}, it is
clear that only the symmetric part of it contributes. The reason that we have included a factor of $(\mathds{1}-F)$ in $\Phi((\mathds{1}-F)M_{2})$ is to implement folding trick, as discussed around Figure \ref{fig:defectpics}. Another way to see that it should be there is to notice that if $FM_{2} = M_{2}$ then the
defect $\cC_{F}(X_4)$ should be transparent to $M_{2}$, and hence $\Phi(M_{2})$ should drop out of the condensate. 

We now fix the bilinear pairing $\ev{\cdot,\cdot}_{F}$ by evaluating $\Phi(\widetilde{M_{2}}) \times \cC_{F}(X_4)$, which gives 
\beaa
  \label{eq:B.7.5}
  {|H^1(X_4, \ZZ_p^{2g})|\over |H^0(X_4, \ZZ_p^{2g})|}&&\hspace{-0.25in}\Phi(\widetilde{M_{2}}) \times \cC_{F}(X_4) \nonumber \\
  &\hspace{-1.5in}= &\hspace{-0.8in}\sum_{M_{2} \in H_{2}(X_{4} , \mathbb{Z}_{p}^{2g})} \exp(\frac{2\pi i}{2p}\ev{M_{2},M_{2}}_{F}) \Phi(\widetilde{M_{2}}) \times \Phi((\mathds{1}-F)M_{2})  \nonumber \\
   &\hspace{-1.5in}= &\hspace{-0.8in}\sum_{M_{2} \in H_{2}(X_{4} , \mathbb{Z}_{p}^{2g})} \exp(\frac{2\pi i}{2p}\left(\ev{M_{2},M_{2}}_{F} + \ev{\widetilde{M_{2}} , (\mathds{1}-F)M_{2}}\right))\Phi(\widetilde{M_{2}} + (\mathds{1}-F)M_{2})  \nonumber \\
   &\hspace{-1.5in}= &\hspace{-0.8in}\sum_{M_{2} \in H_{2}(X_{4} , \mathbb{Z}_{p}^{2g})} \exp(\frac{2\pi i}{2p}\left(\ev{M_{2},M_{2}}_{F} + \ev{\widetilde{M_{2}} ,(\mathds{1}-F) M_{2}}\right))\nonumber \\
   &\hspace{-1.5in} \vphantom{.}&\hspace{1.2in} \times\Phi((\mathds{1}-F) \widetilde{M_{2}} + (\mathds{1}-F)M_{2} + F\widetilde{M_{2}})  \nonumber \\
                   &\hspace{-1.5in}= &\hspace{-0.8in} \sum_{M_{2} \in H_{2}(X_{4} , \mathbb{Z}_{p}^{2g})} \mathrm{exp}\left(\frac{2\pi i}{2p}\left(\ev{M_{2},M_{2}}_{F} + \ev{\widetilde{M_{2}} ,(\mathds{1}-F) M_{2}} -  \ev{(\mathds{1}-F)(\widetilde{M_{2}} + M_{2}) , F\widetilde{M_{2}}}\right)\right)\no\\
              % &\hspace{-1.5in} \vphantom{.}&\hspace{1in}   \left.\left. - \nonumber \\
                 & \hspace{-1.5in} \vphantom{.}&\hspace{1.2 in}\times \Phi((\mathds{1}-F) (\widetilde{M_{2}} + M_{2})) \times \Phi(F\widetilde{M_{2}}) \nonumber \\
                   &\hspace{-1.5in}= &\hspace{-0.8in} \sum_{M_{2} \in H_{2}(X_{4} , \mathbb{Z}_{p}^{2g})} \exp\left(\frac{2\pi i}{2p}\left(\ev{M_{2} - \widetilde{M_{2}},M_{2} - \widetilde{M_{2}}}_{F} +\ev{\widetilde{M_{2}} , (\mathds{1}-F)(M_{2} - \widetilde{M_{2}})} \right.\right. \nonumber \\
                   &\hspace{-1.5in} \vphantom{.}&\hspace{0.7 in} -\ev{(\mathds{1}-F)M_{2} , F\widetilde{M_{2}}}\Big)\bigg) \Phi((\mathds{1}-F)M_{2}) \times \Phi(F\widetilde{M_{2}}) ~. \no
\eeaa
We now demand that the left-hand side is proportional to $ \cC_{F}(X_4) \times \Phi(F\widetilde{M_{2}})$. If we assume that $\mathrm{ker}(\mathds{1}-F)$ is trivial, then this simply amounts to the requirement that 
\begin{align}
  \label{eq:B.7.7}
  \ev{M_{2} - \widetilde{M_{2}},M_{2} - \widetilde{M_{2}}}_{F} + \ev{\widetilde{M_{2} }, (\mathds{1}-F)(M_{2} - \widetilde{M_{2}})} - \ev{(\mathds{1}-F)M_{2} , F\widetilde{M_{2}}} = \ev{M_{2},M_{2}}_{F} ~.
\end{align}
To solve for $\ev{\cdot, \cdot}_{F}$ we first consider the case $\widetilde{M_{2}} = M_{2}$. This gives
\begin{align}
  \label{eq:B.7.8}
  \ev{\widetilde{M_{2}} , \widetilde{M_{2}}}_{F} &= -\ev{\widetilde{M_{2}} , F \widetilde{M_{2}}} = \ev{F\widetilde{M_{2}} , \widetilde{M_{2}}}~,
\end{align}
which in turn motivates
\begin{align}
  \label{eq:B.7.9}
  \ev{M_{2} , \widetilde{M_{2}}}_{F} := \ev{F M_{2} , \widetilde{M_{2}}}~.
\end{align}
This pairing is not symmetric, but only the symmetric part contributes to \eqref{eq:B.7.3}. 
By making use of (\ref{eq:B.7.6}) and (\ref{eq:Falphabetarel}), it can be shown that this definition matches that appearing in (\ref{eq:CFfinalans}) and (\ref{eq:RFij}).

Inserting this definition into the more general \eqref{eq:B.7.7}, we see that this is satisfied only if
\begin{align}
  \label{eq:B.7.10}
  \ev{FM_{2},F\widetilde{M_{2}}} &= \ev{M_{2} , \widetilde{M_{2}}} ~,
\end{align}
i.e if $F$ is a symplectic matrix. It can also be checked that $\cC_{F}(X_4) \times \overline{\cC_{F}}(X_4)= \Phi(0) = \mathbbm{1}$ if we define $\overline{\cC_{F}}(X_4) = \chi(X_4, \ZZ_p)^{-2g}\, \cC_F(X_4)^\dagger$. 

More care is required in the case that $\mathrm{ker}(\mathds{1}-F)$ is nontrivial. There are two cases to be considered here. First, consider the case in which $\mathds{1}-F$ has no non-trivial Jordan blocks with eigenvalue 1. In this case we have the splitting 
\bea
\label{eq:H2splittingimp}
H_2(X_4, \ZZ_p^{2g}) = \mathrm{ker}(\mathds{1}-F) \oplus \mathrm{im}(\mathds{1}-F)~,
\eea
and we see that requiring $ \Phi(\widetilde{M_{2}})\times  \cC_{F}(X_4) =  \cC_{F}(X_4) \times \Phi(F\widetilde{M_{2}})$ no longer imposes (\ref{eq:B.7.7}), but rather the seemingly looser requirement 
\beaa
\sum_{N_2 \in \mathrm{ker}(1-F)} e^{{2 \pi i \over 2 p} \langle M_2 + N_2, M_2 + N_2 \rangle_F} &=& \sum_{N_2 \in \mathrm{ker}(\mathds{1}-F)} e^{{2 \pi i \over 2 p}  \ev{M_{2} +N_2- \widetilde{M_{2}},M_{2}+N_2 - \widetilde{M_{2}}}_{F}}
\\
&\vphantom{.}&\hspace{0.8 in}\times\, e^{{2 \pi i \over 2 p}\left( \ev{\widetilde{M_{2} }, (\mathds{1}-F)(M_{2} - \widetilde{M_{2}})} - \ev{(\mathds{1}-F)M_{2} , F\widetilde{M_{2}}} \right)}~,
\no
\eeaa
obtained by comparing the coefficients of $\Phi((\mathds{1}-F)M_2)$ on both sides. Similar to before, in order to determine the pairing $\langle \cdot, \cdot \rangle_F$ we first restrict to $\widetilde{M_2} = M_2 + N_2$, in which case we obtain 
\beaa
\sum_{N_2 \in \mathrm{ker}(\mathds{1}-F)} e^{{2 \pi i \over 2 p} \langle M_2 + N_2, M_2 + N_2 \rangle_F} &=& e^{{2\pi i \over 2 p}\langle FM_2, M_2\rangle} \sum_{N_2 \in \mathrm{ker}(\mathds{1}-F)} e^{{2\pi i \over 2 p} \langle N_2, (\mathds{1}-F)M_2 \rangle}
\no\\
&=& |\mathrm{ker}(\mathds{1}-F)|\,e^{{2\pi i \over 2 p}\langle FM_2, M_2\rangle}
\eeaa
where we have used the fact that any $M_2$ satisfies $ \langle N_2, (\mathds{1}-F)M_2 \rangle = 0$. Thus in this case we obtain the form of the defect 
\bea
 \cC_{F}(X_4) = {|H^0(X_4, \ZZ_p^{2g})| \over |H^1(X_4, \ZZ_p^{2g})| } |\mathrm{ker}(\mathds{1}-F)|\sum_{M_{2} \in \mathrm{im}(\mathds{1}-F)} \exp(\frac{2\pi i}{2p}\ev{F M_{2},M_{2} }) \Phi((\mathds{1}-F)M_{2}) ~.
\eea

On the other hand, if there exists a non-trivial Jordan blocks with eigenvalue 1, then the splitting in (\ref{eq:H2splittingimp}) is no longer possible, and the above manipulations are no longer valid.\footnote{We can see that the splitting fails in the case of a non-trivial Jordan block of size $k>1$ by considering the vector $v^{(k-1)}$ in the notation of
\eqref{eq:B.41}, which cannot be written as a sum of an element in the image and an element of the kernel. } In that case, an alternative way to proceed (which works as long as $p \neq 2$) is to work in a basis where $F$ is given by
\begin{align}
  \label{eq:B.7.14}
  F =
  \begin{pmatrix}
    F' & 0 \\
    0  & \widetilde{F}
  \end{pmatrix}
\end{align}
where $F'$ is a block containing all Jordan blocks of eigenvalue $1$ and $\widetilde{F}$ contains the rest of the blocks. Then instead of starting with $F$, we start with
the matrices,
\begin{align}
  F^{-} :=
  \label{eq:B.7.15}
  \begin{pmatrix}
    -F'& 0 \\
    0  & \widetilde{F}
  \end{pmatrix} ~,
       &&
          I^{-} := \begin{pmatrix}
                      - \mathbbm{1} & 0 \\
                      0  & \mathbbm{1}
                    \end{pmatrix}~.
\end{align}
Since $F^{-}$ and $I^{-}$ both have no non-trivial Jordan blocks of eigenvalue $1$, we can write down the defects for both of them. We then define the defect for $F$ in a roundabout manner as\footnote{When $p=2$ this trick no longer works. For a discussion of this case in the example of $g=1$ and $F = \sfS$, see \cite{Kaidi:2022cpf}. } 
\begin{align}
  \label{eq:B.7.16}
  \mathcal{C}_{F}(X_4) := \mathcal{C}_{F^{-}}(X_4) \times \mathcal{C}_{I^{-}}(X_4) ~.
\end{align}

We note that in practice this is not an irrelevant subtlety: non-trivial Jordan blocks of eigenvalue $1$ appear in our discussion whenever the characteristic polynomial of $F$ is $\phi_{p}(x)$ and the finite field we are working over is $\mathbb{Z}_{p}$. A concrete example is the case of the $\ZZ_5 \subset \ZZ_{10}$ symmetry appearing at genus 2, discussed in Section \ref{eq:genus2invglob}, when $p=5$.

\end{appendix}

\bibliographystyle{ytphys}
\bibliography{dualitydefects2.bib}

\end{document}